\documentclass[aps,superscriptaddress, showpacs,preprintnumbers, superscriptaddress, nofootinbibt,twocolumn]{revtex4-2}
\usepackage{eurosym}
\usepackage{dcolumn}
\usepackage{bm}
\usepackage{enumerate}
\usepackage{float}
\usepackage{epstopdf}
\usepackage{amsmath}
\usepackage{bm}
\usepackage{amsfonts}
\usepackage{amssymb}
\usepackage{graphicx}
\usepackage{alphalph,mathtools}
\usepackage{etoolbox}
\usepackage{color}

\def\be{\begin{equation}}
\def\ee{\end{equation}}
\def\bea{\begin{eqnarray}}
\def\eea{\end{eqnarray}}

\newcommand{\R}{\ensuremath{\mathbb{R}}}
\newcommand{\de}{\mathrm{d}}

\begin{document}

\title{Cosmological evolution and dark energy in osculating Barthel-Randers geometry}
\author{Rattanasak Hama}
\email{rattanasak.h@psu.ac.th}
\affiliation{Faculty of Science and Industrial Technology, Prince of Songkla University, Surat Thani Campus, Surat Thani, 84000, Thailand,}
\author{Tiberiu Harko}
\email{tiberiu.harko@aira.astro.ro}
\affiliation{Department of Theoretical Physics, National Institute of Physics
and Nuclear Engineering (IFIN-HH), Bucharest, 077125 Romania,}
\affiliation{Astronomical Observatory, 19 Ciresilor Street,
	Cluj-Napoca 400487, Romania,}
\affiliation{Department of Physics, Babes-Bolyai University, Kogalniceanu Street,
	Cluj-Napoca, 400084, Romania,}
\affiliation{School of Physics, Sun Yat-Sen University, Guangzhou, 510275, People's
	Republic of China,}
\author{Sorin V. Sabau}
\email{sorin@tokai.ac.jp}
\affiliation{School of Biological Sciences, Department of Biology, Tokai University, Sapporo 005-8600, Japan,}
\affiliation{Graduate School of Science and Technology, Physical and Mathematical Sciences, Tokai University, Sapporo 005-8600, Japan,}
\author{Shahab Shahidi}
\email{s.shahidi@du.ac.ir}
\affiliation{School of Physics, Damghan University, Damghan, 41167-36716, Iran}

\begin{abstract}
We consider the cosmological evolution in an osculating point Barthel-Randers type geometry, in which to each point of the
space-time manifold an arbitrary point vector field is associated. This Finsler type geometry is assumed to describe the physical
properties of the gravitational field, as well as the cosmological dynamics. For the Barthel-Randers geometry the connection is given by the Levi-Civita connection of the associated Riemann metric. The generalized Friedmann equations in the Barthel-Randers geometry are obtained by considering that the background Riemannian metric in the Randers line element is of Friedmann-Lemaitre-Robertson-Walker type. The matter energy balance equation is derived, and it is interpreted from the point of view of the thermodynamics of irreversible processes in the presence of particle creation. The cosmological properties of the model are investigated in detail, and it is shown that the model admits a de Sitter type solution, and that an effective cosmological constant can also be generated. Several exact cosmological solutions are also obtained. A comparison of three specific models with the observational data and with the standard $\Lambda$CDM model  is also performed by fitting the observed values of the Hubble parameter, with the models giving a satisfactory description of the observations.
\end{abstract}

\date{\today }
\maketitle
\tableofcontents


\section{Introduction}

General relativity, one of the most successful physical theories ever proposed, was born as a result of the fruitful interaction between physics, represented by the equivalence principle, and the deformation of the Minkowski metric, and mathematics, represented by the differential geometric theory of the Riemann spaces, introduced, and initially developed in \cite{r1}.  In Riemannian geometry the properties of the spacetime are described via a metric tensor $g_{\mu \nu}$, and an affine connection $\Gamma _{\mu \nu}^{\lambda}$, which in turn is determined by the metric tensor. From these quantities one can construct the Riemann curvature tensor $R_{\alpha \beta \gamma}^{\delta}$, as well as its contractions $R_{\alpha \beta \gamma}^{\beta}$, and $R=R_{\alpha }^{\alpha}$, and it turns out that the gravitational properties of the spacetime are determined by the Einstein tensor $G_{\mu \nu}$, and the gravitational field equations $G_{\mu \nu}=R_{\mu \nu}-(1/2)Rg_{\mu \nu}=\kappa ^2 T_{\mu \nu}$, where $T_{\mu \nu}$ is the matter energy-momentum tensor, $\kappa ^2 =8\pi G/c^4$ is the gravitational coupling constant \cite{r2,r3,r4}. The Einstein gravitational field equations can also be derived from the Hilbert-Einstein variational principle by varying the action $S=\int{\left(R/2\kappa +L_m\right)\sqrt{-g}d^4x}$, where $L_m$ is the matter Lagrangian, with respect to the metric tensor \cite{r3}. In Riemann geometry the metric tensor satisfies the condition $\nabla _\lambda g_{\mu \nu}=0$, where $\nabla _{\lambda}$ denotes the covariant derivative with respect to the Levi-Civita affine connection. General relativity did have a tremendous impact on the development of physics, astrophysics, and cosmology, leading to a completely new understanding of the gravitational interaction, and providing  natural and unified explanations to such different phenomena as the precession of the perihelion of planet Mercury, the deflection of light by the Sun, or the expansion of the Universe \cite{Gron}.

Almost immediately of the proposal of general relativity, inspired by the possibility of unifying gravity and electromagnetism, Weyl proposed an extension of the Riemannian geometry,  in which the covariant derivative of the metric does not vanish, and it is given by $\nabla _{\lambda}g_{\mu \nu}=Q_{\lambda \mu \nu}$, where $Q_{\lambda \mu \nu}$ denotes the nonmetricity of the space-time \cite{r5}. The Weyl geometry represented the basic framework for the first unified field theory, and, despite its sharp criticism by Einstein, and initial rejection in the scientific community, it become an important field of study. In particular, some physical extensions of the theory were proposed by Dirac \cite{W1,W2}, in which a scalar field is introduced together with the tensor electromagnetic tensor $F_{\mu \nu}$ coming from Weyl geometry. Recently, the physical implications of Weyl gravity were considered, from the elementary particle physics point of view, in \cite{Gh1,Gh2,Gh3,Gh4,Gh5,Gh6,Gh7}. In particular, in \cite{Gh1} it was showed that the gauged Weyl gravity action, quadratic in the scalar curvature and in the Weyl tensor ${\tilde{C}}_{\mu \nu  \rho \sigma}$ of the Weyl conformal geometry has spontaneous symmetry breaking in which the Weyl gauge field $\omega _{\mu}$ becomes massive. Hence, from Weyl conformal gravity one can recover the Einstein-Hilbert action with a positive cosmological constant, and the Proca action for the massive Weyl gauge field.

An interesting approach based on Weyl geometry is the so-called symmetric teleparallel gravity approach, in which the gravitational field is described by the nonmetricity alone \cite{Nester}. This approach was extended in the form of the $f(Q)$ theory in \cite{Q1}, where $Q$ is the nonmetricity scalar, and further analyzed and extended in \cite{Q2,Q3,Q4,Q5,Q6,Q7, Q8,Q9}.

 A few years after the introduction of Weyl geometry the concept of torsion was introduced by Elie Cartan \cite{r6}, leading to another important developments in both differential geometry, and general relativity. The extension of general relativity including torsion is called
the Einstein-Cartan theory \cite{r7,r8,r9}, and in this theory, from a physical point
of view  the torsion field $T_{\mu \nu}^{\lambda}\neq 0$ is interpreted as the spin density of the matter \cite{r10}.

An important development of the gravitational field theories took place through the applications of the mathematical results that appeared
in the work of Weitzenb\"{o}ck \cite{r11}, who introduced the geometries presently known as the
Weitzenb\"{o}ck spaces, characterized  by the properties $\nabla_ {\alpha }g_{\mu \nu}=0 $, $T_{\mu \nu}^{\alpha}\neq 0$, and $R_{\alpha \beta \gamma}^{\lambda}= 0$, respectively.  In a Weitzenb\"{o}ck space the Riemann curvature tensor identically vanishes, and therefore these spaces have the property of distant parallelism, also known as teleparallelism or absolute parallelism.  Einstein was the first to use Weitzenb\"{o}ck type teleparallel geometries for developing a unified theory of electromagnetism and gravitation \cite{r12}.  This approach was further generalized to the so-called teleparallel equivalent of General Relativity (TEGR) \cite{r13,r14,r15}, also known as the $f(\cal{T})$ gravity theory, where $\cal{T}$ is the torsion scalar. Weitzenb\"{o}ck geometries in the presence of nonmetricity, as well as their physical implications, were studied in \cite{r16,r17}. In the framework of Riemannian geometry several extensions of general relativity involving geometry-matter coupling were proposed and investigated in \cite{e1,e2,e3,e4,e5,e6,e7}. For a detailed review of the theories with nonminimal couplings between matter and geometry see \cite{e8} and \cite{e9}, respectively.

In the same year Weyl proposed his extension of Riemann geometry, another important mathematical development took place through the publication of Paul Finsler's dissertation \cite{F1}, in which a new class of differential geometric objects was introduced. For short, Finsler geometry can be defined as "... just Riemannian geometry without the quadratic restriction" \cite{F2}. In fact, Finsler geometry was anticipated by Riemann \cite{r1}, who introduced in a general space a metric structure defined as $ds=F\left(x^1,...,x^n; dx^1,...,dx^n\right)=F(x,dx)$, where for $y\neq 0$, $F(x, y)$ is a positive function
on the tangent bundle $TM$, and is homogeneous of degree one in $y$, $F(x,\lambda dx)=\lambda F(x,dx)$. When $F^2=g_{ij}(x)dx^idx^j$ we obtain the important case of the Riemann geometry \cite{F2}. The Finsler metric function $F$ can be written in terms of the canonical coordinates of
the tangent bundle $(x, y) = \left(x^I,y^I\right)$, where $y = y^I\left(\partial /\partial x^I\right)$ is a tangent vector at $x$. Hence in a general Finsler space we can write the arc element as $ds^2=g_{IJ}dx^Idx^J$. In a general Finsler space one can define three kinds of curvature tensors $\left(R_{\nu \lambda \mu}^{\kappa}, S_{\nu \lambda \mu}^{\kappa},P_{\nu \lambda \mu}^{\kappa} \right)$, and five torsion tensors \cite{Bao}, indicating a much richer mathematical structure as compared to Riemannian geometry.

Even that Finsler geometry attracted the interest of mathematicians quite early \cite{F3,F4,F5}, and it is now a well-established and important field of research in mathematics, the physical applications of the theory did develop at a much slower pace. A unified theory of gravity and electromagnetism, proposed by Randers \cite{Rand} within the framework of five dimensional general relativity provided an example of a Finsler geometry, with $ds=\left(\alpha +\beta\right)du$, with $\alpha =\left[g_{ik}(x)y^iy^k\right]^{1/2}$, and $\beta =b_i(x)y^i$, with $y^i=dx^i/du$. The nonlocal field theory proposed by Yukawa \cite{Yuk1,Yuk2}, in which two sets of space-time coordinates
$\left(x^{\kappa}, x^{\lambda}\right)$, $\kappa, \lambda  = 1,2,3,4$ are adopted as the independent variables, can also be investigated by using the methods of Finsler geometry \cite{Hor,Tak, Ik1}.  Recently, a Finslerian type geometrization of the hydrodynamical formulation of quantum mechanics was introduced in \cite{Quant}.  Extension of the Randers-Finsler geometry  to the Nambu-Goto action were considered in \cite{string}.

One of the first attempts to construct a relativistic theory of gravitation by using Finsler geometry is the work by Horv\'{a}th \cite {Hor1}, and Horv\'{a}th  and Mo\'{o}r \cite{Hor2}, which was later extended in \cite{Tak1} and \cite{Tak2}, respectively. The Finslerian type field equations proposed in these works can be written as
\be\label{h1}
R_{\mu \nu}-\frac{1}{2}g_{\mu \nu}R+\lambda g_{\mu \nu}=\chi T_{\mu \nu},
\ee
\be\label{h2}
K_{\mu \nu}-\frac{1}{2}g_{\mu \nu}K+\lambda g_{\mu \nu}=\chi T_{\mu \nu},
\ee
and
\be\label{h3}
S_{\mu \nu}-\frac{1}{2}g_{\mu \nu}S-\lambda^{(i)} g_{\mu \nu}=-\chi ^{(i)}T_{\mu \nu}^{i},
\ee
respectively, where $R_{\mu \nu}$, $R$, $K_{\mu \nu}$, $K$ are the contracted third curvatures, $S_{\mu \nu}$ and $S$ are the v-Ricci curvature tensor, and the v-scalar curvature, $T_{\mu \nu}$ is the energy-momentum
tensor, $T_{\mu \nu}^{i}$ is the internal energy-momentum
tensor, $\chi$ is the gravitational constant, $\lambda$ is the cosmological constant, and $\lambda ^{(i)}$ and $\chi ^{(i)}$ are the internal cosmological and gravitational constants, respectively. Eq.~(\ref{h3}) describes the $y$-dependence of the metric tensor. A Finslerian extension of general relativity was proposed in \cite{As0},  and examined with particular emphasis on the Finslerian generalization of the equation of motion in a gravitational field. The construction of a gravitational Lagrangian density by substituting the osculating Riemannian metric tensor in the Einstein density was also studied.  For an attempt to present in a systematic way the general relativity principles together with the development of Finsler geometry as a metric generalization of Riemannian geometry, as well as the extensions of general relativity on the basis of Finsler geometry see \cite{As1}. The Finslerian extensions of the Schwarzschild metric were considered in \cite{As2,As3}.

A system of Einstein type field equations was introduced, by adopting the vector bundle point, in \cite{Miron}. In this approach the $y$-field is regarded as a fibre at the point $x$ of the base $x$-field, and the total space of this vector bundle is considered a unified field between the $x$ and $y$ fields \cite{Ikeda}. For this unified field one introduces the adapted frame defined as
\bea
X_A&=&\left(\frac{\delta}{\delta x^{\lambda}}=\frac{\partial}{\partial x^{\lambda}}-N_{\lambda}^i\frac{\partial}{\partial y^i}, \frac{\partial}{\partial y^i}\right),\nonumber\\
X^A&=&\left(dx^{\kappa},\delta y^i=dy^i+N_{\lambda}^idx^{\lambda}\right),
\eea
where the first line is the adapted basis of the tangent
space $T_xM$, while the second line is the adapted cobasis in the cotangent
space $T_x^*M$, $A,B=\left(\kappa,i\right)=0,1,2,3,...,7$, $\lambda, \nu= 0,1,2,3$, and $N_{\lambda}^i$ is the nonlinear connection. The adapted frame is assumed to be adapted to the metric $G\left(=G_{AB}\right)$, given by
\be
G=g_{\lambda \kappa}(x,y)dx^{\kappa}dx^{\lambda}+g_{ij}(x,y)\delta y^i\delta y^j.
\ee
The field equations on the total space are defined as $\mathcal{R}_{AB}-(1/2)\mathcal{R}G_{AB}=\tau _{AB}$, where $\tau _{AB}$ is the energy-momentum tensor, and can be decomposed as
\be
R_{\lambda \nu}-\frac{1}{2}(R+S)g_{\lambda \nu}=\tau _{\lambda \nu}, \overset{1}{P}_{i \lambda}=\tau _{i\lambda},  \overset{2}{P}_{ \lambda i}=-\tau _{\lambda i},
\ee
\be
S_{i j}-\frac{1}{2}(R+S)g_{i j}=\tau _{ij}.
\ee

An interesting perspective of the gravitational field equations was proposed in \cite{Rutz}, where it was assumed that the Einstein vacuum equations in a Finsler geometry are given by $H=H_i^i=0$, where
\be
H_k^i=2\frac{\partial G^i}{\partial x^k}-\frac{\partial ^2G^i}{\partial x^j \partial \dot{x}^k}\dot{x}^j+2\frac{\partial ^2G^i}{\partial \dot{x}^j\dot{x}^k}G^j-\frac{\partial G^i}{\partial \dot{x}^j}\frac{\partial G^j}{\partial \dot{x}^k},
\ee
and $G^l=\gamma _{jk}^l\dot{x}^j\dot{x}^k/2$, with $\gamma _{jk}^l$ denoting the Christoffel symbols of second kind of the Finsler metric. These generalised field equation reduce to the Einstein equations for Riemannian metrics, and also admits non-Riemannian solutions.

In the Berwald-Finsler space a gravitational field equation was introduced in \cite{Lixin} as
\begin{eqnarray}
\hspace{-0.3cm}\left[Ric_{\mu\nu}-\frac{1}{2}g_{\mu\nu}S\right]+\left\{\frac{1}{2}
B^{~\alpha}_{\alpha~\mu\nu}+B^{~\alpha}_{\mu~\nu\alpha}\right\}=8\pi
G T_{\mu\nu} ,
\end{eqnarray}
where $B_{\mu\nu\alpha\beta}=-A_{\mu\nu\lambda}R^{~\lambda}_{\theta~\alpha\beta}y^\theta/F$, and $
A_{\lambda\mu\nu}\equiv\frac{F}{4}\frac{\partial}{\partial y^\lambda}\frac{\partial}{\partial y^\mu}\frac{\partial}{\partial y^\nu}(F^2)$ is the Cartan tensor, that can be regarded as measuring the deviation of the geometry from the Riemannian manifold.

For the vacuum dynamics of gravitational fields  the following field equations have been proposed in \cite{Voicu1},
\be
2R - \frac{L}{3}g^{Lij}R_{\cdot i \cdot j} + \frac{2L}{3}g^{Lij}\left[ (\nabla P_{i})_{\cdot j} + P_{i|j} - P_{i}P_{j}\right]= 0,
\ee
where $L$ is the Finsler-Lagrange function, $g^L_{ij}=\frac{1}{2}\frac{\partial ^{2}L}{\partial \dot{x}^{i}\partial \dot{x}^{j}}=\frac{1}{2} L_{\cdot i\cdot j}$, $R_{.i.j}$ is the geodesic derivation operator, $R$ is its trace, and $P$ is the Landsberg tensor. These field equations can be obtained by varying with respect to $L$ the action $S[L] = \int_{\Sigma\subset TM} \mathrm{vol}(\Sigma) R_{|\Sigma}$, where $\Sigma = \{(x,\dot x)\in TM|F(x,\dot x) = 1\}$ denotes the unit tangent bundle, while $\mathrm{vol}(\Sigma)$ is the volume form on $\Sigma$, obtained by using the Finsler metric. An approach to investigate general relativistic kinetic gas theory by using methods from Finsler geometry was introduced in \cite{Voicu2}.

The recent measurements by the Planck satellite of the temperature fluctuations of the Cosmic Microwave Background Radiation \cite{1g,1h}, as well as the observations of the distant supernovae \cite{Riess} have confirmed that the Universe is in a state of accelerating expansion, and that its matter content consists of only 5\% baryonic matter, while 95\% of matter-energy resides in two mysterious components, called dark energy and dark matter, respectively. To explain the cosmological observations, the $\Lambda$CDM model was introduced, which is essentially based on the introduction in the field equations of the cosmological constant, first postulated by Einstein in 1917 \cite{Ein}. Even that the $\Lambda$CDM paradigm provides and excellent fit to the observational data, its theoretical foundations are problematic due to the lack of a firm theoretical basis, related to the numerous problems raised by $\Lambda$. Hence, in order to obtain a physically and mathematically consistent picture of the Universe, two different approaches have been proposed, called the dark components model, and the dark gravity model, respectively. The dark components model \cite{Rev1,Rev2,Rev3,Rev4, Rev5} assumes that the Universe is filled with two (still mysterious) components, dark energy, and dark matter, respectively, components for which many proposals have been advanced. In the dark gravity approach it is assumed that the nature of the gravitational force changes on astrophysical (galactic) and cosmological scales, and that the standard Einstein equations, so successful at the level of the Solar System, must be replaced by a novel theory of gravity, like, for example, theories with geometry-matter coupling \cite{e1,e4}, or gravitational models built upon more general geometries than the Riemannian one. In this latter direction Finsler type cosmological models represent an attractive possibility of explaining/replacing dark energy, and perhaps even dark matter.

There are many attempts to apply Finsler geometry for understanding the dynamics and evolution of the Universe \cite{Fc1,Fc2,Fc3,Fc4,Fc5,Fc6,Fc7,Fc8,Fc9,Fc10,Fc11,Fc12,Fc13,Fc14,Fc15,Fc16,Fc17,Fc18,Fc19,Fc19a, Fc20,Fc21,Fc22,Fc23, Fc24, Fc24a}.  Many of these approaches use a Finsler-Randers geometry, in which the generalized Friedmann equations are obtained. For example, in \cite{Fc23}, the following generalization of the Friedmann equations in a Randers-Finsler geometry was proposed,
\be
\dot{H}+H^2+\frac{3}{4}HZ_t=-\frac{4\pi G}{3}\left(\rho +3p\right),
\ee
\be
\dot{H}+3H^2+\frac{11}{4}HZ_t=4\pi G\left(\rho -p\right),
\ee
where $H$, $\rho$, $p$ denotes the Hubble function, energy density and pressure, respectively, and $Z_t=\dot{u}_0$, where $u_0$ is the time component of the four-velocity $u_{\mu}$. From the above equations one can obtain the relation $3H^2+3HZ_t=8\pi G \rho$. The new term $HZ_t$, coming from the Finsler-Randers geometry, can induce new phases in the cosmological history of the Universe. The same generalized Friedmann equations were used in \cite{Fc20} to investigate particle creation processes due to the Finslerian structure of the space-time.

The cosmological implications of scalar-tensor theories that arise effectively from the Lorentz fiber bundle of a Finsler-like geometry were investigated in \cite{Fc19a}. The considered action in the presence of matter is
\begin{equation}
\mathcal S = \frac{1}{16\pi G} \int \sqrt{|\det\mathbf G|}\,\mathcal L_G dx^{(N)}+ \int
\sqrt{|\det\mathbf G|}\,\mathcal L_M dx^{(N)},
\end{equation}
where $dx^{(N)} = d^4x\wedge\de\phi^{(1)}\wedge\de\phi^{(2)}$. Several Lagrangian densities were considered,  with
$\mathcal L_G = \mathcal{\tilde R} - \frac{1}{\phi}V(\phi)$,
where $V(\phi)$ is a potential for the scalar $\phi$, and
$\mathcal{\tilde R} = R - \frac{2}{\phi}\square \phi +
\frac{1}{2\phi^2}\partial_\mu\phi\partial^\mu\phi$, and $\mathcal{\tilde R}$
is the curvature for the specific case of a holonomic basis $
[X_M,X_N] = 0 $. For  a non-holonomic basis a Lagrangian density of the form
$\overline{\mathcal L}_G = \tilde R$ was considered. The two sets of Friedmann equations are given by
\begin{align}
    & 3H^2 = 8\pi G\rho_m - \frac{\dot\phi^2}{4\phi^2} +\frac{1}{\phi}\left( \frac{V(\phi)}{2}
-
3H \dot\phi\right)  \label{hol fried 1}, \\
    & \dot H = -4\pi G(\rho_m+P_m)  + \frac{\dot\phi^2}{4\phi^2} + \frac{1}{2\phi}\left(
H\dot\phi- \ddot\phi\right)\label{hol fried 2},
\end{align}
\bea
     \ddot\phi + 3H\dot\phi &=& - 16\pi G \phi \mathcal \rho_m - 6\phi\left(\dot H +2H^2
\right) \nonumber\\
&&+ \phi V'(\phi) + \frac{\dot\phi^2}{2\phi},
\eea
and
\begin{eqnarray}
  &&
  \!\!\!\!  \!\!\!\!\!\!\!\!
  3H^2 = 8\pi G\rho_m - (1+A)\frac{\dot\phi^2}{4\phi^2} - 3H
\frac{\dot\phi}{\phi}
\label{nonhol fried 1},\\
&&\!\!\!\!  \!\!\!\!\!\!\!\!
    \dot H = \! -4\pi G(\rho_m\!+\!P_m) + (1\!+\!A)\frac{\dot\phi^2}{4\phi^2} +
\frac{1}{2\phi}
\left(H\dot\phi\! - \!\ddot\phi\right)\!,
\\
  && \!\!\!\! \!\! \!\! \!\!\!\!(1+A)\left( \ddot\phi + 3H\dot\phi \right) =  - 16\pi G
\phi \mathcal \rho_m -
6\phi\left(\dot H +2H^2 \right) \nonumber\\
    &&\,
    \ \ \     \ \ \     \ \ \     \ \ \     \ \ \     \ \ \     \ \ \     \  \,
    + \frac{\dot\phi^2}{2\phi}\left( 1+A+\phi A' \right) - \dot\phi \dot A
     \label{nonhol fried 3},
\end{eqnarray}
respectively, where $A(\phi)$ is a real function of $\phi$.  In the late time limit it can be shown that one can reproduce the
thermal history of the Universe, including the succession of matter and dark-energy dominated epochs. Interestingly enough, the parameter of the effective dark energy equation of state parameter can be phantom or quintessence-like, or it can lead to a phantom-divide crossing during the cosmological evolution.

The most general spatially
homogeneous and isotropic Berwald spacetimes, defined by a Finsler Lagrangian built from a
zero-homogeneous function on the tangent bundle, which encodes the velocity dependence of the Finsler
Lagrangian, were obtained in \cite{Fc22}, and where it was also suggested that the cosmological Berwald geometries may be used for the
description of the geometry of the Universe.

Generalized scalar-tensor theories arising from vector
bundle constructions were investigated in \cite{Fc24a}, where their kinematic, dynamical and cosmological consequences were considered.  A fiber structure with two scalar fields was defined over a pseudo-Riemannian space-time base manifold. The resulting space is a 6-dimensional vector bundle endowed with a non-linear connection. The geodesics and the Raychaudhuri and general field equations were obtained in both
Palatini and metrical approaches. This geometrical structure generates new terms in the modified Friedmann equations, thus leading to the appearance of an
effective dark energy sector. Moreover, an interaction of the dark mater sector with the metric is also present.

It is the goal of the present paper to consider a systematic investigation of the application of the Finsler geometry for the description of the gravitational interaction by using the mathematical description introduced initially in \cite{Bar1,Bar2}, and later developed in \cite{Ing0,Ing1,Ing2,Ing2a}, where also some physical applications were suggested.  In the approach introduced by Barthel one can consider
a Finsler space as an $n$-dimensional {\it point space}, which is locally Minkowskian but, in general, it is not locally Euclidean. A Minkowski space is flat, homogeneous, but anisotropic, while a general Finsler space is both inhomogeneous and anisotropic.

In order to developed a theory of gravitation based on the point Finsler spaces we shall first assume that the gravitational field can be represented by a Riemannian metric $g(x)$, from which the Einstein gravitational field equations can be derived via the Hilbert-Einstein variational principle. As a next step we non-localize (anisotropize) the gravitational field, by attaching to each point $x$ ($=x^{I }$), $I =0,1,2,3$, an internal variable $y$ ($=y^{J}$), $J =0,1,2,3$. By assuming that $y$ is a vector, the nonlocal gravitational field can be described by a Finsler type geometry,
realized in a Finsler space $F^{4}$ (or by the geometry on a general vector bundle), with a metric tensor that depends on both $x$ and $y$, $\hat{g}=%
\hat{g}(x,y)$.

Generally, in many  realistic physical situations the internal variable $y$ becomes a function of the position, so that $%
y=Y(x)$, with the Finslerian metric given by $\hat{g}=\hat{g}\left(
x,Y(x)\right) $. Therefore, in this approach the Finslerian metric tensor
becomes a function of $x$ alone. The manifold defined in this way is called
the osculating Finsler manifold. On this {\it point Finsler space} we introduce {\it the Barthel connection} \cite{Bar1,Bar2,Ing0}, to which generally a torsion tensor is also associated. We further restrict our study to the case of the $(\alpha, \beta)$ metrics, and in this case it turns out that {\it the Barthel connection is the Levi-Civita connection of the Riemannian metric} $\hat{g}_{ij}(x,y(x))$, a remarkable mathematical result that is the starting point of our investigations of {\it the point Finsler theories of gravitation}. In order to obtain a specific theory we consider the case of the Randers geometry, in which we construct in a systematic way the cosmological evolution equations, by adopting for the background Riemann metric the Friedmann-Lemaitre-Robertson-Walker form. In this Barthel-Randers geometric framework we obtain the generalized Friedmann equations, and we investigate their properties. An interesting result is related to the fact that the matter energy-momentum tensor is not conserved anymore. We discuss in detail the thermodynamic interpretation of this result, which could be related to the open (from a thermodynamic point of view) of the Barthel-Randers Universe, which implies the existence of particle creation processes.

The generalized Friedmann equations of the Barthel-Randers geometry allow the construction of a number of cosmological scenarios, describing very diverse evolutions. In particular, the model admits the de Sitter solution, which describes an accelerating Universe. An effective cosmological constant can also be generated. Decelerating solutions also do exist, and they may offer some alternatives descriptions to the radiation dominated epoch of standard cosmology. For three theoretical models a detailed comparison with the observational data is performed, and it turns out that they can give a satisfactory description of the cosmological dynamics.

The present paper is organized as follows. In Section~\ref{sect1} we review the basics of the Finsler geometry, we introduce the Barthel connection and the osculating Riemann metric, and we present the definitions and the basic properties of the curvature tensors. The Einstein field equations, describing the properties of the gravitational interaction in the Finslerian geometric framework, are also introduced. In Section~\ref{sect2} we specialize our investigations by adopting for the Finsler metric a Randers type form. The geometric properties of the osculating Barthel-Randers space are studied in detail, and we obtain the basic quantities describing the metric properties, as well the connection coefficients.   Section~\ref{sect3} is devoted to a systematic construction of the cosmology of the Barthel-Randers space-time geometry. By adopting for the Riemannian metric of the Randers line element the Friedmann-Lemaitre-Robertson-Walker form, we obtain the connection coefficients, the curvature tensors, and the generalized Friedmann equations in the osculating Barthel-Randers geometry. The energy balance equation for ordinary matter is also derived, and we discuss in detail the physical interpretation of the model in the framework of the thermodynamic of open systems, with irreversible particle creation.  A number of exact cosmological solutions of the Barthel-Randers model are obtained in Section~\ref{sect4}, where the de Sitter solution is also derived, and the possibilities of generating a cosmological constant are also considered. A comparison of three theoretical models with the recent observational data is performed in Section~\ref{sect5}. Finally, we discuss and conclude our results in Section~\ref{sect6}.  In Appendix~\ref{appa} we present the details of the calculation of the connection coefficients of the Barthel-Randers geometry.  The derivation of the standard Friedmann equations in general relativity is presented, for the sake of comparison with the Barthel-Randers theory, in Section~\ref{appb}. The details of the calculations of the curvatures and of the generalized Friedmann equations are given in Section~\ref{appc}.

\section{The geometry of Finsler spaces, the Barthel connection, and
the Einstein gravitational field equations}\label{sect1}

In the following sub Sections we briefly introduce the basic properties of the conic Finsler spaces, and of the Barthel connection \cite{Bar1,Bar2,Ing0}, which we will use to construct the Finslerian model of gravitation. In our exposition  we will closely follow the approach developed in  \cite{Ing1,Ing2}.

\subsection{Finsler connection}

Let us consider {\it a geometrical structure} $\left( M^n,F\Gamma \right) $, which
consists of an $n$-dimensional manifold $M^n$ equipped with a Finsler
connection $F\Gamma $ (see \cite{Ing3} for the general theory of Finsler connections). We recall here that a  Finsler connection $F\Gamma $ of $M^n$ is defined as
a pair $(\Gamma ,N)$ of a connection $\Gamma $ in the Finsler bundle $F(M)$ of $M$ and a nonlinear
connection $N$ in the tangent bundle $TM$, where  $F(M)$ is the
induced bundle from the linear frame bundle $L(M)$ by the projection of the
tangent bundle $T(M)$.

Given a Finsler vector field $X(x,y)$ its covariant
derivatives $\nabla ^{h}X$ and $\nabla ^{v}X$ are defined as
\begin{equation}
\nabla ^{h}X:X_{|j}^{i}=\frac{\partial X^{i}}{\partial x^{j}}-\frac{\partial
X^{i}}{\partial y^{r}}N_{j}^{r}+F_{rj}^{i}X^{r},
\end{equation}
\begin{equation}
\nabla ^{v}X:X^{i}|_{j}=\frac{\partial X^{i}}{\partial y^{j}}%
+V_{rj}^{i}X^{r}.
\end{equation}

{\it The functions $F_{jk}^{i}(x,y)$, $N_{j}^{i}(x,y)$, and $V_{jk}^{i}(x,y)$ are
the connection coefficients}, and $F\Gamma $ denotes the triad $\left(
F_{jk}^{i}(x,y),N_{j}^{i}(x,y),V_{jk}^{i}(x,y)\right) $. In the following we
assume that $F\Gamma $ satisfies {\it the $D$-condition} $y^{j}F_{jk}^{i}=N_{k}^{i}
$ and the {\it $V_{1}$-condition} $y^{j}V_{jk}^{i}=0$. Then the absolute
differential of a vector $y^{i}$ is given by
\begin{equation}
Dy^{i}=dy^{i}+N_{j}^{i}\left( x,y\right) dx^{j},
\end{equation}%
while the absolute differential of a Finsler vector field $X^{i}(x,y)$ is
obtained as
\begin{equation}
DX^{i}(x,y)=dX^{i}+\left( F_{jk}^{i}(x,y)dx^{k}+V_{jk}^{i}(x,y)dy^{k}\right)
X^{j}(x,y).
\end{equation}

\subsection{An induced linear connection}

If $(M,F\Gamma )$ is a structure given by a smooth manifold $M$ and a Finsler connection $F\Gamma$, then observe that the Finsler connection $F\Gamma$ is actually a connection on the tangent bundle and not on the base manifold.

It is possible to construct an affine connection on the base manifold $M$ from $(M,F\Gamma )$ by considering $Y(x)$ a nowhere vanishing vector field on $M$, provided such a vector field exists. Of course, there are topological restrictions to the existence of an $Y(x)\neq 0$ everywhere on a smooth manifold $M$. Indeed, we can introduce a new  structure $(M^n,F\Gamma ,Y(x))$, given by evaluating all geometrical objects of $(M,F\Gamma)$ at the tangent vector field $Y(x)$. The connection form $\omega $ of the
connection $\Gamma $ of $F\Gamma $ gives rise to {\it the induced connection form
$\underset{-}{\omega }$ on the linear frame bundle $L(M)$}, and we obtain the connection $\Gamma
\left( Y\right) $ corresponding to $\underset{-}{\omega }$, which {\it is called
the linear $Y$ -connection associated to $F\Gamma $ by $Y(x)$}. The
connection coefficients of $\Gamma _{jk}^{i}$ of $\Gamma \left( Y\right) $
are written as
\begin{equation}
\Gamma _{jk}^{i}(x)=F_{jk}^{i}(x,Y)+V_{jr}^{i}(x,Y)Y_{k}^{r}(x),
\end{equation}%
where
\begin{equation}
Y_{k}^{r}(x)=\frac{\partial Y^{r}(x)}{\partial x^{k}}+N_{k}^{r}(x,Y).
\end{equation}

$\Gamma (Y)$ has a {\it torsion tensor} $\bar{T}$ with components given by
\begin{equation}
\bar{T}_{jk}^{i}(x)=T_{jk}^{i}(x,Y)+V_{jr}^{i}\left( x,Y\right)
Y_{k}^{r}(x)-V_{kr}^{i}\left( x,Y\right) Y_{j}^{r}(x),
\end{equation}%
where $T_{jk}^{i}(x,y)=F_{jk}^{i}(x,y)-F_{kj}^{i}(x,y)$ are the components
of the $(h)h$-torsion tensor $T$ of $F\Gamma $.

Given {\it a Finsler tensor field} $K(x,y)$ we obtain {\it an ordinary tensor field} $%
\bar{K}\left( x\right) =K\left( x,Y\right)$, which is called {\it the $Y$ -
tensor field}. With respect to $\Gamma \left( Y\right) $,  the covariant
derivative $\bar{\nabla}\bar{K}$ is given by
\begin{equation}
\bar{\nabla}\bar{K}:\bar{K}_{j|k}^{i}=K_{j|k}^{i}\left( x,Y\right)
+K_{j}^{i}|_{r}Y_{k}^{r}(x),
\end{equation}
where
\begin{equation*}
    \begin{split}
        K_{j|k}^{i} &:= \frac{\partial K^i_j}{\partial x^k}-
        \frac{\partial K^i_j}{\partial y^r}N^r_k+K^r_jF^i_{rk}-K^i_rF^r_{jk},\\
        K_{j}^{i}|_{k} &:=\frac{\partial K^i_j}{\partial y^k}+
        K_j^rV^i_{rk}-K^i_rV^r_{jk},
    \end{split}
\end{equation*}
are the $h$- and $v$-covariant derivatives, respectively (for details see for instance \cite{Bao}, p. 44).

The absolute differential $DX$ of a tangent vector field $X(x)$ with respect
to  $\Gamma (Y)$ is defined as  $DX^{i}:=dX^{i}+\Gamma _{jk}^{i}X^{j}dx^{k}$.

Observe that if $%
F\Gamma $ satisfies the $D$ and $V_{1}$ conditions, then $%
DY^{i}=dY^{i}+N_{j}^{i}\left( x,Y\right) dx^{j}=Y_{j}^{i}(x)dx^{j}$.

\subsection{The Barthel connection}

If $(M,F)$ is a Finsler space, and $Y(x)\neq 0$ on $M$,
we introduce now a specific  structure $(M^n,F(x,y),Y(x))$, that is, a Finsler space %
$(M^n,F(x,y))$ {\it having a tangent vector field $Y(x)$}.
The fundamental function {$F(x,y)$} is also called the length function, and it defines the
length of a piece of a smooth curve with directions in $C_{x}$, $x=x(t)\in
M^{n}$, $a\leq t\leq b$, $s\left( a,b\right) =\int_{a}^{b}F\left( x(t),\dot{x%
}\left( t\right) \right) dt$, $\dot{x}\left( t\right) \in C_{x(t)}$.

From the fundamental function $F(x,y)$ we construct the fundamental tensor $\hat{g%
}(x,y)$ having the components
\begin{equation}\label{eq9}
\hat{g}_{ij}(x,y)=\frac{1}{2}\frac{\partial ^{2}F^{2}}{\partial y^{i}\partial
y^{j}},
\end{equation}
and the Cartan tensor $C(x,y)$, with components
\begin{equation}
\hat{C}_{ijk}=\frac{1}{2}\frac{\partial \hat{g}_{ij}\left( x,y\right) }{\partial
{y}^{k}}.
\end{equation}

The notion of Finsler metrics that we have recalled here belong to the class
of \textit{classic Finsler metrics}, in other words, at each point $x\in M$,
the function $F_x:T_xM\to \mathbb{R} $ is a function defined on the tangent
space $T_xM$ of a differentiable manifold $M$ satisfying the following
conditions
{\it

\begin{enumerate}
\item[(i)] $F_x$ is $C^\infty$ on $\widetilde{T_xM}:=T_xM\setminus\{0\}$,

\item[(ii)] $F_x$ is 1-positive homogeneous: $F_x(\lambda y)=\lambda F_x(y)$%
, for all $\lambda>0$ and $y\in T_xM$,

\item[(iii)] for each $x\in M$, the Hessian matrix \eqref{eq9} is
positive defined in $\widetilde{T_xM}$.
\end{enumerate}
}
At each point $x\in M$, the indicatrix $\{y\in T_xM:F_x(y)=1\}$ is a closed,
strictly convex, smooth hypersurface around the origin of $T_xM$.

A more general geometric notion is the concept of \textit{conic Finsler metrics} that
is, {\it Finsler norms defined only on a conic domain of $T_xM$}. Let us recall
that $A_x\subset T_xM$ is called a \textit{conic domain} of $T_xM$ if $A_x$
is an open, non-empty subset of $T_xM$ such that if $v\in A_x$, then $%
\lambda v\in A_x$, for all $\lambda>0$. We remark that the origin of $T_xM$
does not belong to $A_x$ except for the case $A_x=T_xM$.

We can now {\it define a
Finsler norm defined only on a conic domain $A_x\subset T_xM$ with the
properties (i)-(iii) given above for all $y\in A_x$}. At each point $x\in M$,
the indicatrix $S_x:=\{y\in A_x\subset T_xM:F_x(y)=1\}$ is a hypersurface
embedded in $A_x$ as a closed subset.

Let $A\subset TM$ be an open subset of the tangent bundle $\pi:TM\to M$ such
that $\pi(A)=M$, and $A$ is conic in $TM$, that is for each $x\in M$, the
set $A_x:=A\cap T_xM$ is a conic domain in $T_xM$.

A function $F:A\to
\mathbb{R} $ is called a {\it conic Finsler metric} if its restriction $F_x:A_x\to
\mathbb{R} $ satisfies the conditions (i)-(iii) above, for each $x\in M$.
The local and global geometry of conic Finsler spaces can be now developed
in a similar way with the case of classical Finsler metrics (see \cite{JS}, \cite{YS1}
and references therein).

{\it The Finslerian metric $\hat{g}(x,y)$ gives rise to the $Y$-Riemann metric $%
\hat{g}_{Y}(x)=\hat{g}(x,Y)$}, provided $Y$ is nowhere vanishing.

Next we recall the notion of  {\it generalized Cartan connection}
 $C\Gamma(T)=(F^i_{jk},N^i_k,V^i_{jk})$, where $T$ is the $(h)$$h$-torsion $T^i_{jk}=F^i_{jk}-F^i_{kj}$.

\bigskip
{\bf Definition 1.}  {\it Let $(M,F)$ be a Finsler space. The $N
$-linear connection $C\Gamma (T)$ uniquely determined by the following four
conditions,

$\left( C\Gamma (T)1\right) $ $h$-metrical,

$\left( C\Gamma (T)2\right) $ $v$-metrical,

$\left( C\Gamma (T)3\right) $ $(v)v$-torsion
$S^{1}=0,S^{1}:=S_{jk}^{i}=V_{jk}^{i}-(j|k)$,

$\left( C\Gamma (T)4\right) $ Deflection $D=0$, $%
D_{j}^{i}=y^{r}F_{rj}^{i}-N_{j}^{i}$,\\
is called a generalized Cartan connection,}  where $(j|k)$ means interchange of $j$ and $k$ in the preceding term.

If the torsion tensor $T$
vanishes, then the generalized Cartan
connection reduces to the usual {\it Cartan connection} $C\Gamma=(\hat{F}^i_{jk},\hat{N}^i_j,\hat{C}^i_{jk}) $, where
\begin{equation}\label{formula F}
\begin{split}
    \hat{F}^i_{jk}(x,y)&=\frac{1}{2}\hat{g}^{il}\left(
    \frac{\delta \hat{g}_{lk}}{\delta x^j}+ \frac{\delta \hat{g}_{lj}}{\delta x^k}-\frac{\delta \hat{g}_{jk}}{\delta x^l}
    \right)\\
    &=\hat{\gamma}^i_{jk}(x,y)-\hat{g}^{si}\left(
    \hat{C}_{sjr}\hat{N}^r_k+\hat{C}_{skr}\hat{N}^r_j
    -\hat{C}_{jkr}\hat{N}^r_s
    \right),
\end{split}
\end{equation}

\bea\label{gamma}
\hat{\gamma} _{jk}^{i}\left( x,y\right)& =&\frac{1}{2}\hat{g}^{il}\left( x,y\right)
\Bigg( \frac{\partial \hat{g}_{lk}\left( x,y\right) }{\partial x^{j}}+\frac{%
\partial \hat{g}_{lj}\left( x,y\right) }{\partial x^{k}}\nonumber\\
&&-\frac{\partial \hat{%
g}_{jk}\left( x,y\right) }{\partial x^{l}}\Bigg) ,
\eea

\begin{equation*}
    \frac{\delta}{\delta x^i}=\frac{\partial}{\partial x^i}-\hat{N}^j_i(x,y)\frac{\partial}{\partial y^j},
\end{equation*}
\begin{equation*}
\begin{split}
    \hat{N}^i_j(x,y) &= \hat{\gamma} _{jk}^{i}\left( x,y\right)y^k-\hat{C}^i_{jk}\hat{\gamma}^k_{rs}y^ry^s, \\
\hat{C}_{ijk}(x,y) &=
\frac{1}{2}\frac{\partial \hat{g}_{ij}(x,y)}{\partial y^k},\quad
\hat{C}^i_{jk} = \hat{g}^{il}\hat{C}_{ljk}.
\end{split}
\end{equation*}

We can now introduce
the definition of {\it the Barthel connection} as follows:

\textbf{Definition 2.} {\it The linear $Y$ -connection $\Gamma (Y)$ associated to
the generalized Cartan connection $C\Gamma (T)$ is called the Barthel
connection $B\Gamma ^{\ast }(T)=(b_{jk}^{i}(x))$.}

The coefficients of the Barthel connection \cite{Bao,Ing3} are given by
\bea\label{eq12}
b_{jk}^{i}(x)&=&F_{jk}^{i}\left( x,Y(x)\right) +V_{jr}^{i}\left( x,Y(x)\right)
Y_{k}^{r}(x),
\eea
while the components of the torsion tensor are given by
\begin{equation}\label{eq13}
\left( T^{(B)}\right) _{jk}^{i}(T)=\left[ T_{jk}^{i}+\left(
V_{jr}^{i}Y_{k}^{r}-V_{kr}^{i}Y_{j}^{r}\right) \right] _{y=Y(x)}.
\end{equation}

In the case of the Cartan connection $C\Gamma$ defined above, the associated Barthel connection has the coefficients
\begin{equation}\label{Barthel 2}
    b_{jk}^{i}(x)=\hat{F}_{jk}^{i}\left( x,Y(x)\right) +\hat{C}_{jr}^{i}\left( x,Y(x)\right)
Y_{k}^{r}(x),
\end{equation}
where
\begin{equation*}
Y_{k}^{r}(x)=\frac{\partial Y^{r}(x)}{\partial x^{k}}+\hat{N}_{k}^{r}(x,Y),
\end{equation*}
with torsion
\begin{equation*}
\left( T^{(B)}\right) _{jk}^{i}=\left[\left(
\hat{C}_{jr}^{i}Y_{k}^{r}-\hat{C}_{kr}^{i}Y_{j}^{r}\right) \right] _{y=Y(x)}.
\end{equation*}

{\bf Remark 1. }
Let us remark that the functions (\ref{Barthel 2}) are indeed the coefficients of an affine connection because they obey the right transformation laws with respect to a coordinate change on $M$.

{\bf Remark 2.} Observe that the Barthel connections with coefficients given by equations (\ref{eq12}) and (\ref{Barthel 2}) are different in the sense that they are induced by the generalized Cartan connection $C\Gamma(T)$ with torsion $T$  and the usual Cartan connection $C\Gamma$ with vanishing torsion $T=0$, respectively. {\it Strictly speaking, the Barthel connection in (\ref{eq12}) should be called $T$-Barthel connection, and (\ref{Barthel 2}) should be called $(T=0)$-Barthel
    connection}, but {\it for the sake of simplicity we call both of them Barthel connection}. In the rest of the paper, {\it the naming Barthel connection always {\it indicates  $(T=0)$-Barthel connection}} (\ref{Barthel 2}).

\subsection{The  $\left( \alpha ,\beta \right) $ metrics}

Let $\left( M,F\left( \alpha ,\beta \right) \right) $ be a Finsler space
with an $\left( \alpha ,\beta \right) $ metric $F\left( \alpha ,\beta
\right) $, where $F(\alpha ,\beta )$ {\it is a positively homogeneous function of
first degree in two variables} $(\alpha ,\beta )$.  Here, $\alpha $ is a
Riemannian metric $\sqrt{a_{ij}(x)y^{i}y^{j}}$, and $\beta $ is a
differential form $\beta =A_{i}(x)y^{i}$. If the Hessian matrix $\hat{g}_{ij}(x,y)=\frac{1}{2}\frac{\partial ^2 F^2}{\partial y^i\partial y^j}$ is positive definite, then $(M,F)$ is a classical Finsler space.

Denoting $L=F^{2}/2$, we obtain
for the fundamental metric tensor and the Cartan tensor $C$ the expressions
\bea
\hat{g}_{ij}(x,y)&=&\frac{L_{\alpha }}{\alpha }h_{ij}+\frac{L_{\alpha \alpha }%
}{\alpha ^{2}}y_{i}y_{j}+\frac{L_{\alpha \beta }}{\alpha }\left(
y_{i}A_{j}+y_{j}A_{i}\right) \nonumber\\
&&+L_{\beta \beta }A_{i}A_{j},
\eea
and
\begin{equation}
2\hat{C}_{ijk}=\frac{L_{\alpha \beta }}{\alpha }\left(
h_{ij}p_{k}+h_{jk}p_{i}+h_{ki}p_{j}\right) +L_{\beta \beta \beta
}p_{i}p_{j}p_{k},  \label{Cijk}
\end{equation}
respectively (see for instance \cite{Ing2}, p. 47), where 
\begin{equation}
\begin{split}
    h_{ij}&:=\alpha\frac{\partial ^{2}\alpha\left( x,y\right)}{\partial y^{i}\partial y^{j}} =a_{ij}-\frac{y_{i}y_{j}}{\alpha ^{2}},\nonumber\\
y_{i}&=a_{ij}y^{j},p_{i}=A_{i}-%
\frac{\beta }{\alpha ^{2}}y_{i},
\end{split}
\end{equation}
 the indices $\alpha $, $\beta $ of $L$ indicate partial differentiation
with respect to $\alpha $ and $\beta$, and $h_{ij}(x,y)$ is {\it the angular metric of the Riemannian space} $(M,\alpha)$, respectively.

{\bf Remark 3.}
In the case of Randers and Kropina metrics, $F=\alpha+\beta$ and $F=\frac{\alpha^2}{\beta}$, respectively,  the Cartan torsion tensor is $C$-reducible, namely
\begin{equation}
    \hat{C}_{ijk}=\frac{1}{L\left( x,y\right) }\frac{1}{n+1}\left(
h_{ij}C_{k}+h_{jk}C_{i}+h_{ki}C_{j}\right),
\end{equation}
where 
$C_{k}:=\hat{g}^{ij}\hat{C}_{ijk}$ (for details in the Randers case see for instance \cite{Bao}, p. 291).

\subsection{The osculating Riemannian metric}

In this Subsection we will introduce the osculating Riemannian metric associated to a Finsler metric $(M,F)$.

As mentioned already, the fundamental geometrical objects of Finsler geometry are defined on the total space $TM$ of the tangent bundle $\pi_M:TM\to M$ regarded as $2n$-dimensional differentiable manifold with the canonical coordinates $(x,y)$. Here $x=(x^i)$ and $y=(y^i)$ are obviously independent variables. For instance $\hat{g}_{ij}:TM\setminus O\to \R$, where $O$ denotes the zero section of the tangent bundle.

On the other hand, since $\pi_M:TM\to M$ is a fiber bundle, we can consider a local section $Y:U\to TU$, where $U\subset M$ is an open neighborhood on $M$ such that $Y(x)\neq 0$, for all $x\in U$. We have $\pi_M(Y(x))=x$ on $U$.

If we fix such a local section $Y$ of $\pi_M:TM\to M$, all geometrical objects defined on the manifold $TM$ can be pulled back to $M$, for instance $\hat{g}_{ij} \circ Y$ is a function on $U$, hence we can define
\begin{equation}\label{Y-riem}
 \hat{g}_{ij}(x):=\hat{g}_{ij}(x,y)|_{y=Y(x)},\quad x\in U.
\end{equation}

The pair $(U,\hat{g}_{ij})$ is a Riemannian manifold and this $\hat{g}_{ij}$ is called the $Y$-{\it osculating Riemannian metric} associated to $(M,F)$.

The Christoffel symbols of the first kind  of the osculating Riemannian metric \eqref{Y-riem} are defined as
\begin{equation*}
\begin{split}
  \hat{\gamma}_{ijk}(x)&:=\frac{1}{2}\left(\frac{\partial}{\partial x^j}\left[\hat{g}_{ik}(x,Y(x))\right]
+\frac{\partial}{\partial x^k}\left[\hat{g}_{ij}(x,Y(x))\right]\right.\\
&\left.-\frac{\partial}{\partial x^i}\left[\hat{g}_{jk}(x,Y(x))\right]\right)
\end{split}
\end{equation*}
and by using the derivative law of composed functions we get
\begin{equation}\label{Christ for g_Y}
\begin{split}
\hat{\gamma}_{ijk}(x)&=\left.\hat{\gamma}_{ijk}(x,y)\right|_{y=Y(x)}\\
& +2\left.\left(\hat{C}_{ijl}\frac{\partial Y^l}{\partial x^k}+\hat{C}_{ikl}\frac{\partial Y^l}{\partial x^j}-\hat{C}_{jkl}\frac{\partial Y^l}{\partial x^i} \right)\right|_{y=Y(x)}.
\end{split}
\end{equation}

If $Y$ is a non-vanishing global section of $TM$, i.e. $Y(x)\neq 0$, for all $x\in M$, then we can define the osculating Riemannian manifold $(M,\hat{g}_{ij})$. However, observe that the existence of globally non-vanishing sections of $TM$ depends on the topology of $M$. For instance in the case of a 2-dimensional sphere, such sections do not exists. It is known that all noncompact manifolds admits non-vanishing global vector fields. Compact manifolds admits non-vanishing global vector fields if and only if its Euler characteristic vanishes (see for instance \cite{St}, p. 207). We will always assume that non-vanishing global vector fields exist on our differential manifold $M$.

With the assumption above, in the case of an
$(\alpha,\beta)$-metric, let us consider the vector field $Y=A$
having the components $A^{i}=a^{ij}A_{j}$. The vector field $A$ is globally non-vanishing on $M$ is equivalent with the fact that $\beta $ has
no zero points. With these notations, we consider the $A$-osculating Riemannian manifold $(M,\hat{g}_{ij})$, where $\hat{g}_{ij}(x):=\hat{g}_{ij}(x,A)$.

By denoting by $\tilde{a}$
the length of $A$ with respect to $\alpha $, we have $\tilde{a}%
^{2}=A_{i}A^{i}=\alpha ^{2}\left( x,A\right) $, $Y_{i}\left( x,A\right)
=A_{i}$, and the $A$-Riemannian metric takes the form
\begin{equation}
\begin{split}
   \hat{g}_{ij}\left( x\right) & =\left.\frac{L_{\alpha }}{\tilde{a}}\right|_{y=A(x)}a_{ij}\\
   & +\left.\left(
\frac{L_{\alpha \alpha }}{\tilde{a}^{2}}+2\frac{L_{\alpha \beta }}{\tilde{a}}%
+L_{\beta \beta }-\frac{L_{\alpha }}{\tilde{a}^{3}}\right)\right|_{y=A(x)} A_{i}A_{j}.
\end{split}
\end{equation}

Furthermore, we have $\beta \left( x,A\right) =\tilde{a}^{2}$, $p_{i}\left(
x,A\right) =0$, and consequently from  Eq. (\ref{Cijk}) it follows that $%
\hat{C}_{ijk}\left( x,A\right) =0$. Moreover, Eq. (\ref{eq13}) indicates that the
torsion $T^{(B)}$ vanishes.

On other hand, in the case of $Y=A$, observe that \eqref{formula F} implies
\begin{equation*}
     b^i_{jk}(x)=\left.\hat{F}^i_{jk}(x,y)\right|_{y=A(x)}=\left.\hat{\gamma}^i_{jk}(x,y)\right|_{y=A(x)},
\end{equation*}
and furthermore, from \eqref{Christ for g_Y} we get
$$
\hat{\gamma}_{ijk}(x)=\left.\hat{\gamma}_{ijk}(x,y)\right|_{y=A(x)}.
$$

Hence, we obtain the fundamental result that for {\it a
Finsler space with $(\alpha ,\beta )$-metric the linear $A$-connection
associated to the Cartan connection by $A=(a^{ij}A_{j})$, that is, the Barthel
connection is the Levi-Civita connection of the $A$-Riemannian space}.

{\bf Remark 4.} The result above also results by observing that $
\hat{C}_{ijk}\left( x,A\right) =0$ implies that the Riemannian metric \eqref{Y-riem} is metrical with respect to the Barthel connection and that the torsion of the Barthel connection vanishes. The fundamental Theorem of Riemannian geometry implies that  the Barthel
connection is the Levi-Civita connection of the $A$-Riemannian space.

\subsection{The definition of the curvature}

We have seen already in Subsection C that the Barthel connection is an affine connection on $M$ with the local coefficients (\ref{Barthel 2}). Indeed, it can be seen that
 the {\bf torsion tensor} of the Barthel connection is
$$
T_{jk}^i=b_{jk}^i-b_{kj}^i=C_{jr}^i\Big|_{y=Y(x)}Y_k^r-C_{kr}^i\Big|_{y=Y(x)}Y_j^r,
$$
where we have used the result that the $(h)h$-torsion of the Cartan connection vanishes.

The Barthel connection is  $g_Y$-metrical, since the generalized Cartan connection $C\Gamma(T)$, as well as the Cartan connection $C\Gamma$, are  $g$-metrical, see  Definition 1, for $(C\Gamma(T)1)$ and  $(C\Gamma(T)2)$, respectively.  Indeed, the covariant derivative of $\hat{g}(x)$ with respect to the Barthel connection, i.e.,
\begin{equation}
  \hat{g}_{ij;k}=\frac{\partial \hat{g}_{ij}(x)}{\partial x^k}-b^l_{ik}\hat{g}_{lj}-b^l_{jk}\hat{g}_{li},
  \end{equation}
after some elementary computations, can be written as
  \begin{equation}
\hat{g}_{ij;k}=  \left[g_{ij|k}(x,y)+g_{ij}|_r(x,y)Y^r_k\right]_{y=Y(x)}=0,
\end{equation}
where ``$\ _|$'' and ``$|$'' are the horizontal and vertical covariant derivatives with
respect to the Cartan connection $C\Gamma$.

{\bf Remark 5.} The Barthel connection coefficients can be written as
\bea\label{bart coeffs}
  b_{jk}^i(x)&=&\left.\left[\gamma_{jk}^i-C_{jr}^iN_k^r-C_{kr}^iN_j^r+C_{jkr}N_s^rg^{si}\right]\right|
  _{y=Y(x)}\nonumber\\
  &&+C_{jr}^i\Big|_{y=Y(x)} Y_k^r(x).
\eea

The curvature of the Barthel connection can be {\it defined as the curvature of any affine connection}. By using Eq.~\eqref{bart coeffs}, after some long but not complicated computations, it can be seen that
{\bf the curvature tensor} of the Barthel connection can be written as
\begin{eqnarray}
\hat{R}^{\ i}_{j\ km}(x)&=&\Bigg[R_{j\ km}^{\ i}+\left(P_{j\ ks}^{\ i}
  Y_m^s-P_{j\ ms}^{\ i}Y_k^s\right)\nonumber\\
 && +S_{j\ rs}^{\ i}Y_k^rY_m^s\Bigg]_{y=Y(x)},
\end{eqnarray}
where $R_{j\ km}^{\ i}, P_{j\ km}^{\ i}, S_{j\ km}^{\ i}$ are the local coefficients of the curvature of Cartan connection $C\Gamma$ (see also \cite{Ing2}, p. 43).

{\bf Remark 6.}
If we denote
$$
R_{ijkh}=g_{js}R_{i\ kh}^{\ s};\ P_{ijkh}=g_{js}P_{i\ kh}^{\ s};
\ S_{ijkh}=g_{js}S_{i\ kh}^{\ s},
$$
then it is well-known that the curvature tensors of Cartan connection $C\Gamma(N)$ have the properties:

\begin{equation*}
\begin{split}
&\begin{cases}
R_{ijhk}+R_{jihk}=0,\\
R_{ijkh}+R_{ijhk}=0,
\end{cases};\
P_{ijhk}+P_{jihk}=0,\\
&\begin{cases}
S_{ijkh}+S_{jikh}=0,\\
S_{ijkh}+S_{ijhk}=0.
\end{cases}
\end{split}
\end{equation*}

From the {\bf Ricci identities} (see for instance \cite{Ing3}, p. 82) we get
\begin{equation*}
\begin{cases}
R_{s\ hk}^{\ i}y^s=R_{hk}^i; \ P_{ijk}=C_{ijk|r}y^r,\\
P_{s\ hk}^{\ i}y^s=P_{hk}^i;\ \sum_{(ijk)}R_{ijk}=0,\\
S_{s\ hk}^{\ i}y^s=0,
\end{cases}
\end{equation*}
where $\sum_{(ijk)}$ means cyclic sum for $(i,j,k)$.

For the Barthel connection curvature tensor $\hat{R}^{\ i}_{j\ km}$ we can denote
\bea
\hat{R}_{ijkh}&:=&\hat{g}_{js}\hat{R}_{i\ kh}^{\ s}=\Bigg[R_{ijkh}+\left(P_{ijks}Y_h^s-P_{ijhs}Y_k^s\right)\nonumber\\
&&+S_{ijrs}Y_k^rY_h^s\Bigg]_{y=Y(x)},
\eea
and hence we have
$$
\hat{R}_{ijkh}=-\hat{R}_{jikh}.
$$

The {\bf Ricci tensors} of the Barthel connection can be defined as
$$
\hat{R}_{jk}(x):=\hat{R}^{\ i}_{j\ ik}.
$$

Moreover, on the Riemannian osculator space $(M,\hat{g})$ we can define the {\bf Ricci scalar} of the Barthel connection by
$$
\hat{R}=\hat{g}^{ij}\hat{R}_{ij}.
$$

The absolute differential of $Y$ with respect to the Barthel connection is
\begin{equation}
DY^{i}=dY^{i}+Y^{k}b_{kh}^{i}( x) dx^{h},  \label{dif}
\end{equation}%
where $b_{kh}^{i}( x) $ are the Barthel connection
coefficients.


The covariant derivative of $Y$  with respect to the Barthel connection being given as  
\begin{equation}
Y^{i}_{\ ;j}(x)=\frac{\partial Y^{i}(x)}{\partial x^{j}}%
+b_{kj}^{i}( x) Y^{k}(x).
\end{equation}

The vector field $Y^{i}(x)$ is {a parallel vector field}  with respect to the Barthel connection if
$Y^{i}_{\ ;j}(x)=0$,  that is,
\begin{equation}\label{dYi}
\frac{\partial Y^{i}(x)}{\partial x^{j}}=
-b_{kj}^{i}
( x) Y^{k}(x).
\end{equation}

In the case the vector field $Y(x)$ is {parallel} with respect to the Barthel connection induced by the Cartan connection $C\Gamma=(\hat{F}^i_{jk},\hat{N}^i_j,\hat{C}^i_{jk})$, Eq.~(\ref{Barthel 2}) simplifies to
\begin{equation}
\label{Barthel symple}
b_{jk}^{i}(x)=\hat{\gamma} _{jk}^{i}\left( x,Y(x)\right) -\hat{\gamma} _{js}^{r}\left(
x,Y(x)\right) Y^{s}(x)\hat{C}_{rk}^{i}\left( x,Y(x)\right).
\end{equation}

Let us observe that the original definition (\ref{eq12})  of the Barthel connection show that this is an affine connection on the base manifold $M$. On the other hand, the right hand side of the formula \eqref{Barthel symple} suggests that in the case when $Y$ is parallel with respect to the Barthel connection, we can define
on the tangent bundle $TM$ the functions
\begin{equation}
    \mathbf{b}^i_{jk}(x,y):=\hat{\gamma} _{jk}^{i}\left( x,y\right) -\hat{\gamma} _{js}^{r}\left(
x,y\right) y^{s}\hat{C}_{rk}^{i}\left( x,y\right),
\end{equation}
and hence
\begin{equation}
    b^i_{jk}(x)=\mathbf{b}^i_{jk}(x,Y(x)).
\end{equation}

Therefore, Eq.~\eqref{dYi} can be written as
\begin{equation}\label{parallel cond2}
\frac{\partial Y^{i}(x)}{\partial x^{j}}=
-\mathbf{b}_{kj}^{i}
( x,Y(x)) Y^{k}(x).  
\end{equation}

By taking the derivative of Eq.~(\ref{parallel cond2}) we immediately find
\begin{eqnarray}
\frac{\partial ^{2}Y^{i}(x)}{\partial x^{l}\partial x^{k}}&=&\Bigg[ -\frac{%
\partial \mathbf{b}_{jk}^{i}\left( x,y\right) }{\partial x^{l}}\Bigg|_{y=Y(x)}  \notag \\
&&+\frac{\partial \mathbf{b}_{jk}^{i}\left( x,y\right) }{\partial y^{s}}\Bigg|_{y=Y(x)}%
\mathbf{b}_{pl}^{s}\left( x,Y(x)\right) Y^{p}(x)  \notag \\
&&+\mathbf{b}_{rk}^{i}\left( x,Y(x)\right) \mathbf{b}_{jl}^{r}\left( x,Y(x)\right) \Bigg] %
Y^{j}\left( x\right) .
\end{eqnarray}

Thus,
\begin{equation}
\frac{\partial ^{2}Y^{i}(x)}{\partial x^{l}\partial x^{k}}-\frac{\partial
^{2}Y^{i}(x)}{\partial x^{k}\partial x^{l}}=\hat{R}_{jkl}^{i}\left(
x,Y\left( x\right) \right) Y^{j}\left( x\right) ,
\end{equation}%
where (see also \cite{Ing2a})
\bea\label{eq25}
&&\hat{R}_{jkl}^{i}\left( x,Y\left( x\right) \right) =\frac{\partial b_{jl}^{i}\left( x,Y(x)\right) }{\partial x^{k}}-\frac{\partial
  b_{jk}^{i}\left( x,Y(x)\right) }{\partial x^{l}}\nonumber\\
&&+
\left[ \frac{\partial
b_{jk}^{i}\left( x,y\right) }{\partial y^{s}}b_{pl}^{s}\left(
x,y\right) -\frac{\partial b_{jl}^{i}\left( x,y\right) }{\partial y^{s}%
}b_{pk}^{s}\left( x,y\right) \right]_{y=Y(x)}
Y^{p}(x)\nonumber\\
&&+\left[b_{rk}^{i}\left(
x,y\right) b_{jl}^{r}\left( x,y\right) -b_{rl}^{i}\left( x,y\right)
b_{jk}^{r}\left( x,y\right)\right]_{y=Y(x)}
\eea
is precisely the curvature tensor of the Barthel connection for the parallel
vector field $Y(x)$, and  it can be checked that indeed $\hat{R}_{jkl}^{i}\left(x,Y\left( x\right) \right) Y^{j}\left( x\right)=0$.

\subsection{The gravitational Einstein equations}

The contractions of the curvature tensor lead to the generalized Ricci
tensor, and Ricci scalar, respectively, given by
\begin{equation}
\hat{R}_{jk}=\hat{R}_{jik}^{i},\quad\hat{R}_{j}^{i}=\hat{g}^{ik}\hat{R}_{kj},
\end{equation}%
and
\begin{equation}
\hat{R}=\hat{R}_{i}^{i},
\end{equation}%
respectively.

\textit{We postulate that the Einstein gravitational field equations can be
formulated in a Barthel geometry as}
\begin{equation}
\hat{R}_{jk}-\frac{1}{2}\hat{g}_{jk}\hat{R}=\kappa ^2 \hat{T}_{jk},  \label{Eineq}
\end{equation}%
where $\kappa^2=8\pi G/c^4 $ is a constant, with $G$ and $c$ denoting the Newtonian gravitational constant, and the speed of light, respectively, and $\hat{T}_{jk}$ is the matter
energy-momentum tensor, constructed with the help of the usual thermodynamic
quantities, and of the Finslerian metric tensor $\hat{g}_{ik}$.

\section{Geometry of the osculating Barthel-Randers space}\label{sect2}

In the following we will investigate the physical implications of the
generalized gravitational type equations (\ref{Eineq}) by assuming that the
length function $F\left( x,y\right) $ is given by a Randers \cite{Rand} type
form,
\begin{equation}
F\left( x,y\right) =\sqrt{\epsilon \,g_{ij}(x)y^{i}y^{j}}+A_{k}(x)y^{k}=%
\alpha \left( x,y\right) +\beta (x,y),
\end{equation}%
where $g_{ij}(x)=g_{ji}(x)$ is {\it a Riemannian metric}, $A_{k}(x)$ is {\it an
arbitrary vector field},
\be
\alpha \left( x,y\right) =\sqrt{\epsilon%
\,g_{ij}(x)y^{i}y^{j}},
\ee
and
\be
\beta (x,y)=A_{k}(x)y^{k},
\ee
respectively.
Also, $\epsilon=\pm1$ for a time-like/space-like vector $y^i$, respectively.

\subsection{Geometric quantities}

As a first step in constructing the generalized Einstein equations for the
Barthel connection we need to calculate the components of the Finslerian
metric tensor $\hat{g}_{ij}(x)$, given by \cite{Mats, Pey}

\begin{equation}
  \hat{g}_{ij}(x,y)=\frac{F}{\alpha }h_{ij}  +\left( A_{i}+\epsilon\,\frac{y_{i}}{\alpha }\right) \left( A_{j}+\epsilon\,%
\frac{y_{j}}{\alpha }\right) ,
\end{equation}
where
\begin{equation}\label{angular metric}
    h_{ij}=\epsilon\,g_{ij}(x)-\frac{y_{i}}{%
\alpha }\frac{y_{j}}{\alpha }
\end{equation} is the angular metric of the Riemannian space $(M,\alpha)$, $y_{i}=g_{ij}y^{j}$, and $A^{i}=g^{ij}A_{j}$. Moreover, we denote $%
A^{2}(x)=g_{ij}(x)A^{i}(x)A^{j}(x)=A_{i}A^{i}$, while the relation between
the determinants of $\hat{g}_{ij}(x,y)$ and $g_{ij}(x)$ is given by the
relation $\det $ $\hat{g}_{ij}(x,y)=\left( F/\alpha \right) ^{n+1}\det
g_{ij}(x)$, For the inverse of the Finsler metric tensor we obtain
\begin{equation}
\hat{g}^{ij}(x,y)=\frac{\epsilon\,\alpha }{F}g^{ij}(x)-\frac{%
\epsilon\,\alpha }{F^{2}}\left( A^{i}y^{j}+A^{j}y^{i}\right) +\frac{%
A^{2}\epsilon\,\alpha +\beta }{F^{3}}y^{i}y^{j}.
\end{equation}

But, as we have shown in the previous Section, {\it in the case of the Barthel connection of a Finsler space with Randers type $(\alpha, \beta)$ metric we are considering $ C^i_{jk}(x,A)=0$}.
It is  important to note that for a Finsler metric the Cartan tensor depends on $(x,y)$. This means that the components of the Cartan tensor in other tangent directions different from $A_i$ are obviously non-vanishing, and for $y\neq A$ the considered $(\alpha, \beta)$ metric has no reason to be Riemannian everywhere.

Explicitly, from \eqref{Christ for g_Y}, we obtain the Christoffel symbols of the first kind as
\begin{eqnarray}\label{eq36}
&&\hat{\gamma}_{ijk}(x)=\nonumber\\
&&\Bigg[ \epsilon \left( 1+\frac{\beta }{\alpha }%
\right) \Gamma _{ijk}(x)-\frac{1}{2}\left( \frac{\partial A_{m}}{\partial x^{i}}\tilde{%
l}^{m}-\epsilon \frac{\beta }{\alpha }{\Gamma}_{nni}\right) h
_{jk}\nonumber\\
&&+\frac{1}{2}\left(\mathcal{M}_{ijk}+\mathcal{M}_{ikj}\right)\nonumber\\
&&+2\Bigg( \hat{C}_{ijl}\frac{%
\partial Y^{l}}{\partial x^{k}}+\hat{C}_{ikl}\frac{\partial Y^{l}}{\partial
x^{j}}-
\hat{C}_{jkl}\frac{\partial Y^{l}}{\partial x^{i}}\Bigg) \Bigg]
_{y=Y(x)},
\end{eqnarray}
where
\begin{eqnarray}\label{eq37}
\hspace{-0.5cm}\mathcal{M}_{ijk} &=&\left( \frac{\partial A_{m}}{\partial x^{k}}\tilde{l}%
^{m}-\epsilon \frac{\beta }{\alpha }{\Gamma}_{nnk}\right) h
_{ij}+\left( \Gamma _{jnk}+\Gamma _{njk}\right) \xi _{i}\nonumber\\
\hspace{-0.5cm}&&-\left( \Gamma
_{kni}+\Gamma _{nki}\right) \xi _{j}
+\left( \Gamma
_{inj}+\Gamma _{nij}\right) \xi _{k}\nonumber\\
\hspace{-0.5cm}&&
-\epsilon {\Gamma}_{nnk}\left(
\tilde{l}_{i}\xi _{j}+\tilde{l}_{j}\xi _{i}\right) \nonumber\\
\hspace{-0.5cm}&&+\epsilon {\Gamma%
}_{nni}\tilde{l}_{j}\xi _{k}+
\frac{\partial A_{j}}{\partial x^{k}}\left( A_i+\epsilon\tilde{l}_{i}\right) \nonumber\\
\hspace{-0.5cm}&&+\left( \frac{\partial A_{i}}{\partial x^{k}}-\frac{\partial
A_{k}}{\partial x^{i}}\right) A_{j}+\epsilon \left( \frac{\partial A_{i}}{\partial x^{k}}-\frac{%
\partial A_{k}}{\partial x^{i}}\right) \tilde{l}_{j},
\end{eqnarray}
and
\begin{equation}
\tilde{l}_{i}:=\frac{y_{i}}{\alpha },y_{i}:=g_{ij}(x)y^{j},
\end{equation}
\begin{equation}
\xi _{i}:=\epsilon A_{i}-\frac{\beta }{\alpha }%
\tilde{l}_{i},
\end{equation}

\begin{equation}
{\Gamma}_{nnj}:=\Gamma _{stj}\tilde{l}^{s}\tilde{l}^{t}=\Gamma _{stj}%
\frac{y^{s}}{\alpha }\frac{y^{t}}{\alpha },
\end{equation}
respectively. Here $\Gamma_{ijk}(x)$  are the first kind Christoffel symbols of the Riemannian metric $g_{ij}(x)$, and $h_{ij}$ is the angular metric \eqref{angular metric}. The full details of the derivation of Eq.~(\ref{eq36}) are presented in Appendix~\ref{appa}.

In the following we will consider a particular model of the
Barthel-Randers geometry, based on a specific choice of the vector $Y^i$.

\subsection{The model  $Y^{i}=A^{i}$}

In the present Barthel-Randers type geometrical model we substitute the vector
$y$ by $y^{i}=Y^{i}=A^{i}$, so that the Finslerian metric tensor becomes $%
\hat{g}^{ij}(x,y)=\hat{g}^{ij}(x,A(x))$. Moreover, we normalize the vector $%
A(x)$ so that $\alpha ^{2}=\epsilon\,g_{ij}(x)A^{i}A^{j}=\epsilon\,A^{2}$, $%
\alpha =\sqrt{\epsilon\,A^{2}}$, $\beta =A^{2}=\epsilon\,\alpha ^{2}$, and $%
F\left( x,A(x)\right) =\alpha\left( x,A(x)\right) +\epsilon\,\alpha\left( x,A(x)\right) ^{2} $. Hence for the
Finsler metric tensor we find
\begin{align}
&\hat{g}_{ij}(x,A(x)) =\left( \epsilon\,+\alpha \right) \left[ g_{ij}(x)+%
\frac{1}{\alpha }A_{i}(x)A_{j}(x)\right] ,
\end{align}
while for the inverse of the metric tensor we obtain
\begin{align}
\hat{g}^{ij}(x,y)=\frac{1}{\epsilon\,+\alpha }\left[ g^{ij}(x)-\frac{%
\epsilon\,}{\alpha \left( \epsilon\,+\alpha \right) }A^{i}(x)A^{j}(x)\right]
.
\end{align}

{\bf Lemma 1.}

{\it
Let $(M,F=\alpha+\beta)$ be a Randers space with
$\alpha^2=\epsilon g_{ij}(x)y^iy^j$ and $\beta=A_i(x)y^i$. In the case $y^{i}=A^{i}(x)=g^{ij}(x)A_j$,
we have
\begin{enumerate}
    \item $\hat{C}_{ijk}(x,y)|_{y=A(x)} =0$
\item $\xi_i|_{y=A(x)}=0$.
\end{enumerate}

}
 This result follows from the obvious relation $\left[
\epsilon A_{j}-\left( \beta /\alpha \right) \left( y_{j}/\alpha \right) \right]
\left. {}\right\vert _{y_{j}=A_{j}(x)}=0$. Therefore, we directly arrive at the
important result that \textit{the Barthel connection of the $L=\alpha +\beta
$ metric, with $\beta =A_{k}y^{k}$, coincides with the Levi-Civita
connection for $\hat{g}_{ij}\left( x,A(x)\right) $.}

Hence for the components of the Barthel connection we obtain the expressions
\begin{equation}\label{eq46}
    \begin{split}
       & \hat{\gamma}_{ijk}(x) =\left( \epsilon +\alpha \right) \Gamma _{ijk}(x)\\
       &-\frac{\epsilon}{2\alpha}\left( \frac{\partial A_m}{\partial x^i}A^m-\epsilon\frac{\beta}{\alpha^2}\Gamma_{sti}A^sA^t
       \right)
       \left( g_{jk}(x)-\frac{1}{A^2}A_jA_k
       \right)\\
       &+ \frac{1}{2}\left(\mathcal{M}_{ijk}+\mathcal{M}_{ikj}\right),
    \end{split}
\end{equation}
where
\bea
        \mathcal{M}_{ijk}&=& \frac{\epsilon}{\alpha}
        \left(\frac{\partial A_m}{\partial x^k}A^m-
        \epsilon\frac{\beta}{\alpha^2}\Gamma_{stk}A^sA^t
        \right)\left( g_{ij}(x)-\frac{1}{A^2}A_iA_j
       \right)
        \nonumber\\
        &&+\frac{\alpha+\epsilon}{\alpha}\left[
        \frac{\partial A_j}{\partial x^k}A_i+
        \left(\frac{\partial A_i}{\partial x^k}-
        \frac{\partial A_k}{\partial x^i}
        \right)A_j
        \right],
\eea
and

\begin{equation}
b_{hk}^{i}(x)=\hat{g}^{il}\left( x,A\right) \hat{\gamma}_{lhk}\left( x,A\right)
,
\end{equation}%
respectively.
Vanishing of $\hat{C}_{ijk}(x,A(x))$ means that the
Finslerian metric becomes $A$-Riemannian in this case.

\section{Cosmological evolution in Barthel-Randers geometry}\label{sect3}

We assume now for the Riemannian metric in the Barthel-Randers length
function $L=\alpha +\beta $ the flat and isotropic
Friedmann-Lemaitre-Robertson-Walker form, in which the interval $ds$ between
two neighboring points in the space-time manifold with coordinates $\left(
x^{0}=ct,x^{1}=x,x^{2}=y,x^{3}=z\right) $ is given by
\begin{equation}  \label{metr}
ds^{2}=\left(dx^0\right)^{2}-a^{2}\left(x^0\right)\left( dx^{2}+dy^{2}+dz^{2}\right) ,
\end{equation}%
where  $a^{2}\left(x^0\right)$ is the cosmological scale
factor. In the next Sections we use the Landau-Lifshitz metric conventions \cite{Land}. Hence the non-vanishing components of the Riemannian metric tensor $g_{IJ}$, $I,J\in\{0,1,2,3\}$, are
given by $g_{II}=\left( 1,-a^{2}\left(x^0\right),-a^{2}\left(x^0\right),-a^{2}\left(x^0\right)\right) $.

{\it In the
following we denote by a dot the derivative with respect to the cosmological
time $t$, and by a prime the derivative with respect to the coordinate $%
x^{0} $, namely $a'=\dfrac{d a}{d x^0}=\dfrac{d a}{dt}\dfrac{dt}{d x^0}=\dfrac{1}{c}\dfrac{d a}{dt}=\dfrac{1}{c}\dot{a}$.
}

\subsection{Brief review of standard cosmology}

Due to the assumed homogeneity of the space-time, all physical and
geometrical quantities can be only functions of the cosmological time $t$.
We also suppose that the matter content of the Universe can be described as
a barotropic fluid, characterized by two thermodynamic parameters, the
pressure $p$ and the density $\rho $, obeying the equation of state $%
p=p(\rho )$. Hence the matter energy-momentum tensor is given by
\be
T_{AB}=\left( \rho c^{2}+p\right) u_{A}u_{B}-pg_{AB},
\ee
where $u_{A}$ are the
components of the four-velocity of the cosmological fluid. Usually one
adopts a frame comoving with the cosmological matter, which fixes the
components of the four-velocity as $u^{A}=\left( 1,0,0,0\right) $. Hence the
components of the energy-momentum tensor become
\be
T_{A}^{A}=\left( \rho c^2,-p,-p,-p\right) .
\ee

 Moreover, we introduce the {\it Hubble functions} $H(t)$ and $\mathcal{H}\left(x^0\right)$, defined as
 \be
 H(t)=\frac{\dot{a}(t)}{a(t)},
 \ee
 and
 \be
 \mathcal{H}\left(x^0\right)=\frac{a'\left(x^0\right)}{a\left(x^0\right)}=\frac{1}{a\left(x^0\right)}\frac{da\left(x^0\right)}{dx^0}=\frac{1}{c}H(t),
 \ee
 respectively.

In standard general relativity the Riemannian metric tensor $g_{AB}$, which
encodes all the properties of the gravitational field, satisfies the
Einstein equations \cite{e8, Land},
\begin{equation}
R_{AB}-\frac{1}{2}g_{AB}+\Lambda g_{AB}=\kappa ^2T_{AB},
\end{equation}%
where $\Lambda$ is the cosmological constant.

In the case of the cosmological metric (\ref{metr}), the Einstein
gravitational field equations reduce to the celebrated Friedmann equations, representing the theoretical foundations of modern cosmology, and which are
given by \cite{e8}
\begin{equation}
3H^2(t)=8\pi G \rho (t)+\Lambda c^2,  \label{Fri1}
\end{equation}
and%
\begin{equation}
2\dot{H}(t)+3H^2(t)=-\frac{8\pi G}{c^2}p(t)+\Lambda c^2, \label{Fri2}
\end{equation}
respectively.

The derivation of the geometric part of the Friedmann equations is presented
in detail in Appendix~\ref{appb}. Eqs.~(\ref{Fri1}) and (\ref{Fri2}) did
predict the existence of the large scale expansion of the Universe, and of
the cosmological singularity.

To describe the dynamical nature of the cosmological evolution we introduce
the deceleration parameter $q$, defined as
\begin{equation}
q=\frac{d}{dt}\frac{1}{H}-1=-\frac{\dot{H}}{H^2}-1.
\end{equation}

Positive values of $q$ indicate a decelerating expansion, while for $q<0$
the expansion of the Universe is accelerating. To facilitate the comparison
of the theoretical predictions of the cosmological Barthel-Randers models
with the cosmological observations, we introduce as independent variable the
redshift $z$, defined as
\begin{equation}\label{redshift1}
1 + z =\frac{1}{a}.
\end{equation}

In the above definition we have normalized the scale factor according to the
relation $a(0) = 1$. Hence the derivatives with respect to the time can be
replaced in the cosmological models with the derivatives with respect to $z$%
, by taking into account the relation
\begin{equation}\label{redshift2}
\frac{d}{dt} =\frac{dz}{dt}\frac{d}{dz} = -(1 + z)H(z)\frac{d}{dz} .
\end{equation}

In terms of $z$ the deceleration parameter $q$ is given by
\begin{equation}
q(z) = (1 + z)\frac{1}{H(z)}\frac{dH(z)}{dz} - 1.
\end{equation}

We also assume that the cosmological matter satisfies a linear barotropic
equation of state of the form $p=(\gamma -1)\rho c^2$, where $\gamma $ is a
constant, and $1\leq \gamma \leq 2$.

Recently, especially due to the study of the Cosmic Microwave Background
Radiation by the Planck satellite \cite{1g, 1h}, a large number of high
precision cosmological data have been obtained, which have drastically
modified our views on the Universe. In the following we adopt the
simplifying hypothesis that the matter content of the late Universe contains
only dust matter only. Therefore the matter in the present day Universe has
negligible thermodynamic pressure. Hence, the energy conservation equation
\be
\dot{\rho}+3H\left(\rho+\frac{p}{c^2}\right)=0,
\ee
of standard cosmology gives for the time variation of
the energy density of the dust  matter with $p=0$ the simple expression $\rho= \rho
_0/a^3=\rho _0 (1+z)^3$, where $\rho _0$ is the present day matter density.
The time evolution of the Hubble function is given, as a function of the
scale factor, by \cite{e8}
\begin{equation}
H=H_0\sqrt{\left(\Omega _b+\Omega _{DM}\right)a^{-3}+\Omega _{\Lambda}},
\end{equation}
where by $\Omega _b$, $\Omega _{DM}$,and $\Omega _{\Lambda}$ we have denoted
the density parameters of the baryonic matter, of the cold (pressureless)
dark matter, and of the dark energy (modeled by a cosmological constant),
respectively, while $H_0$ is the present day value of the Hubble function.
The three density parameters obey the relation $\Omega _b+\Omega
_{DM}+\Omega _{\Lambda}=1$, which shows that the geometry of the Universe is
flat.

The deceleration parameter of standard general relativistic cosmology is
given by
\begin{equation}
q(z)=\frac{3 (1+z)^3 \left(\Omega _{DM}+\Omega _b\right)}{2 \left[\Omega
_{\Lambda}+(1+z)^3 \left(\Omega _{DM}+\Omega _b\right)\right]}-1.
\end{equation}

In order to compare the predictions of the Barthel-Randers type cosmological models with observations  for the matter density parameters we adopt the values $\Omega
_{DM}=0.2589$, $\Omega _{b}=0.0486$, and $\Omega _{\Lambda}=0.6911$%
, respectively, which follow from the Planck data  \cite{1h}. The total matter
density parameter $\Omega _m=\Omega _{DM}+ \Omega _b$ has then the numerical
value $\Omega _m=0.3089$. With the help of the density parameters we obtain
for the present day value of the deceleration parameter the value $q(0)=-0.5381$,
which indicates that presently the Universe is in an accelerating phase.

The mathematical representation of the Friedmann equations can be
significantly simplified by introducing a set of dimensionless variable $%
\left(\tau, h,r,P,\lambda,\right)$, given by
\begin{eqnarray}
\hspace{-0.7cm}\tau &=& H_0t, H=H_0h, \rho =\frac{H_0^2}{8\pi G}r,p=\frac{c^2H_0^2}{8\pi G}P, \lambda =\frac{\Lambda c^2}{H_0^2},
\end{eqnarray}
where $H_0$ is the present day value of the Hubble function. The Hubble function can be obtained in a dimensionless form as a
function of the redshift as $H(z)=H_0h(z)$, with
\begin{equation}  \label{63}
h(z)=\sqrt{\left(\Omega _{DM}+\Omega _b\right)\left(1+z\right)^{3}+\Omega
_{\Lambda}}.
\end{equation}

The Friedmann equations take the dimensionless form
\be
3h^2=r+\lambda, 2\frac{dh}{d\tau}+3h^2=-P+\lambda.
\ee

\subsection{Cosmology of the Barthel-Randers Model}

In order to investigate the cosmological implications of the Barthel-Randers
geometry, given by the Barthel-Randers model, {\it we assume that the Riemannian metric $g_{ij}(x)$
is given by the Friedmann-Lemaitre-Robertson-Walker metric} (\ref{metr}).

In order to simplify the mathematical formalism, and to obtain a clear
physical interpretation of the results, we adopt the following
approximations:

\paragraph{The components of the Finsler metric are functions of $x^0$ only.}

We adopt the \textit{cosmological principle} that requires the
homogeneity of the Universe, which implies that \textit{the geometrical and
physical properties of the Universe depend globally and on the large scale
on the cosmological time only}. The cosmological homogeneity assumption
requires that $A_I=A_I\left(x^0\right)$.

\paragraph{The space-like components of $A$ vanish.}

The homogeneity of the Universe as well as the diagonal nature of the metric
imposes another mathematical condition on the components of the vector $A$,
namely, the requirement that its space-like components vanish, $A_1=A_2=A_3=0
$. If this condition does not hold we can perform a spatial rotation, and
then we obtain a preferred direction, for example in the $x$ coordinates.
Such a behavior would contradict the large scale spatial isotropy of the
Universe. Hence in the present model we assume that \textit{the vector $A$
is characterized by one independent component only}, $A_0\left(x^0\right)$. As a result,
we define the 1-form field as
\begin{align}\label{special A_i}
(A_{I})=(a\left(x^0\right)\eta\left(x^0\right),0,0,0)=(A^{I}).
\end{align}

\paragraph{A frame comoving with matter does exist.}

We will assume that, similarly to the Riemannian geometric case, we can introduce in the Barthel-Randers geometry a {\it comoving frame} in which observers move along with the Hubble flow, defined by the metric $g_{ij}(x)$.

\paragraph{Thermodynamic properties.}

We also infer that the properties of the cosmological matter can be described by {\it two thermodynamical quantities only, the energy density $\rho c^2$, and the thermodynamic pressure} $p$, respectively, defined in the usual way. Assumptions {\it c} and {\it d}  allow us to define the components of the matter energy-momentum tensor as $\hat{T}_0^0=\rho c^2$, and $\hat{T}_A^A=-p$, respectively.

From the above assumptions it follows:
\begin{enumerate}
    \item[(i)] $\epsilon=1$;
    \item[(ii)] $\dfrac{\partial A_I}{\partial x^J}-
    \dfrac{\partial A_J}{\partial x^I}=0$;
    \item[(iii)] $(g_{IJ})=\begin{pmatrix}
    1 & 0 & 0& 0\\
    0 & -a^2\left(x^0\right) & 0 & 0 \\
    0 & 0 &-a^2\left(x^0\right) & 0\\
    0 & 0 & 0 &-a^2\left(x^0\right)
    \end{pmatrix};
    $
    \item[(iv)] $\left.\alpha \right|_{y=A(x)}=a\left(x^0\right)\eta\left(x^0\right)$;
    \item[(v)] $\left.\beta \right|_{y=A(x)}=\left[a\left(x^0\right)\eta\left(x^0\right)\right]^2$;
    \item[(vi)] $\left.(h_{IJ}\right|_{y=A(x)})=\begin{pmatrix}
    0 & 0 & 0 & 0\\
    0 & -a^2\left(x^0\right) & 0 & 0 \\
    0 & 0 &-a^2\left(x^0\right) & 0\\
    0 & 0 & 0 &-a^2\left(x^0\right)
    \end{pmatrix};
    $
\end{enumerate}
where $I,J\in \{0,1,2,3\}$, and  $h_{IJ}(x,y):=g_{IJ}-\frac{y_I}{\alpha}\frac{y_J}{\alpha}$ is the angular metric of $(M,\alpha)$.

From the above results, it turns out that the Finsler metric is diagonal.  This fact follows naturally from the construction of the Randers metric. In this way we conserve in the Finslerian extension of general relativity one of the essential features of the cosmological Friedmann-Lemaitre-Robertson-Walker metric (\ref{metr}).

In the following we denote
\be
\phi \left(x^0\right):=1+a\left(x^0\right)\eta\left(x^0\right),
\ee
from  which we obtain
\be
\phi '\left(x^0\right)=a\left(x^0\right)\left[\eta '\left(x^0\right)+\mathcal{H}\left(x^0\right)\eta \left(x^0\right)\right].
\ee

{\bf Lemma 2.}
{\it
a) The non-vanishing components of $\hat{g}_{IJ}$ are
\begin{equation}\label{mF}
\begin{split}
    \hat{g}_{IJ}=\begin{cases}
    \hat{g}_{00}=\left[1+a\left(x^0\right)\eta \left(x^0\right)\right]^2 =\phi ^2\left(x^0\right),\\
    \hat{g}_{ij}= -a^2\left(x^0\right)\left[1+a\left(x^0\right)\eta \left(x^0\right)\right]\\
    \;\;\;\;\;=-a^2\left(x^0\right)\phi \left(x^0\right)\delta_{ij}.
    \end{cases}
\end{split}
\end{equation}

b) The non-vanishing components of $\hat{g}^{IJ}$ are
\begin{equation}
\begin{split}
    \hat{g}^{IJ}=\begin{cases}
    \hat{g}^{00}=\frac{1}{\left[1+a\left(x^0\right)\eta\left( x^0\right)\right]^2}=\frac{1}{\phi^2\left(x^0\right)},\\
    \hat{g}^{ij}= -\frac{1}{a^2\left(x^0\right)\left[1+a\left(x^0\right)\eta\left(x^0\right)\right]}\delta^{ij}=-\frac{1}{a^2\left(x^0\right)\phi \left(x^0\right)}\delta^{ij},
    \end{cases}
\end{split}
 \end{equation}
 \noindent where  $\delta_{ij} $ is the Kronecker delta symbol,  $\delta ^{ij} $ is the inverse of $\delta _{ij}$, and $i,j\in\{1,2,3\}$.
}

\subsection{The generalized Friedmann equations}

After establishing the basic geometrical foundations of the Barthel-Randers cosmology, we can proceed now to derive the cosmological evolution equations of the theory. For the Christoffel symbols of the first kind we obtain
\begin{equation}\label{eq 65}
\begin{split}
       & \hat{\gamma}_{IJK}(x)=(1+a\eta)\Gamma_{IJK}\\
       & -\left.\frac{1}{2a\eta}\left(
       \frac{\partial A_M}{\partial x^I}A^M-\Gamma_{STI}A^SA^T
       \right)h_{JK}\right|_{y=A(x)}\\
       &+\frac{1}{2}\left(\mathcal{M}_{IJK}+\mathcal{M}_{IKJ}\right),
\end{split}
\end{equation}
where
\begin{equation}
    \begin{split}
       & \mathcal{M}_{IJK}=\left. \frac{1}{a\eta}  \left(
       \frac{\partial A_M}{\partial x^K}A^M-\Gamma_{STK}A^SA^T
       \right)h_{IJ}\right|_{y=A(x)}\\
       &+\frac{a\eta+1}{a\eta}\left[
        \frac{\partial A_J}{\partial x^K}A_I+
        \left(\frac{\partial A_I}{\partial x^K}-
        \frac{\partial A_K}{\partial x^I}
        \right)A_J
        \right].
    \end{split}
\end{equation}


The choice $y=A(x)$ implies
\begin{equation}
    \Gamma_{nnI}(x,y)|_{y=A(x)}=0, \quad I\in \{0,1,2,3\},
\end{equation}
see Lemma C1. Moreover, we get
\begin{equation}
    \mathcal{M}_{000}= (1+a\eta)(a\eta)',
\end{equation}
and thus
\bea
    \mathcal{M}_{000}&=&(1+a\eta)(a'\eta+a\eta')= a(1+a\eta)(\eta '+\mathcal{H}\eta),\nonumber\\
    \mathcal{M}_{0ij}&=&0,\quad i,j\in\{1,2,3\}.
\eea

By substituting the adopted expression of the 1-form field into %
Eq.~\eqref{eq 65}, one obtains
\begin{align}
\hat\gamma_{000}=(1+a\eta)(a'\eta+a\eta')=
a(1+a\eta)(\eta '+\mathcal{H}\eta).
\end{align}

Hence all the components of $\hat{\gamma}_{ijk}$ can be obtained by means of the following
\smallskip

{\bf Lemma 3.}
{\it The non-vanishing components of $\hat{\gamma}_{IJK}(x)$ are
\begin{equation*}
\hat\gamma_{000}=a(1+a\eta)(\eta '+\mathcal{H}\eta)=\phi \phi ',
\end{equation*}
\begin{eqnarray*}
\hat\gamma_{0ij}=-\hat\gamma_{i0j}=-\hat\gamma_{ij0}
&=&
\frac{a^2}{2}\left[a\eta '+(2+3a\eta)\mathcal{H}\right]\delta_{ij}\\
& =&\frac{a^2}{2}\left( \phi '+2\phi \mathcal{H}\right)\delta_{ij}.
 \end{eqnarray*}
\smallskip
}
\smallskip

{\bf Lemma 4.}
{\it The non-vanishing components of the Christoffel symbols of the second kind of the osculating Riemannian metric 
are
\begin{equation}
    \hat{\gamma}^A_{BC}=
    \begin{cases}
    \hat{\gamma}^0_{00}=\dfrac{a}{(1+a\eta)} (\eta '+\mathcal{H}\eta)=\dfrac{\phi '}{\phi},  \\
    \ \\
    \hat{\gamma}^0_{ij}= \dfrac{a^2}{2(1+a\eta)^2}\left[a\eta '+
    (2+3a\eta)\mathcal{H}
    \right]\delta_{ij}  \\
   \;\;\;\;\;\; =\dfrac{a^2}{2\phi^2}\left(\phi '+2\phi \mathcal{H} \right)\delta_{ij},
    \\
    \ \\
    \hat{\gamma}^i_{0j}=\dfrac{1}{2(1+a\eta)}\left[a\eta '+
    (2+3a\eta)\mathcal{H}
    \right]\delta ^{i}_{j}  \\
   \;\;\;\;\;\;\; =\dfrac{1}{2\phi}\left(\phi'+2\phi \mathcal{H} \right)\delta^{i}_{j} 
    . \\
    \end{cases}
\end{equation}
}

\subsubsection{The curvature tensors}

We will compute now the components of the Ricci tensor. The curvature tensor of an affine connection with local coefficients $\left(\Gamma _{BC}^A(x)\right)$ is given by
\begin{equation}
R^A_{BCD}=\dfrac{\partial \Gamma^A_{BD}}{\partial x^C}-
    \dfrac{\partial \Gamma^A_{BC}}{\partial x^D}+\Gamma^E_{BD}\Gamma^A_{EC}
    -\Gamma^E_{BC}\Gamma^A_{ED},
    \end{equation}

The Barthel connection with local coefficients $\left(b_{BC}^A(x)\right)$ is an affine connection, and hence its curvature tensor must be given by the above formula,  with $\left(\Gamma _{BC}^A(x)\right)=\left(b_{BC}^A(x)\right)$. As we have already discussed in detail, in the case of the Randers metric $F=\alpha +\beta$, the Barthel connection coincides with the Levi-Civita connection of the osculating metric $\hat{g}_{AB}(x)=g_{AB}\left(x,A(x)\right)$, where $A_I(x)$ are the components of $\beta$, and $g_{AB}$ is the fundamental tensor of $F$. Hence, since $b_{BC}^A=\hat{\gamma}_{BC}^A$, where $\hat{\gamma}_{BC}^A$ are the Levi-Civita coefficients, we obtain for the curvature tensors the expressions
\begin{equation}
\hat{R}^A_{BCD}=\dfrac{\partial \hat{\gamma}^A_{BD}}{\partial x^C}-
    \dfrac{\partial \hat{\gamma}^A_{BC}}{\partial x^D}+\hat{\gamma}^E_{BD}\hat{\gamma}^A_{EC}
    -\hat{\gamma}^E_{BC}\hat{\gamma}^A_{ED},
    \end{equation}
    and
    \begin{equation}
\hat{R}_{BD}=
    \displaystyle\sum_A\left[\dfrac{\partial \hat{\gamma}^A_{BD}}{\partial x^A}-\dfrac{\partial \hat{\gamma}^A_{BA}}{\partial x^D}
    +\sum _E\left(\hat{\gamma}^E_{BD}\hat{\gamma}^A_{EA}-\hat{\gamma}^E_{BA}\hat{\gamma}^A_{ED}\right)\right],
 \end{equation}
respectively, where $A,B,C,D,E\in \{0,1,2,3\}$. For details of the definitions of the affine connections and of the curvature tensors see \cite{Chern1}.

From here we obtain the non-vanishing components of the Ricci tensor $\hat{R}_{AB}$ in the Barthel-Randers geometry as
\begin{equation}\label{Ric1}
   \hat{R}_{00}=-3\mathcal{H}^{\prime }-3\mathcal{H}^{2}-\frac{3}{2}\frac{\phi ^{\prime \prime }}{%
\phi }+\frac{9}{4}\frac{\phi ^{\prime 2}}{\phi ^{2}},
\end{equation}
and
\begin{equation}\label{Ric2}
   \hat{R}_{ij}=
   \frac{a^{2}}{2\phi ^{2}}\left[ \phi ^{\prime \prime }-\frac{\phi ^{\prime
2}}{2\phi }+2\left( \mathcal{H}^{\prime }+3\mathcal{H}^{2}\right) \phi +4\mathcal{H}\phi ^{\prime }\right]\delta_{ij},
\end{equation}
respectively. See Appendix~\ref{appc} for the computations of these Ricci curvatures.

From mathematical point of view, observe that the derivatives of $\phi$ do also include $\mathcal H$, hence, it is possible to rewrite  Eqs.~(\ref{Ric1}) and (\ref{Ric2}) as
 \begin{equation}
    \hat{R}_{00}=-\frac{3(3\phi-1)}{2\phi}\left( \mathcal{H}'+\mathcal{A}_1\mathcal{H}^2
      +\mathcal{A}_2\mathcal{H}+\mathcal{A}_3
      \right),
    \end{equation}
    where
\begin{equation}
      \begin{split}
        \mathcal{A}_1&:= \frac{3\phi^2+4\phi-3}{2\phi(3\phi-1)}\\
        \mathcal{A}_2&:=-\frac{(\phi-1)(\phi-3)}{\phi(3\phi-1)}\cdot\frac{\eta'}{\eta} \\
        \mathcal{A}_3&:=\frac{(\phi-1)}{(3\phi-1)}\left[\frac{\eta''}{\eta}
          -\frac{3}{2}\frac{(\phi-1)}{\phi} \left(\frac{\eta'}{\eta} \right)^2
          \right],
        \end{split}
      \end{equation}
      and
\begin{equation}
        \hat{R}_{ij}=\frac{(\phi-1)^2(3\phi-1)\delta_{ij}}{2\phi^2\eta^2}\left( \mathcal{H}'+\mathcal{B}_1\mathcal{H}^2
       +\mathcal{B}_2\mathcal{H}+\mathcal{B}_3  \right),
   \end{equation}
 where
    \begin{equation}
      \begin{split}
        \mathcal{B}_1&:= \frac{21\phi^2-8\phi-1}{2\phi(3\phi-1)}\\
        \mathcal{B}_2&:=\frac{(\phi-1)(5\phi+1)}{\phi(3\phi-1)}\cdot \frac{\eta'}{\eta}
         \\
          \mathcal{B}_3&:=\frac{(\phi-1)}{(3\phi-1)}\left[\frac{\eta''}{\eta}- \frac{(\phi-1)}{2\phi}
            \left(\frac{\eta'}{\eta} \right)^2
        %
          \right],
        \end{split}
      \end{equation}
 respectively, provided $\phi\neq 0$, $3\phi-1\neq 0$, everywhere.

For the Ricci scalar we find
\begin{eqnarray}
\hspace{-0.5cm}\hat{R} =-\frac{1}{\phi ^{2}}\left( 6\mathcal{H}^{\prime }+12\mathcal{H}^{2}+3\frac{\phi ^{\prime
\prime }}{\phi }-3\frac{\phi ^{\prime 2}}{\phi ^{2}}+6\mathcal{H}\frac{\phi ^{\prime }%
}{\phi }\right).
\end{eqnarray}

\subsubsection{The generalized Friedmann equations}

Now, we can obtain {\it the generalized Friedman cosmological evolution equations} as
\begin{equation}\label{feq1}
3H^{2}=8\pi G\phi ^2 \rho -\frac{3}{4}\frac{\dot{\phi} ^{ 2}}{\phi ^{2}}-3H\frac{%
\dot{\phi}}{\phi },
\end{equation}
and
\begin{equation}\label{feq2}
2\dot{H}+3H^{2}=-\frac{8\pi G}{c^{2}}\phi ^{2}p-\frac{\ddot{\phi}}{\phi }+%
\frac{5}{4}\frac{\dot{\phi}^{2}}{\phi ^{2}}-H\frac{\dot{\phi}}{\phi },
\end{equation}
respectively.

From the above equations we immediately obtain
\be
\dot{H}=-4\pi G \phi^2\left(\rho +\frac{p}{c^2}\right)-\frac{\ddot{\phi}}{2\phi}+\frac{\dot{\phi}^2}{\phi^2}+H\frac{\dot{\phi}}{\phi}.
\ee

Eq.~(\ref{feq1}) can be reformulated as
\be
3\left(H+\frac{\dot{\phi}}{2\phi}\right)^2=8\pi G \phi ^2 \rho.
\ee

By denoting
\be
\tilde{H}=H+\frac{\dot{\phi}}{2\phi},
\ee
the generalized Friedmann equations in the Barthel-Randers cosmology take the form
\be
3\tilde{H}^2=8\pi G \phi ^2 \rho,
\ee
and
\be
2\dot{\tilde{H}}+3\tilde{H}^2=-\frac{8\pi G}{c^2}\phi ^2p+\frac{\dot{\phi}^2}{\phi ^2} +2H\frac{\dot{\phi}}{\phi},
\ee
respectively.

\subsection{The energy conservation equation}

The conservation equation of the matter content of the
Universe can be obtained from the relation $\hat\nabla_\mu T^{\mu\nu}=0$, where the covariant derivative is calculated with the connection $\hat\gamma^\mu_{~\nu\alpha}$. In our case, the conservation equation can be written as
\begin{equation}\label{cons}
\dot{\rho}+3\left(H+\frac{\dot\phi}{2\phi}\right)\left( \rho +\frac{p}{c^2}\right) =0,
\end{equation}
or, equivalently,
\be
\frac{d}{dt}\left(\rho a^3\right)+\frac{p}{c^2}\frac{d}{dt}a^3+\frac{3}{2}a^3\left(\rho +\frac{p}{c^2}\right)=0.
\ee

It should be noted that the conservation equation is not independent, and can be obtained from the Friedman equations. In fact, by taking the derivative of Eq.~\eqref{feq1} and substituting $\dot H$ from Eq.~\eqref{feq2}, one obtains
\begin{align}
&8\pi G\phi^2\left(\dot\rho+\frac{3H p}{c^2}\right)-\frac{3}{8\phi^3}(\dot\phi^3-2H\phi\dot\phi^2-20H^2\phi^2\dot\phi^2)\nonumber\\&+4\pi G\phi\dot\phi(3p+4\rho)+9H^3=0.
\end{align}

Now, substituting $H^2$ from the Friedman equation \eqref{feq1}, one obtains again Eq.~\eqref{cons}.

\subsection{Thermodynamical interpretation of the Barthel-Randers cosmology}

Eq.~(\ref{cons}) shows that, as opposed to the standard general relativistic case, in the present Barthel-Randers type cosmological model the matter content of the Universe is not conserved. This raises the question of the physical interpretation of this result, and of its cosmological implications. One possibility for obtaining a physical insight into the energy nonconservation is to interpret it in the framework of the thermodynamics of irreversible processes as describing {\it particle creation/annihilation}. In the following we briefly present the thermodynamic
interpretation of the Barthel-Randers type cosmological theories. The nonconservation of the energy-momentum appears in several approaches to gravity involving the presence of geometry-matter coupling, like, for example, in $f\left(R,L_m\right)$ and $f(R,T)$ theories \cite{Ha14}.
The non-conservation of the matter energy-momentum tensor, as shown by Eq.~(\ref{cons}), suggests
that due to the presence of the Finslerian geometric effects,  particle creation processes may
occur during the cosmological evolution in the Barthel-Randers geometry, corresponding to a creation of matter from geometry. Particle creation does  also appear in quantum field theories in curved space-times, as initially discussed in
\cite{Parker,Parker1,Zel,Parker2}, and it is a direct  consequence of the temporal variation
of the gravitational field. In \cite{Zel} particle creation in an anisotropic Bianchi type I metric was considered, and the renormalized value of the energy-momentum tensor of a quantum scalar field with a non-zero mass was first, and correctly obtained. Therefore the Barthel-Randers type gravity theory, in which matter
creation processes also occur, could lead to the possibility of a semiclassical effective description of the
quantum effects in a gravitational field.

\subsubsection{Matter creation in irreversible thermodynamics}

Eq.~(\ref{cons}) shows that the covariant divergence of
the basic equilibrium quantities of a thermodynamic system consisting of ordinary matter, and described by the energy-momentum tensor,  is different from zero. This result implies that similar effects must appear at the level of other thermodynamical quantities, including the particle and entropy fluxes. Consequently, in the presence of particle creation all the balance equilibrium equations must be modified in order to include this effect
 \cite{P-M,Lima,Su}. In the following {\it we will investigate the physical consequences of energy nonconservation from the cosmological perspective of the Riemann space}, with metric $g_{AB}(x)$, and we assume that the Finslerian effects can be interpreted as some physical event in this space. Therefore all geometrical and physical quantities will be defined with the use of the FLRW metric (\ref{metr}) only.

 In the presence of particle creation, one must modify the balance equation for the particle flux
$N^{A} \equiv nu^A$, where $n$ is the particle number density, and $u^{A}$ is the four-velocity defined in the Riemann space, according to
\begin{equation}
\nabla _{A}N^{A}=\dot{n}+3Hn=n\Psi,
\end{equation}
where $\nabla _A$ is the covariant derivative defined with respect to the Levi-Civita connection associated to the metric (\ref{metr}), while $\Psi $ is {\it the matter creation rate}.

We assume in the following that all quantities are functions of the cosmological time only. If $\Psi \ll H$, one can neglect the source term in the particle balance equation. The entropy flux vector is defined according to $S^{A} \equiv \tilde{s}u^{A} = n\sigma u^{A}$, where $\tilde{s}$ is {\it the entropy density}, while $%
\sigma $ denotes {\it the entropy per particle}. By taking the divergence of the entropy flux
we obtain
\begin{equation}  \label{62b}
\nabla _{A}S^{A}=n\dot{\sigma}+n\sigma \Psi\geq 0,
\end{equation}
where the positivity condition follows from the second law of thermodynamics. If the entropy per particle  $\sigma $ is a constant, then we obtain the condition
\begin{equation}
\nabla _{A}S^{A}=n\sigma \Psi =\tilde{s}\Psi\geq 0,
\end{equation}
which indicates that the change of the entropy is exclusively due to the matter production processes by the gravitational field. Since always $\tilde{s}>0$, it follows that the matter production rate must satisfy the important condition $\Psi \geq 0$. This condition can be interpreted physically as allowing the gravitational fields to create matter, but forbidding the inverse process. In the presence of matter  production the
energy-momentum tensor of cosmological fluid  must also be modified through the inclusion of the irreversible effects related to the second law of thermodynamics, and it can be represented generally as \cite{Bar}
\begin{equation}\label{64}
T^{AB}=T^{A B}_\text{eq}+\Delta T^{AB},
\end{equation}
where $T^{AB}_\text{eq}$ represents the standard equilibrium component~\cite{Bar}, while $\Delta T^{AB}$ corresponds to the adjustments necessary due to
matter creation. Since we assume that the space-time is homogeneous and isotropic, $\Delta T^{AB}$, giving the supplementary
contribution to $T^{AB}$, can be generally written in the form
\begin{equation}
\Delta T_{\; 0}^0=0, \quad \Delta T_{\; A}^B=-p_c\delta_{\; A}^B,
\end{equation}
where the quantity $p_c$ represents the \textit{creation pressure}, an effective quantity that  describes
{\it phenomenologically} in a macroscopic physical system the thermodynamical effects of matter production. The tensor $\Delta T^{AB}$ can be represented covariantly as \cite{Bar}
\begin{equation}
\Delta T^{A B}=-p_ch^{AB}=-p_c\left(g^{A B}-u^{A}u^{B}\right),
\end{equation}
where $h^{AB}$ is the projection operator. Hence, we can obtain immediately the relation  $u_{A}\nabla _{B}\Delta T^{A B}=3HP_c$.
Therefore, in the presence of matter creation, the scalar component of the energy balance
equation $u_{A}\nabla _{B}T^{A B}=0$, which follows from Eq.~\eqref{64}, gives the temporal variation of the energy density of the cosmological fluid in the form
\begin{equation}\label{cons1}
\dot{\rho}+3H\left[\rho+\frac{1}{c^2}\left(p+p_c\right)\right]=0.
\end{equation}

The basic thermodynamic parameters of the cosmological fluid must also satisfy the Gibbs law, which is given by \cite{Lima}
\begin{equation}\label{Gibbs}
n T \mathrm{d} \left(\frac{\tilde{s}}{n}\right)=nT\mathrm{d}\sigma=\mathrm{d}\rho -\frac{\rho+p/c^2}{n}\mathrm{d}n,
\end{equation}
where by $T$ we have denoted the thermodynamical temperature of the Barthel-Randers Univerze.

\subsubsection{Irreversible thermodynamics and Barthel-Randers cosmology}

The matter energy balance equation~\eqref{cons} of the Barthel-Randers cosmology can be reformulated after some simple algebraic transformations as
\begin{equation}\label{cons3}
\dot{\rho}+3H\left[\rho +\frac{p}{c^2}+\frac{c^2\dot{\phi}}{2c^2H\phi}\left(\rho +\frac{p}{c^2}\right)\right] =0.
\end{equation}%

The simple comparison of Eqs.~(\ref{cons1}) and (\ref{cons3}) gives the expression of the creation pressure in the Barthel-Randers cosmological model in the presence of matter creation as
\be\label{pc}
p_{c}=\frac{\dot{\phi}}{2H\phi}\left(\rho c^2+p\right)=\frac{c^2\rho \dot{\phi}}{2H\phi}\left(1+w\right),
\ee
where we have introduced the notation $w=p/\rho c^2$. Then the energy density balance Eq.~\eqref{cons}
can be derived by taking the divergence in the Riemann space with the FLRW metric (\ref{metr}) of the total energy momentum tensor $%
T^{AB}$, given by
\begin{equation}
T^{A B }=\left( \rho c^2 +p+p_{c}\right) u^{A }u^{B }-\left(
p+p_{c}\right) g^{A B }.
\end{equation}

Furthermore, under the important assumption of {\it adiabatic particle production},
which implies $\dot{\sigma}=0$, from the Gibbs law (\ref{Gibbs}) it follows
\begin{equation}
\dot{\rho}
=\left(\rho+\frac{p}{c^2}\right)\frac{\dot{n}}{n}
=\left(\rho+\frac{p}{c^2}\right)\left(\Psi-3H\right),
\end{equation}
which together with the energy balance equation gives immediately the
relation between the particle creation rate and the creation pressure as
\begin{equation}
\Psi=\frac{-3Hp_c}{\rho c^2+p}.
\end{equation}

Therefore, with the use of Eq.~(\ref{pc}) we obtain for the particle  creation rate in the Barthel-Randers cosmology the simple expression
\begin{eqnarray}
\Psi=-\frac{3}{2}\frac{\dot{\phi}}{\phi}=-\frac{3}{2}\frac{\dot{a}\eta+a\dot{\eta}}{1+a\eta}.
\end{eqnarray}

Under the assumption $1+a\eta >0$, the condition $\Psi \geq 0$, which assures the existence of particle creation, imposes the  constraint $\dot{\phi}=\dot{a}\eta+a\dot{\eta}=a\left(\dot{\eta}+H\eta\right)\leq 0$ on the two
physical parameters of the theory, the scale factor $a$ and the vector field $\eta$. This condition depends only indirectly, via the Hubble function, on the equation of state of the ordinary matter. On the other hand, the condition $\dot{\phi}<0$ leads to a negative creation pressure, as seen from Eq.~(\ref{pc}). {\it Hence particle creation can take place only in the presence of a negative creation pressure} $p_c$.

In the Barthel-Randers Universe the particle balance equation can be therefore formulated as
\be
\dot{n}+3Hn=-\frac{3}{2}n\frac{\dot{\phi}}{\phi}.
\ee

In the case of a dust Universe with $p=0$, the creation pressure takes the form
\be
p_{c}=\frac{\dot{\phi}}{2H\phi}\rho c^2,
\ee
and it is directly proportional to the matter density.

As a function of  the creation pressure, the divergence of the entropy flux vector $S^a$ can be written  as
\begin{equation}
\nabla _{A}S^{A}=\frac{-3 n \sigma H p_c}{\rho c^2 +p}=-\frac{3}{2}n\sigma \frac{\dot{\phi}}{\phi}=n\sigma \Psi.
\end{equation}

An important parameter in systems with particle creation is the temperature $T$. To investigate the temperature evolution in a system with particle
creation, and to determine the time dependence of the temperature of the relativistic
fluid in the Barthel-Randers Universe, we consider the general case in which the equations of state of the density and pressure are functions of the particle number and the temperature, and
have the general parametric form $\rho =\rho (n, T )$ and $p=p(n,T)$, respectively. Then
we immediately find
\begin{equation}
\dot{\rho}=\left(\frac{\partial \rho }{\partial n} \right)_T\dot{n}+\left(%
\frac{\partial \rho }{\partial T} \right)_n\dot{T}.
\end{equation}

With the  use of the energy and particle balance equations we obtain
\begin{equation}  \label{78a}
-3H\left(\rho c^2+p+p_c\right)=\left(\frac{\partial \rho }{\partial n}
\right)_T n\left(\Psi-3H\right)
+\left(\frac{\partial \rho }{\partial T} \right)_n\dot{T}.
\end{equation}

Next, by making use of the thermodynamic identity \cite{Bar}
\begin{equation}
T\left(\frac{\partial p}{\partial T}\right)_n=\rho c^2+p-n\left(\frac{\partial
\rho}{\partial n}\right)_T,
\end{equation}
Eq.~\eqref{78a} leads to the temperature evolution of the cosmologic fluid in the Barthel-Randers Universe, and in
the presence of matter creation as
\begin{equation}
\frac{\dot{T}}{T}=\left(\frac{\partial p}{c^2\partial \rho}\right)_n\frac{\dot{n}}{n}=w\frac{\dot{n}}{n}.
\end{equation}

 By taking into account that from the particle balance equation we obtain
 \be
 \frac{\dot{n}}{n}=-3\left(\frac{\dot{\phi}}{\phi}+\frac{\dot{a}}{a}\right),
 \ee
 and assuming that $\dot{\phi}<0$, for the temperature evolution of the newly created matter in the Barthel-Randers Universe we find the expression
 \be
 T=T_0\frac{\phi ^{3w/2}}{a^3}.
 \ee

 Generally, one can assume that $w$ is an arbitrary, varying equation-of-state parameter $w(a)$, which is also a function of the scale factor \cite{r3s}. Moreover, in a consistent cosmological model all physical and geometrical quantities must be regular and well-defined for all values of $w(a)$.

\subsubsection{The case of the exotic matter}

 In the thermodynamical approach discussed in the previous Section we have assumed that particles are created in the form of ordinary baryonic matter, and hence it satisfies the conditions $\rho c^2+p\geq 0$, and  $w\geq 0$, respectively. However, we cannot exclude automatically the case in which the Barthel-Randers Universe is filled with exotic fluids with $w<0$. The thermodynamic approach and the physical interpretation of the Barthel-Randers theory can also be extended to the case $w<0$. In this case the creation pressure is negative for $\dot{\phi}>0$, while the particle creation rate becomes positive. Hence in the presence of exotic matter particle creation processes can also take place in the Barthel-Randers Universe.

 An interesting case is represented by the equation of state $\rho c^2+p=0$, corresponding to $w=-1$, or, from a geometric point of view, to the presence of a cosmological constant. Then Eq.~(\ref{cons1}) immediately gives
 \begin{equation}
  \dot{\rho} = -3H p_c.
\end{equation}

 By assuming adiabatic particle production, with $\dot{\sigma}=0$, from the Gibbs law we obtain
\begin{equation}
\dot{\rho} = (\rho c^2+p)\frac{\dot{n}}{n} = 0.
\end{equation}
 Consequently, the above two equations independently give the result
\begin{equation}
  \dot{\rho} = p_c = 0,
\end{equation}
implying an Universe with constant matter density. But, if, for example, $w=-2$, and $\rho >0$, then $p_c=-\left(c^2\dot{\phi}/2H\phi\right)\rho$,  the particle creation rate is given by $\Psi=(3/2)\dot{\phi}/\phi$, and particle creation takes place if $\dot{\phi}>0$.

Hence the Barthel-Randers geometry can act as a source of exotic matter in the Riemannian spacetime described by the FLRW metric.

\section{Exact cosmological models}\label{sect4}

In the present Section we investigate the general cosmological implications of the generalized Friedmann equations (\ref{feq1}) and (\ref{feq2}), describing the cosmological evolution in the Barthel-Randers geometry, and we also consider some specific models.

\subsection{General properties of Barthel-Randers cosmological models}

In the limit $\phi \rightarrow 1$, the Friedmann equations (\ref{feq1}) and (\ref{feq2}) do reduce to the standard equations of general relativity. Therefore, the function $\phi=1+a\eta $ describes the deviations from the general relativistic background, as induced by the new geometrical structure attached to the standard Friedmann-Lemaitre-Robertson-Walker spacetime.

By representing the function $\phi (t)$ in terms of a new function $\alpha (t)$, defined as $\phi (t)=\phi _0e^{\alpha (t)}$, or $\alpha (t)=\ln \left[(1+a\eta)/\phi _0\right]$, where $\phi _0$ is a constant, the Friedmann equations (\ref{feq1}) and (\ref{feq2}) can be reformulated as
\be
3H^2=8\pi G \phi_0^2e^{2\alpha}\rho -\frac{3}{4}\dot{\alpha}^2-3H\dot{\alpha}=8\pi G_{eff}\left(\rho+\rho_{eff}\right),
\ee
and
\bea
2\dot{H}+3H^2&=&-\frac{8\pi G \phi_0^2}{c^2}e^{2\alpha}p-\ddot{\alpha}+\frac{1}{4}\dot{\alpha}^2-H\dot{\alpha}\nonumber\\
&=&-\frac{8\pi G_{eff}}{c^2}\left(p+p_{eff}\right),
\eea
respectively, where
\be
G_{eff}=\phi _0^2Ge^{2\alpha},
\ee
\be
\rho_{eff}=-\frac{1}{\phi _0^2G}\left(\frac{3}{4}\dot{\alpha}^2+3H\dot{\alpha}\right)e^{-2\alpha},
\ee
and
\be
p_{eff}=\frac{1}{\phi _0^2G}\left(\ddot{\alpha}-\frac{1}{4}\dot{\alpha}^2+H\dot{\alpha} \right)e^{-2\alpha},
\ee
respectively.

From the generalized Friedmann equations, after eliminating $3H^2$, we immediately obtain
\be
2\dot{H}=-\frac{8\pi G\phi _0^2}{c^2}\left(\rho c^2+p\right)e^{2\alpha}-\ddot{\alpha}+\dot{\alpha}^2+2H\dot{\alpha}.
\ee

For the deceleration parameter of the model we find the expression
\begin{equation}
q=\frac{\left( 4\pi \phi _{0}^{2}G/c^{2}\right) \left( \rho c^{2}+p\right)
e^{2\alpha }+(3/2)\ddot{\alpha}-(3/4)\dot{\alpha}^{2}}{8\pi \phi
_{0}^{2}G\rho e^{2\alpha }-(3/4)\dot{\alpha}^{2}-3H\dot{\alpha}}.
\end{equation}

The effective energy-density generated by the Barthel-Randers geometry is positive if the condition $(3/4)\dot{\alpha}^2+3H\dot{\alpha}<0$, $\forall t\geq 0$. However, there is no guarantee that this conditions is satisfied for all times. Negative energy densities also do appear in the framework of the so-called phantom field cosmological models, initially proposed in \cite{Ph}, where the kinetic energy of the scalar field is negative.

The system of the Friedmann equations can be formulated in a dimensionless form, by introducing the set of the dimensionless variables $\left(\tau, h, r, P\right)$, defined as
\be
H=H_0h, \tau =H_0t,\rho=\frac{H_0^2}{8\pi G\phi _0^2}r,p=\frac{H_0^2c^2}{8\pi G\phi _0^2}P.
\ee
Then we obtain
\be\label{feq3}
3\frac{1}{a^2}\left(\frac{da}{d\tau}\right)^2=e^{2\alpha}r-\frac{3}{4}\left(\frac{d\alpha }{d\tau}\right)^2-3h\frac{d\alpha }{d\tau},
\ee
and
\be\label{feq4}
\frac{2}{a}\frac{d^2a}{d\tau^2}+\frac{1}{a}\left(\frac{da}{d\tau}\right)^2=-e^{2\alpha}P-\frac{d^2\alpha}{d\tau ^2}+\frac{1}{4}\left(\frac{d\alpha}{d\tau}\right)^2-h\frac{d\alpha }{d\tau},
\ee
respectively. By taking into account that $\alpha$ is a function of $a$ and $\eta$, the system of equations (\ref{feq3}) and (\ref{feq4}) can be explicitly written down as
\be\label{feq5}
3\frac{a'^2}{a^2}=(1+a\eta)^2r-\frac{3 \left(\eta  a'+a \eta '\right)^2}{4 (1+a \eta
   )^2}-\frac{3 a' \left(\eta  a'+a \eta '\right)}{a (1+a \eta)},
\ee
and
\bea\label{feq6}
\hspace{-0.8cm}2\frac{a''}{a}+\frac{a'^2}{a^2}&=&-(1+a\eta)^2P+\frac{5 \left(\eta  a'+a \eta '\right)^2}{4 (1+a \eta
   )^2}\nonumber\\
\hspace{-0.8cm} &&  -\frac{a' \left(\eta  a'+a \eta '
   \right)}{a (1+a \eta )}-\frac{\eta  a''+2 a'
   \eta '+a \eta ''}{1+a \eta},
\eea
respectively, where, from now on, {\it a prime denotes the derivative with respect to $\tau$.} Once the equation of state $P=P(r)$ is specified, Eqs.~(\ref{feq5}) and (\ref{feq6}) represents a system of two, strongly nonlinear, ordinary differential equations for $a$ and $\eta$ that must be solved with some appropriately chosen initial conditions $a(0)=a_0$, $\eta (0)=\eta _0$, and $\eta '(0)=\eta _0'$.

The first generalized Friedmann equation can be reformulated as
\be
3\left[h+\frac{1}{2}\frac{\eta a'+a\eta'}{1+a\eta}\right]^2=(1+a\eta)^2r.
\ee

\subsection{The de Sitter solution}

We will look first for the de Sitter type solutions of the Finslerian cosmological model described by Eqs.~(\ref{feq1}) and (\ref{feq2}) for a vacuum Universe, corresponding to $H=H_0={\rm constant}$,  $\rho =p=0$, and $a(t)=e^{H_0\left(t-t_0\right)}$, respectively. Then from Eqs.~(\ref{feq1}) and (\ref{feq2}) it follows that $\phi$ must satisfy the differential equation
\be
\ddot{\phi}-2\frac{\dot{\phi}^2}{\phi}-2H_0\dot{\phi}=0.
\ee

By introducing a new variable $\dot{\phi}=u$, the above equation becomes
\be
\frac{du}{d\phi}-\frac{2}{\phi}u-2H_0=0,
\ee
with the general solution given by
\be
u(\phi)=C\phi ^2-2H_0\phi,
\ee
where $C$ is an arbitrary constant of integration. Hence we obtain
\be
\phi(t)=\frac{2H_0}{e^{2H_0\left(t-t_0\right)}+C},
\ee
where $t_0$ is an integration constant, giving
\begin{equation}
\eta (t)=-\frac{e^{H_{0}\left( t-t_{0}\right) }+\left( C-2H_{0}\right)
e^{-H_{0}\left( t-t_{0}\right) }}{e^{2H_{0}\left( t-t_{0}\right) }+C}.
\end{equation}

\subsection{Generating a cosmological constant}

If the condition
\be\label{150}
-\frac{3}{4}\frac{\dot{\phi}^2}{\phi^2}-3H\frac{\dot{\phi}}{\phi}=8\pi G\Lambda \phi^2,
\ee
holds, where $\Lambda$ is a constant, the first Friedmann equation (\ref{feq1}) can be reformulated as
\be
3H^2=8\pi G \phi ^2 \left(\rho +\Lambda\right),
\ee
while Eq.~(\ref{feq2}) becomes
\be
2\dot{H}+3H^2=-\frac{8\pi G}{c^2}\phi ^2\left(p-\Lambda c^2\right)-\frac{\ddot{\phi}}{\phi}+2H\frac{\dot{\phi}}{\phi}+2\frac{\dot{\phi}^2}{\phi^2}.
\ee

Hence, while there is a positive contribution to the energy density of the ordinary matter in the form of a constant term, the effective cosmological constant term, introduced through the condition (\ref{150}), generates an effective pressure term in the second Friedmann equation. Moreover, we have also taken into account the variation of the gravitational coupling.

Eq.~(\ref{150}) determines the cosmological Barthel-Randers vector $\eta$ as a solution of the ordinary differential equation
\bea
\dot{\eta} &=&\pm\frac{2 (1+a \eta ) \sqrt{3 H^2-8\pi G\Lambda  (1+a \eta )^2}}{\sqrt{3}
   a}\nonumber\\
  && -\frac{\eta  \dot{a}+2  (1+a \eta)H}{a}.
\eea

It is interesting to note that if the solution of the above equation does exist, the Barthel-Randers geometry will generate an effective cosmological constant for all functional forms of the scale factor $a$. Alternatively, one can impose the simpler condition
\be\label{150a}
-\frac{3}{4}\frac{\dot{\phi}^2}{\phi^2}-3H\frac{\dot{\phi}}{\phi}=\Lambda,
\ee
to generate a "simple" cosmological constant, that may allow to write the first Friedmann equation as
\be
3H^2=8\pi G\phi^2 \rho+\Lambda,
\ee
and which gives $\eta$ as a solution of the differential equation
\be\label{150b}
\dot{\eta} =\pm\frac{2 (1+a \eta ) \sqrt{3 H^2-\Lambda  }}{\sqrt{3}
   a}-\frac{\eta  \dot{a}+2  (1+a \eta)H}{a}.
\ee

As an application of the above relation we consider the simple cosmological scenario in which $a(t)=t/t_0$, $H(t)=1/t$, and $q=0$, respectively, where $t_0$ is a constant, which, without any loss of generality,  we will take as one in the following. Hence Eq.~(\ref{150b}), in which we adopt the minus sign,  takes the form
\be
\dot{\eta} = -\frac{t  \left(2 \sqrt{9-3 \Lambda  t^2}+9\right)\eta +2
   \sqrt{9-3 \Lambda  t^2}+6}{3 t^2},
\ee
giving
\be
\eta=\frac{3 c_1 e^{-2 \sqrt{1-\frac{\Lambda  t^2}{3}}} \left(2 \sqrt{9-3 \Lambda
   t^2}-\Lambda  t^2+6\right)}{t^5}-\frac{1}{t},
\ee
where $c_1$ is a constant of integration. By assuming $a\eta <<1$, Eq.~(\ref{150b}) can be approximated as
\be
\dot{\eta}+H\eta=\pm\frac{2 \sqrt{3 H^2-\Lambda  }}{\sqrt{3}
   a}-\frac{2 H}{a}.
\ee

For a scale factor of the form $a=\left(t/t_0\right)^{n}, n>0$, and by adopting the minus sign in the above equation, the functional form of the Barthel-Randers cosmological vector $\eta $ generating a cosmological constant in the first Friedmann equation is given by
\bea
\hspace{-0.6cm}\eta &=&\frac{1}{3} t^{-n} \Bigg[3 c_1-2 t \sqrt{\frac{3 n^2}{t^2}-3 \Lambda }+2 \sqrt{3} n \times \nonumber\\
\hspace{-0.6cm}&&\ln
   \left(t \sqrt{\frac{n^2}{t^2}-\Lambda }+n\right)-2 \left(3+\sqrt{3}\right) n \ln
   (t)\Bigg],
\eea
where $c_1$ is an arbitrary constant of integration, and we have again taken $t_0=1$.

Alternatively, we can generate an effective cosmological constant via the second Friedmann equation, by imposing the condition
\be\label{161}
-\frac{\ddot{\phi}}{\phi}+\frac{5}{4}\frac{\dot{\phi}^2}{\phi ^2}-H\frac{\dot{\phi}}{\phi}=\Lambda,
\ee
which would allow to write the second Friedmann equation as
\be
2\dot{H}+3H^2=-\frac{8\pi G}{c^2}p+\Lambda.
\ee

The first Friedmann equation then becomes
\be
3H^2=8\pi G\phi^2\rho +\Lambda +\frac{\ddot{\phi}}{\phi}-2\frac{\dot{\phi}^2}{\phi ^2}-2H\frac{\dot{\phi}}{\phi},
\ee
and it can be interpreted in terms of an effective energy density satisfying the standard field equation in the presence of a varying gravitational coupling.  By introducing the new variable $u=-\dot{\phi}/\phi$, Eq.~(\ref{161} takes the form of a Riccati equation,
\be
\dot{u}(t)+\frac{1}{4}u^2(t)-H(t)u(t)=\Lambda.
\ee

Once the form of the Hubble function is known, the form of the Finsler vector $\eta$ generating a cosmological constant and an effective energy density can be obtained either analytically, or with the use of numerical methods.

\subsection{Solutions with constant vector field}

As a second exact solution of the cosmological evolution equation in the Barthel-Randers geometry we consider the case in which the vector field $\eta$ is a constant, $\eta =\eta _0={\rm constant}$.  In this case, the cosmological evolution equations reduce to the form
\begin{equation}\label{c1}
3\frac{a^{\prime 2}}{a^{2}}=\left( 1+\eta _{0}a\right) ^{2}r-\frac{3\eta
_{0}^{2}a^{\prime 2}}{4\left( 1+\eta _{0}a\right) ^{2}}-\frac{3\eta
_{0}a^{\prime 2}}{a\left( 1+\eta _{0}a\right) },
\end{equation}
\bea\label{c2}
\frac{2a^{\prime \prime }}{a}+\frac{a^{\prime 2}}{a^{2}}&=&-\left( 1+\eta
_{0}a\right) ^{2}P+\frac{5\eta _{0}^{2}a^{\prime 2}}{4\left( 1+\eta
_{0}a\right) ^{2}}\nonumber\\
&&-\frac{\eta _{0}a^{\prime 2}}{a\left( 1+\eta _{0}a\right) }%
-\frac{\eta _{0}a^{\prime \prime }}{1+\eta _{0}a}.
\eea

\subsubsection{The dust Universe with $P=0$}

As a first example of a solution of the generalized Friedmann equations in a Barthel-Randers geometry with constant vector field we consider the case of a dust Universe, with $P=0$. Then Eq.~(\ref{c2}) takes the form
\be\label{117}
 4 a (1+\eta _0 a) (2+3 \eta _0 a) a''+\left[3
   \eta _0 a (\eta _0 a+4)+4\right] a'^2=0.
\ee

 By introducing a new dependent function $a'=u$, $a''=u(du/da)$, the above equation takes the form
 \be
  4 a (1+\eta _0 a) (2+3 \eta _0 a)\frac{du}{da}+\left[3
   \eta _0 a (\eta _0 a+4)+4\right]u=0,
 \ee
 with the general solution given by
 \be\label{119}
 u= \frac{c_1 (1+\eta _0 a)^{5/4}}{\sqrt{a}
   (2+3 \eta _0 a)},
 \ee
where $c_1$ is a constant of integration. Hence the general solution of Eq.~(\ref{117}) can be obtained by quadratures as
\be\label{120}
\tau-\tau_0=\int{\frac{\sqrt{a}\left(2+3\eta_0a\right)da}{c_1\left(1+\eta _0a\right)^{5/4}}},
\ee
where $\tau _0$ is a constant of integration, or,
\be\label{121}
\tau-\tau_0=\frac{4 \sqrt{a} \left[3 a \eta _0-8 (a \eta _0+1) \,
   _2F_1\left(1,\frac{5}{4};\frac{3}{2};-a \eta _0\right)+8\right]}{5 c_1
   \eta _0 \sqrt[4]{1+\eta _0a }},
\ee
where $_2F_1(a;b;c;z)=\sum_{k=0}^{\infty}{\left[(a)_k(b)_k/(c)_k\right]z^k/k!}$ is the hypergeometric function. When $\eta _0a<<1$, from Eq.~(\ref{120}) it follows that $a(\tau)\propto \tau ^{2/3}$, that is, we reobtain the standard dust solution of general relativity. The general relativistic dust solution is decelerating, with $q=1/2$. In the limit $\eta _0a>>1$, from Eq.~(\ref{120}) we obtain $a(\tau)\propto \tau ^{4/5}$, giving $q=1/4$.

Even that the expansion of the Barthel-Randers Universe is faster than the general relativistic one, it is still decelerating, but with a smaller deceleration parameter. The behavior of the scale factor can be obtained implicitly by performing a series expansion of Eq.~(\ref{121}, which gives
\be
\tau -\tau _0=\frac{4 a^{3/2}}{3 c_1}+\frac{a^{5/2} \eta _0}{5 c_1}-\frac{15 a^{7/2} \eta _0^2}{56 c_1}+O\left(a^{9/2}\right),
\ee
in which the first term corresponds to the general relativistic case, while the higher order terms give the Finslerian corrections.

The energy density of the matter is obtained generally as
\be
r(a)= \frac{3 \left(2+3 \eta _0
   a\right)^2 a'^2}{4 a^2 (1+\eta _0 a
   )^4},
\ee
giving, after the substitution of the expression of $u$,
\be
r(a)=\frac{3c_1^2}{4}\frac{1}{a^3\left(1+\eta _0a\right)^{3/2}}.
\ee

In the limit $\eta _0a<<1$ we reobtain again the standard general relativistic result.

\subsubsection{The stiff fluid Universe}

The stiff fluid equation of state, or the Zeldovich equation of state,  with $P=r$ \cite{stiff1,stiff2},  is assumed to describe the physical properties of very dense matter, having densities as high or higher than ten times the nuclear density, that is, densities greater
than $10^{17}$ g/cm$^3$, corresponding to temperatures of the order of $T = (\rho /\sigma)^{1/4} > 10^{13}$ K, where $\sigma$ is the radiation constant \cite{stiff2}. One of the important features of stiff matter is that the speed of sound $c_s$ equals the speed of light, so that $c_s^2=\partial p/\partial \rho=c^2$. This is one of the attractive features of the Zeldovich equation of state, since in a stiff fluid the speed of the perturbations cannot exceed the speed of light.

For a stiff fluid the generalized Friedmann equations in Barthel-Randers geometry can be reduced to the following single second order differential equation,
\be\label{123}
2 a \left(1+\eta _0 a\right) \left(2+3 \eta _0 a\right) a''+\left[3
   \eta _0 a \left(8+5 \eta _0 a\right)+8\right] a'^2=0.
\ee

By denoting $a'=u$, the above equation takes the form
\be
2 a \left(1+\eta _0 a\right) \left(2+3 \eta _0 a\right) \frac{du}{da}+\left[3
   \eta _0 a \left(8+5 \eta _0 a\right)+8\right] u=0,
\ee
and it has the general solution given by
\be
u=\frac{c_1 \sqrt{1+\eta _0 a }}{a^2 \left(2+3 \eta _0a
   \right)},
\ee
where $c_1$ is a constant of integration.
Hence the solution of Eq.~(\ref{123}) can be obtained as
\be
\tau-\tau_0=\frac{1}{c_1}\int{\frac{a^2\left(2+3\eta_0a\right)da}{\sqrt{1+\eta _0a}}},
\ee
or
\be
\tau (a)-\tau_0=\frac{2 \sqrt{1+\eta _0a } \left(45 \eta _0^3a^3 -12 \eta _0^2a^2 +16 \eta _0 a -32\right)}{105 c_1 \eta _0^3}.
\ee

The matter density can be found from Eq.~(\ref{c1}), and it is given by
\be
r(a)=\frac{3 c_1^2}{4 a^6 \left(1+\eta _0 a\right)^3}.
\ee

At the initial moment $\tau=\tau_0$, the scale factor is obtained from the algebraic equation $45 \eta _0^3a^3 -12 \eta _0^2a^2 +16 \eta _0 a -32 =0$, and it has the finite value $a\left(\tau_0\right)=a_0\approx 38/45\eta _0$. The energy density has the finite value $r\left(\tau _0\right)\approx 0.33c_1^2\eta _0^6$, which sensitively depends on the numerical value of $\eta _0$. In the standard general relativistic scenario the stiff fluid is described by $a(\tau)=\tau ^{1/3}$, and $\rho (a)=\rho _0/a^6$. The solution is singular at $\tau =0$. This singularity is removed in the osculating Barthel-Randers cosmology in the presence of a constant vector,  generated from the Finsler type geometry.

\subsection{The case $(a\eta)'=\beta(1+a\eta)$ }

Next we consider some exact solutions of the generalized Friedmann equations in Barthel-Randers geometry that satisfy the condition
\be
\frac{a'\eta+a\eta'}{1+a\eta}=\beta,
\ee
where $\beta $ is a constant. This condition can be reformulated as a first order differential equation for $a\eta$, given by
\be
(a\eta)'=\beta+\beta(a\eta),
\ee
with the general solution given by
\be
a\eta=C_1e^{\beta \tau}-1.
\ee

Then, by assuming $P=0$, Eq.~(\ref{feq6}) becomes
\be
2h'+3h^2-\beta h-\frac{\beta ^2}{4}=0,
\ee
with the general solution satisfying the initial condition $h(0)=C_0$ given by
\be
h(\tau)=\frac{\beta }{2}-\frac{2 \beta  \left(\beta -2 C_0\right)}{(\beta +6
   C_0)e^{\beta  \tau } +3 \left(\beta-2 C_0\right)}.
\ee

For the scale factor we obtain
\be
a(\tau)=e^{-\frac{\beta  \tau }{6}} \left[e^{\beta  \tau } (\beta +6 C_0)+3 \left(\beta -2
   C_0\right)\right]^{2/3}.
\ee

In the limit of large times the Hubble function tends to $h(\tau)=\beta /2$, while the scale factor is given by an exponential function $a(\tau)=e^{\beta \tau/2}$.  This gives the interpretation of the constant $\beta $ as $\beta =2h_0$, where $h_0$ is the present day value of the Hubble function. The deceleration parameter is given by
\be
q=\frac{8  \left( 2 C_0-\beta\right) \left(\beta +6 C_0\right)e^{\beta  \tau }}{\left[
    \left(\beta +6 C_0\right)e^{\beta  \tau } +2 C_0-\beta\right]^2}-1.
\ee

If the condition  $2C_0h_0+h_0^2<C_0^2$ is satisfied, then, at $\tau =0$, $q(0)>0$, and the Universe begins its expansion from a decelerating phase. At the moment $\tau _{cr}=\ln \left[\left(3+2\sqrt{2}\right)\left(C_0-h_0\right)\left(3C_0+h_0\right)\right]$, the deceleration parameter is zero, $q\left(\tau _{cr}\right)=0$, and for $\tau >\tau_{cr}$ the Universe enters into an accelerating phase that ends with a de Sitter type expansion. The energy density of the matter varies according to
\be
r(\tau)=\frac{12}{C_1^2e^{4h_0\tau}}\left[h_0+\frac{2h_0\left(C_0-h_0\right)}{\left(h_0+3C_0\right)e^{2h_0\tau}-3\left(C_0-h_0\right)}\right]^2,
\ee
and it tends to zero in the limit of large times. As for the vector field $\eta$, its time dependence is obtained as
\begin{equation}
\eta \left( \tau \right) =\frac{e^{h_{0}\tau /3}\left( C_{1}e^{2h_{0}\tau
}-1\right) }{\left[ \left( 2h_{0}+3C_{0}\right) e^{2h_{0}\tau }+6\left(
h_{0}-C_{0}\right) \right] ^{2/3}}.
\end{equation}

\section{Numerical analysis}\label{sect5}

In this Section, we will analyze the cosmological implications of the model by comparing it with the realistic, high precision,  observational data. We assume that the ordinary matter content of the Universe consists of radiation, with energy density $\rho_r$, and pressure $p_r=\rho_r c^2/3$, and of pressureless dust, with energy-density $\rho_m$. We also add to the basic model the cosmological constant $\Lambda$.

In order to simplify the mathematical and numerical formalism  we introduce the following set of dimensionless variables $\left(\tau, h, \Omega _{\Lambda}\bar{\rho}\right)$, defined as
\be
\tau=H_0t, H=H_0h,\Omega_\Lambda=\frac{\Lambda c^2}{3H_0^2},\bar\rho_i=\frac{8\pi G\rho_i}{3H_0^2}, i=r,m,
\ee
where $H_0$ is the current value of the Hubble parameter.
It should be mentioned that the field $\phi$ is dimensionless. 

The conservation equation \eqref{cons} can naturally be decomposed into two equations, governing the conservation of radiations and dust components as
\begin{align}\label{con1}
\frac{d}{d\tau}\bar{\rho}_m+3h\bar\rho_m+\frac32\frac{1}{\phi}\frac{d\phi}{d\tau}\bar\rho_m=0,
\end{align}
and
\begin{align}\label{con2}
\frac{d}{d\tau }\bar{\rho}_r+4h\bar\rho_r+2\frac{1}{\phi}\frac{d\phi}{d\tau}\bar\rho_r=0,
\end{align}
respectively.

The generalized Friedmann  equations can be written in the dimensionless variables introduced above as
\begin{align}
h^2=\phi^2(\bar\rho_m+\bar\rho_r+\Omega_\Lambda)-h\frac{1}{\phi}\frac{d\phi}{d\tau}+\frac{1}{4}\frac{1}{\phi ^2}\left(\frac{d\phi}{d\tau}\right)^2,
\end{align}
\bea
2\frac{dh}{d\tau}+3h^2&=&\phi^2(3\Omega_\Lambda-\bar\rho_r)-\frac{1}{\phi}\frac{d^2\phi}{d\tau ^2}\nonumber\\
&&+\frac54\frac{1}{\phi ^2}\left(\frac{d\phi}{d\tau}\right)^2-h\frac{1}{\phi}\frac{d\phi}{d\tau}.
\eea

It should be noted that in the present Section, we have added the cosmological constant to the cosmological equations. We will see that the cosmological constant can be obtained from the model parameters, and also from the current matter abundances, and, for the simplified models we are considering, it is still necessary to have a solution that is fully compatible with the observational data.

In order to compare the present theory with observational data, we will rewrite the Friedmann equations by using the redshift variable defined in Eq.~\eqref{redshift1}, and with the time derivative obtained by using Eq.~\eqref{redshift2}. In the redshift variable, one can analytically solve the conservation equations \eqref{con1} and \eqref{con2}, with the results
\begin{align}
\bar\rho_m=\frac{\Omega_{m0}(1+z)^3}{\phi^{3/2}},
\end{align}
and
\begin{align}
\bar\rho_r=\frac{\Omega_{r0}(1+z)^4}{\phi^{2}},
\end{align}
respectively. Here $\Omega_{m0}$ and $\Omega_{r0}$ are integration constants. In the following we normalize $\phi$ such that  $\phi(z=0)=1$. In this case $\Omega_{m0}$ and $\Omega_{r0}$ becomes the current values of the density abundances for dust and radiation, respectively.

Using the above relations, one can obtain the Hubble parameter as
\begin{align}
h(z)=\frac{2\phi\sqrt{(1+z)^4\Omega_{r0}+(1+z)^3\sqrt{\phi}\Omega_{m0}+\phi^2\Omega_\Lambda}}{(1+z)\left(d\phi/dz\right)-2\phi}.
\end{align}

Evaluating the Hubble parameter at $z=0$ and taking into account that $h(0)=1$ by definition, one can obtain the cosmological constant as
\begin{align}
\Omega_\Lambda=\left(1-\frac12\phi^\prime(0)\right)^2-\Omega_{m0}-\Omega_{r0},
\end{align}
where here, {\it the prime denotes the derivative with respect to the redshift} $z$.

For the deceleration parameter we obtain
\begin{eqnarray}
&&q(z)=  \nonumber \\
&&-\frac{(1+z)}{4\phi ^{2}\left[ \Omega _{r0}(z+1)^{4}+\Omega _{\Lambda
}\phi ^{2}+\Omega _{m0}(z+1)^{3}\sqrt{\phi }\right] ^{3/2}}  \nonumber \\
&&\times \Bigg\{2\Omega _{m0}(z+1)^{4}\phi ^{3/2}\phi ^{\prime \prime }-%
\frac{5}{2}\Omega _{m0}(z+1)^{4}\sqrt{\phi }\phi ^{\prime 2}  \nonumber \\
&&+4(z+1)\phi ^{2}\left[ 2\Omega _{r0}(z+1)^{2}-\Omega _{\Lambda }\phi
^{\prime 2}\right]   \nonumber \\
&&-2\Omega _{r0}(z+1)^{5}\phi ^{\prime 2}+2\Omega _{\Lambda }\phi ^{3}\Big[%
(z+1)\phi ^{\prime \prime }+3\phi ^{\prime }\Big]  \nonumber \\
&&+2\Omega _{r0}(z+1)^{4}\phi \left[ (z+1)\phi ^{\prime \prime }-\phi
^{\prime }\right]   \nonumber \\
&&+6\Omega _{m0}(z+1)^{2}\phi ^{5/2}\Bigg\}-1.
\end{eqnarray}

The transition from a decelerating state to an accelerating one occurs at a redshift $z=z_{tr}$, satisfying the equation $q\left(z_{tr}\right)=0$.

In the following, we will consider some special choices for the redshift dependence of the function $\phi$, and analyze their cosmological implications. We will also compare the theoretical models with the observational data.

\subsection{Linear case: $\phi(z)=1+\delta z$}

Let us first assume that $\phi$ is a linear function of the redshift $z$, $\phi (z)=1+\delta z$, where $\delta $ is a constant. Because of our normalization, the constant term can be taken to be equal to unity. In order to compare the model with the cosmological observations, we will estimate the best fit values of the parameters $H_0$ and $\delta $ by using the recent observational data on the Hubble parameter \cite{hubble1, hubble2}.

In order to do this, we use the likelihood analysis of the model, based on the data on $H_0$. In the case of independent data points, the likelihood function can be defined as
\begin{align}
L=L_0e^{-\chi^2/2},
\end{align}
where $L_0$ is the normalization constant, and $\chi^2$ is defined as
\begin{align}
\chi^2=\sum_i\left(\frac{O_i-T_i}{\sigma_i}\right)^2.
\end{align}

Here $i$ counts the data points, $O_i$ are the observational values, $T_i$ are the theoretical values, and $\sigma_i$ are the errors associated with the $i$th data from observation. One can then write the Likelihood function as
\begin{align}
L=L_0\,\textmd{exp}\left[-\frac12\sum_i\left(\frac{O_i-T_i}{\sigma_i}\right)^2\right],
\end{align}

By maximizing the likelihood function one can find the best fit values of the parameter $\delta $, and also the numerical value of the current Hubble parameter $H_0$. In Table ~\ref{tab1} we have summarized the results for the linear model. We have also computed the transition redshift of the model, together with its $1\sigma$ and $2\sigma$ intervals. The transition redshift for the $\Lambda$CDM-based observations can be inferred as $z_{tr}=0.65$. We can see that the results of the model are in agreement with the observational data.
\begin{table}[h!]
	\begin{center}
		\begin{tabular}{|c||c|c|c|}
			\hline
			~~~~~~~&~~~Best fit value~~~&~~~$1\sigma$~interval~~~&~~~$2\sigma$~interval~~~\\
				\hline
		$\delta $&0.027&$\pm0.023$&$\pm0.047$\\
		\hline
		$H_0$&67.8&$\pm1.41$&$\pm2.78$\\
			\hline
			$z_{tr}$&0.58&$\pm0.06$&$\pm0.11$\\
			\hline
		\end{tabular}
	\end{center}
	\caption{Best fit values of the model parameters $\delta$, $H_0$ and $z_{tr}$, together with their $1\sigma$ and $2\sigma$ confidence intervals for the linear case $\phi (z)=1+\delta z$.\label{tab1}}
\end{table}

\begin{figure*}[htbp]
	\includegraphics[scale=0.41]{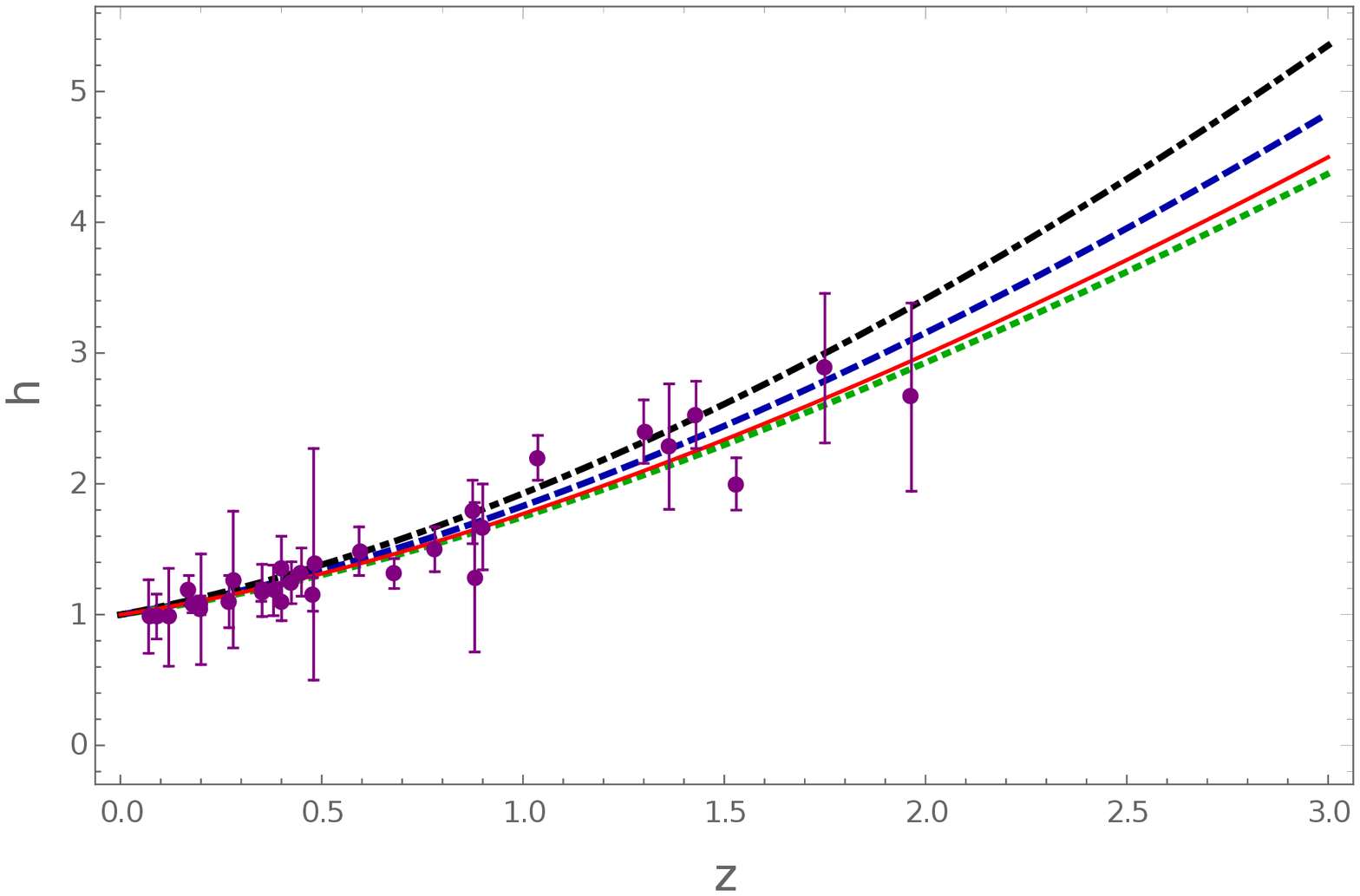}~~\includegraphics[scale=0.41]{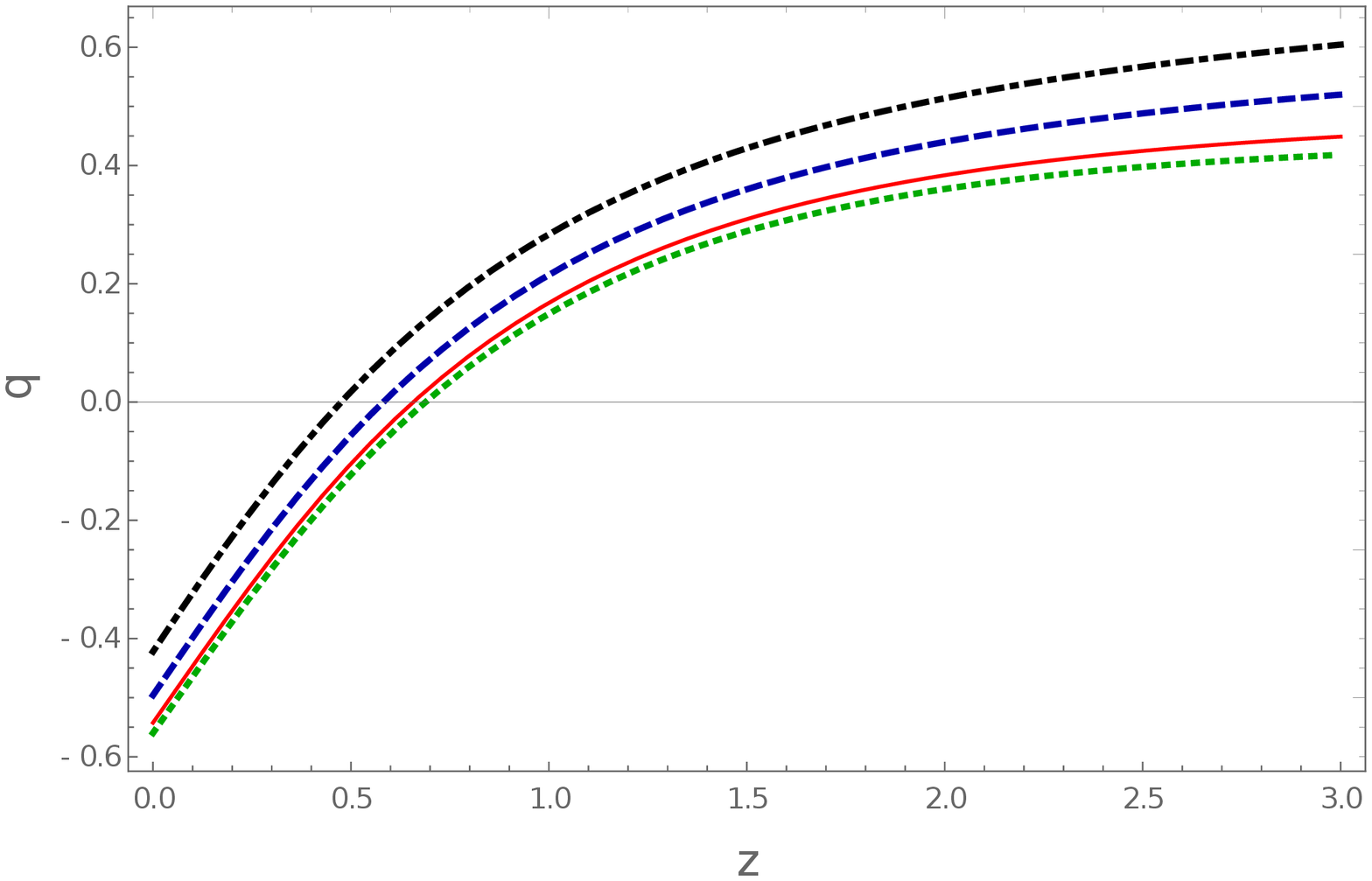}
	\caption{\label{fig1}The evolution of the Hubble parameter (left panel) and of the deceleration parameter (right panel) as a function of the redshift $z$ in the linear case $\phi(z)=1+\delta z$. The dashed, dotted and dot-dashed curves correspond to the best fit and lower/higher $2\sigma$ extreme values of the parameter $\delta $ in the Table~\ref{tab1}. The red solid line corresponds to the $\Lambda$CDM model, and the error bars represent the observational data.}
\end{figure*}

In Fig.~\ref{fig1}, we have plotted the Hubble parameter and the deceleration parameter as a function of the redshift, for the best values of the parameter $\delta $, and also for the extreme $2\sigma$ values. By the red solid line we have represented the predictions of the $\Lambda$CDM model, while the error bars correspond to the observational data.

Since $\phi (z)=1+a(z)\eta (z)=1+\eta (z)/(1+z)$, it follows that in this theoretical model the Barthel-Randers vector $\eta$ increases with the redshift according to
\be
\eta (z)=\delta z\left(1+z\right).
\ee

\subsection{Logarithmic case: $\phi(z)=1+\ln (1+\delta z)$}

Let us consider now a logarithmic ansatz for the function $\phi$ as $\phi (z)=1+\ln(1+\delta z)$, where $\delta $ is a constant. The function $\phi$  is constructed in such a way that at $z=0$, the $\phi$ reduces to unity, as was discussed earlier. In the case of the logarithmic dependence of $\phi$, the best fit value together with their $1\sigma$ and $2\sigma$ confidence intervals for the parameters $H_0$ and $\delta$ are summarized in Table~\ref{tab2}. As one can see from the Table, the present day value of the Hubble function as well as the transition redshift are in agreement with the observational data.

\begin{table}[h!]
	\begin{center}
		\begin{tabular}{|c||c|c|c|}
			\hline
			~~~~~~~&~~~Best fit value~~~&~~~$1\sigma$~interval~~~&~~~$2\sigma$~interval~~~\\
			\hline
			$\delta $&0.029&$\pm0.025$&$\pm0.050$\\
			\hline
			$H_0$&67.75&$\pm1.41$&$\pm2.77$\\
			\hline
				$z_{tr}$&0.57&$\pm0.04$&$\pm0.09$\\
			\hline
		\end{tabular}
	\end{center}
	\caption{Best fit values of the model parameters $\delta$, $H_0$ and $z_{tr} $ together with their $1\sigma$ and $2\sigma$ confidence intervals for the logarithmic model $\phi (z)=1+\ln(1+\delta z)$.\label{tab2}}
\end{table}

\begin{figure*}[htbp]
	\includegraphics[scale=0.41]{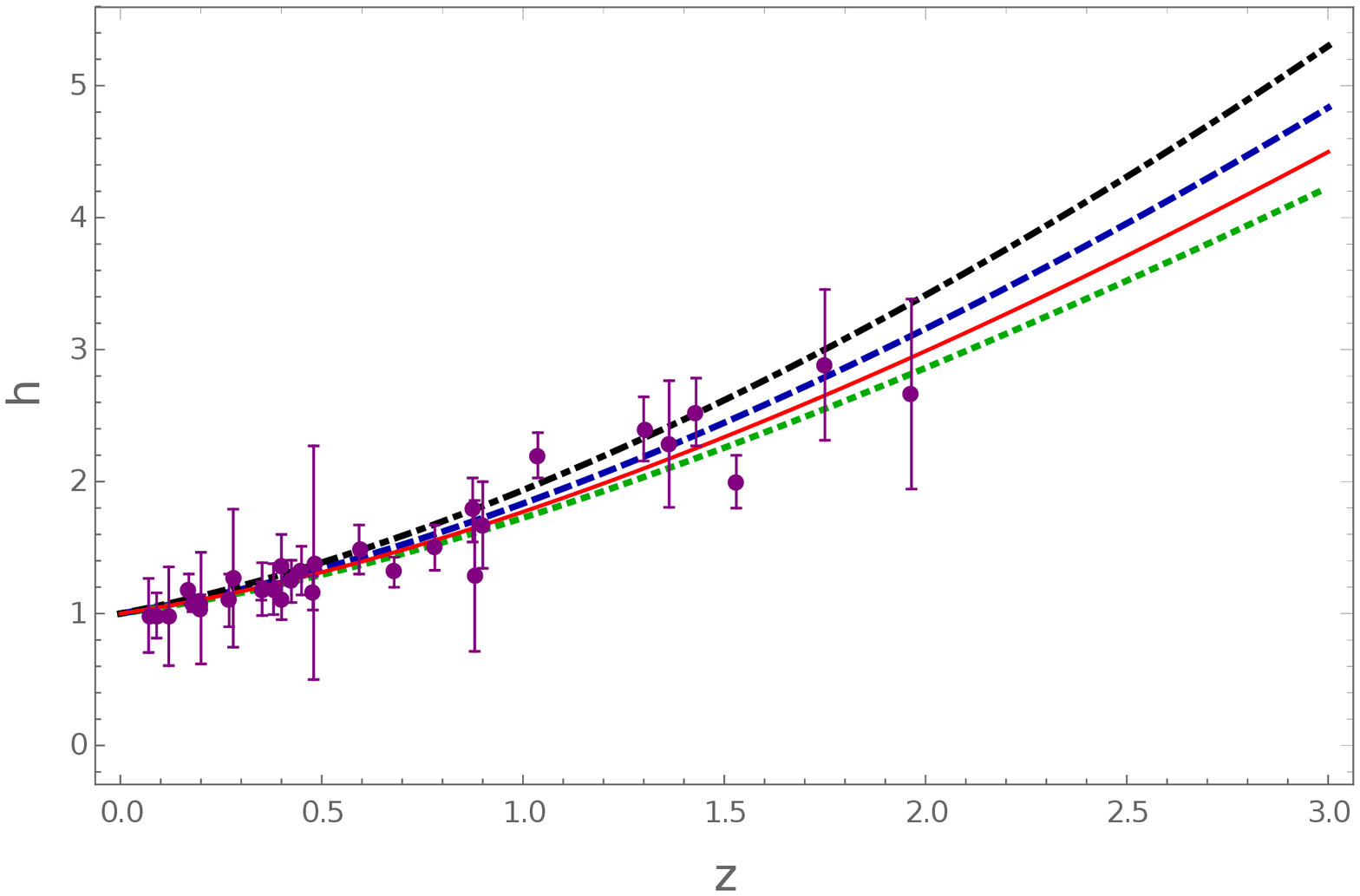}~~\includegraphics[scale=0.41]{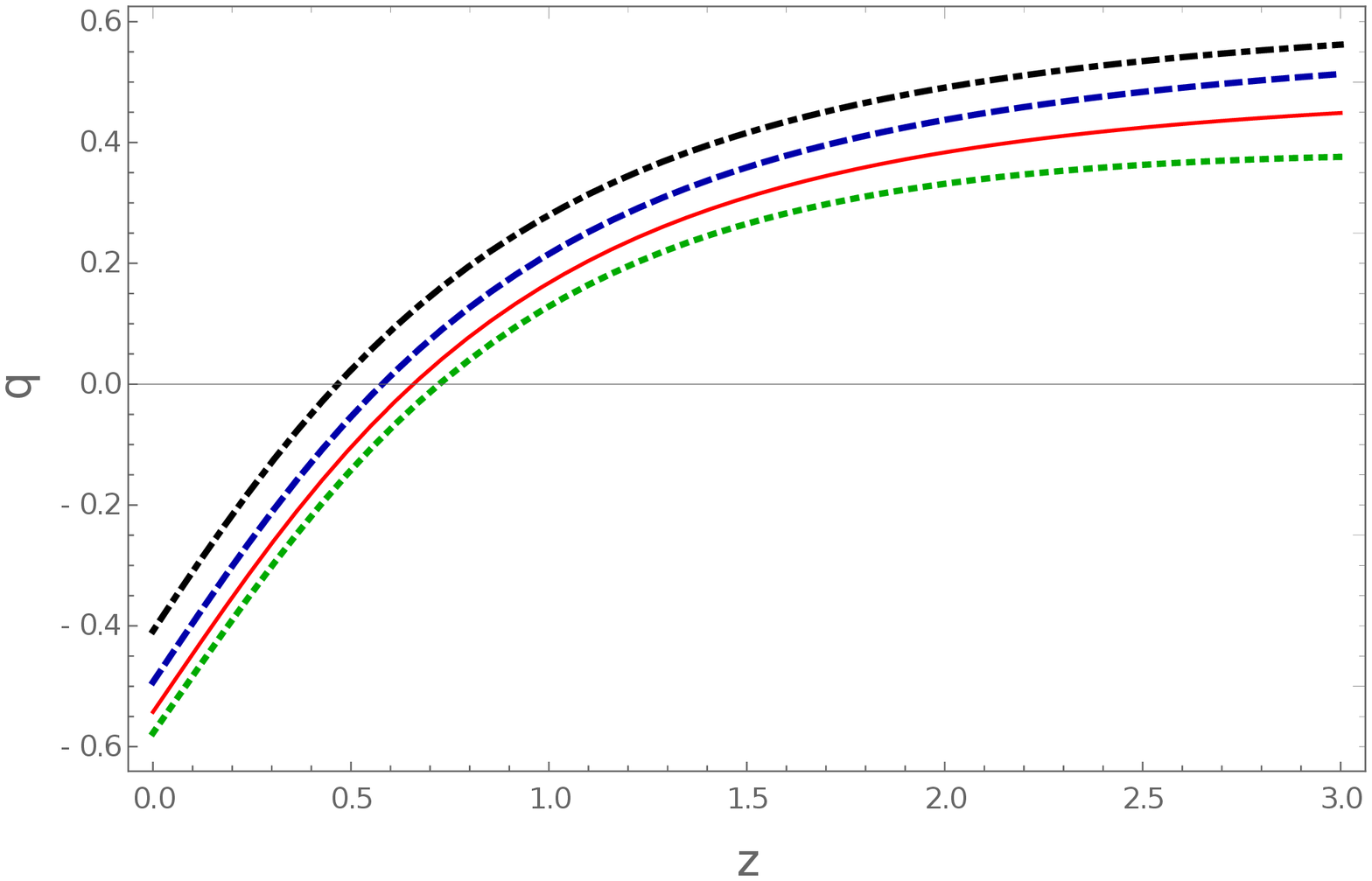}
	\caption{The evolution of the Hubble parameter (left panel) and of the deceleration parameter (right panel) as a function of the redshift $z$ in the logarithmic case $\phi(z)=1+\ln(1+\delta z)$. The dashed, dotted and dot-dashed curves correspond to the best fit and lower/higher $2\sigma$ extreme values of the parameter $\delta $ in Table~\ref{tab2}. The red solid line corresponds to the $\Lambda$CDM model, and the error bars represent the  observational data.}\label{fig2}
\end{figure*}

The evolution of the Hubble and of the deceleration parameters are also depicted in Fig.~\ref{fig2}. One can see that the qualitative behavior of this ansatz for $\phi$ is similar to the linear case.

The Barthel-Randers vector field $\eta$ is given in this model, as a function of redshift, by
\be
\eta (z)=(1+z)\ln (1+\delta z).
\ee

\subsection{Exponential case: $\phi (z)=e^{2\delta z}$}

Let us finally consider the case in which the function $\phi$ has an exponential dependence on the redshift,  $\phi (z)=\exp(2\delta z)$. In this case, the best fit values for $\delta$, $H_0$ and $z_{tr}$ together with their $1\sigma$ and $2\sigma$ confidence intervals, are shown in Table~\ref{tab3}.

\begin{table}[h!]
	\begin{center}
		\begin{tabular}{|c||c|c|c|}
			\hline
			~~~~~~~&~~~Best fit value~~~&~~~$1\sigma$~interval~~~&~~~$2\sigma$~interval~~~\\
			\hline
			$\delta $&0.013&$\pm0.011$&$\pm0.022$\\
			\hline
			$H_0$&67.83&$\pm1.42$&$\pm2.78$\\
			\hline	$z_{tr}$&0.58&$\pm0.04$&$\pm0.09$\\
			\hline
		\end{tabular}
	\end{center}
	\caption{Best fit values of the model parameters $\delta$, $H_0$ and $z_{tr} $ together with their $1\sigma$ and $2\sigma$ confidence intervals for the exponential case $\phi (z)=\exp(2\delta z)$.\label{tab3}}
\end{table}

\begin{figure*}
	\includegraphics[scale=0.41]{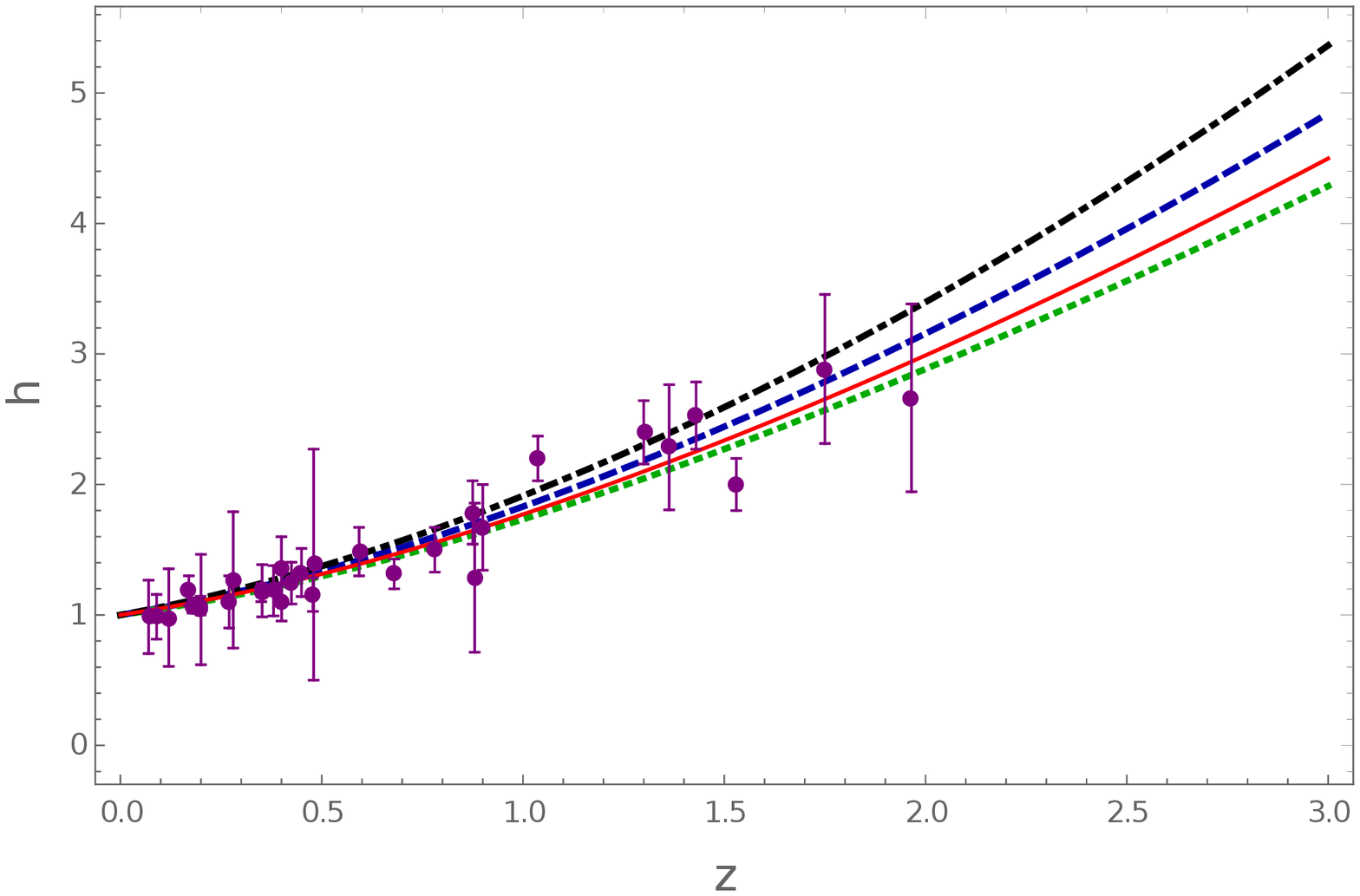}~~\includegraphics[scale=0.41]{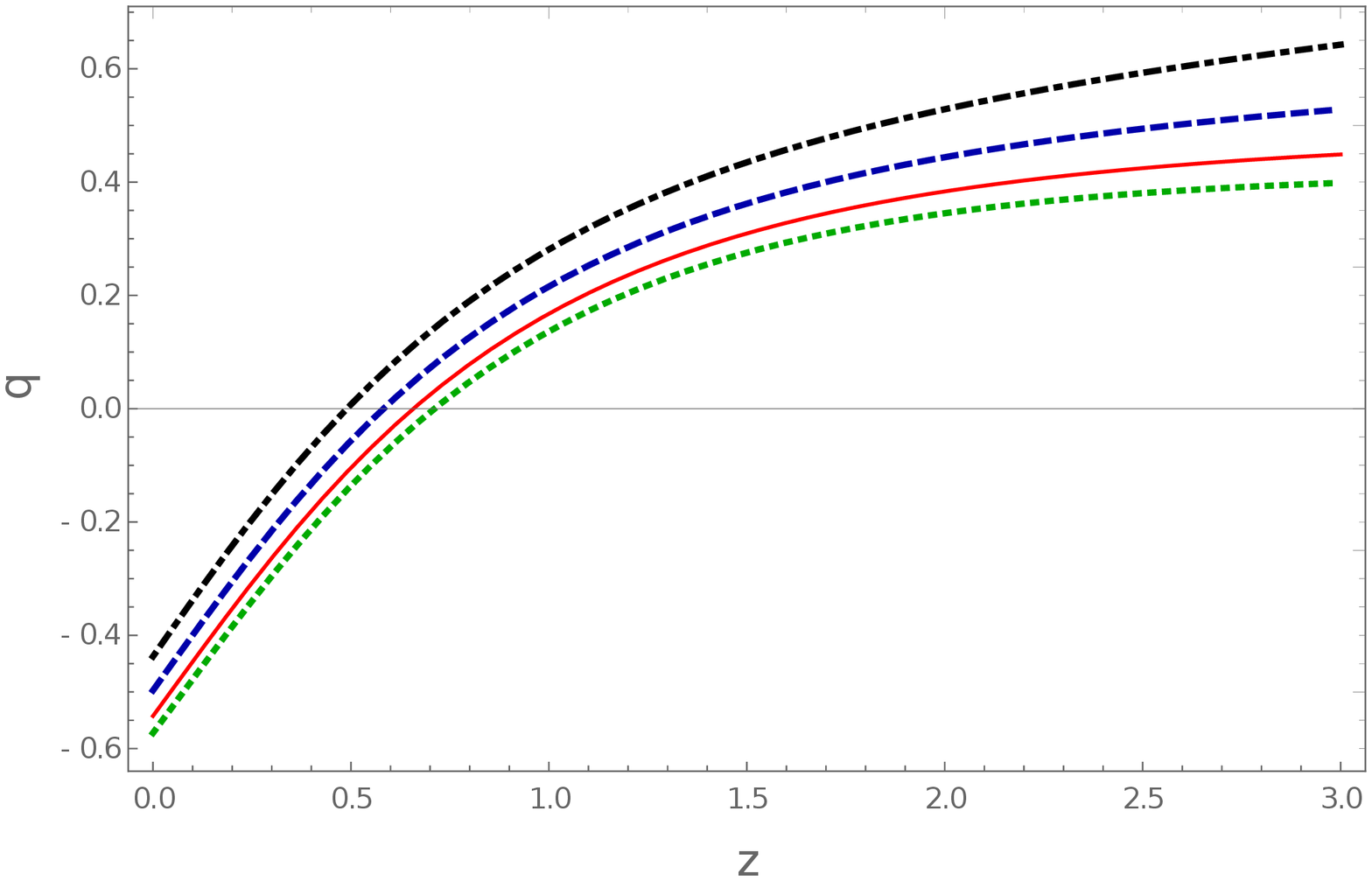}
	\caption{\label{fig3}The evolution of the Hubble parameter (left panel) and of the deceleration parameter (right panel) as a function of the redshift in the exponential case $\phi (z)=\exp(2\delta z)$. The dashed, dotted and dot-dashed curves correspond to the best fit and lower/higher $2\sigma$ extreme values of the parameter $\delta$ in Table~\ref{tab2}. The red solid line corresponds to the $\Lambda$CDM model, and the error bars represent the observational data.}
\end{figure*}

As one can see from Table~\ref{tab3}, the present day value of the Hubble function, as well as the transition redshift, are in agreement with the observational data. Also, in Fig.~\ref{fig3}, we have represented the behavior of the Hubble and of the deceleration parameters as a function of the redshift.

As for the vector field $\eta$, its redshift dependence is given in this model by
\be
\eta (z)=(1+z)\left(e^{2\delta z}-1\right).
\ee

As one can see from the above discussion of the three considered special cases, the Barthel-Randers cosmology can in principle satisfy all the existing basic observational data. The Hubble parameter for the best fit values of the model parameter $\beta$ is for all cases slightly greater than its $\Lambda$CDM counterpart. This shows that, at least for the considered models, the Barthel-Randers cosmological theory predicts smaller universes as compared to the $\Lambda$CDM theory. Also, the present theory predicts lower levels of the acceleration of the Universe, as  compared to the $\Lambda$CDM model. This result is also supported by the dynamical evolution of the Hubble parameter of the models.

\section{Discussions and final remarks}\label{sect6}

There is an almost general consensus in the scientific community that the gravitational interaction can be described only in geometrical terms. However, no such general consensus does exist with respect to which geometry is best fitted to characterize gravity. The initial Riemannian framework was extended to geometric theories containing Weyl-type nonmetricity \cite{W1,W2,Q1}, torsion \cite{r10}, or Weitzenb\"{o}ck type geometries \cite{r16,r17}.

In the present paper we have investigated another geometric perspective on the gravitational phenomena, which is offered by the Finsler geometry. In a Finsler space the independent variable is the line element $(x,y)$, where $y$ is a vector obeying linear transformations,  instead of the point $(x)$ of the Riemannian geometry. Therefore the $y$-dependence essentially characterizes Finsler spaces, and from the physical point of view $y$ plays the role of an internal variable associated with each point $x$. Hence, this dependence can also be viewed in terms of the concepts of anisotropy and nonlocality. However, from a physical point of view one can also assume that under some specific circumstances the vector $y$ becomes a function of $x$. Such a situation may appear, for example, if we interpret $y$ as corresponding to quantum space-time fluctuations \cite{Ikeda}, and we take the average of $y$. The geometry of the Finsler spaces with metric $\left.\hat{g}(x,y)\right|_{y=Y(x)}$ is well developed from mathematical point of view, and leads to interesting physical consequences. In our study we have considered such a geometric structure, constructed over an osculating Riemann geometry, and described by a Barthel connection. Moreover, we have considered a particular $(\alpha, \beta)$ metric, the Randers metric, with the property that the associated Barthel connection reduces to the Levi-Civita connection for $\left.\hat{g}(x,y)\right|_{y=Y(x)}$. In this geometric framework we have obtained the generalized Friedmann equations that describe the evolution of the Universe with Riemann metric $g_{AB}(x)$. The present theory can be interpreted as a two-metrics theory. On the one hand we have the physical Riemann metric $g$, introduced via the term $\alpha$ in the Randers line element, and the associated Finsler metric $\left.\hat{g}(x,y)\right|_{y=Y(x)}$, which also depends on the one-form $A_I$ introduced through the Finsler function $F=\alpha +\beta$. In the cosmological case the one form generates some multiplicative terms in the Finsler metric, with respect to the Riemannian one, as one can see from Eqs.~(\ref{mF}). Since the $y$ dependence is now carried by $A_0$, one could interpret physically $A$ as describing a kind of effective quantum fluctuations of the metric.

Such a physical interpretation may also be supported by the nonconservation of the matter energy-momentum tensor in Eq.~(\ref{cons}). In quantum field theories in curved-space times particle production naturally occur, and it is an intrinsic property of these theories. Therefore, if $y$, and $A$, respectively, can be associated to quantum fluctuations of the space-time, then the appearance of some particle creation processes may be inevitable. We have provided a qualitative interpretation of the balance equation in Barthel-Randers geometry by using the thermodynamics of open systems, in which irreversible particle creation takes place, and we have obtained the basic physical parameters (creation rate, temperature, entropy) that describe such processes. The particle creation rate is essentially determined by the Barthel-Randers vector $\eta$, that originates from the Randers line element, and, as we have earlier discussed, can be interpreted as an averaged space-time fluctuation. In this way particle creation in the present model can be explained as a quantum process.

It is interesting to point out that the nonconservation of the matter energy-momentum tensor does appear in other Finslerian geometrical models. For example, in the scalar-tensor theories that arise effectively from the Lorentz fiber bundle of a Finsler-like geometry, investigated in \cite{Fc19a}, the matter energy-momentum tensor satisfies the balance equation $\nabla _{\mu}T^{\mu}_{\nu}=-\left(\partial _{\nu}\phi/\phi\right)\mathcal{L}_m$, where $\mathcal{L}_m$ is the matter Lagrangian. For a FLRW type geometry, the conservation equations for matter and dark energy can be written in this model  as \cite{Fc19a}
\be\label{ncons1}
\dot{\rho}_m+3H\left(\rho _m+P_m\right)=-\frac{2\dot{\phi}}{\phi}\rho _m,
\ee
and
\be\label{ncons2}
\dot{\rho}_{DE}+3H\left(\rho _{DE}+P_{DE}\right)=\frac{2\dot{\phi}}{\phi}\rho _m,
\ee
respectively. As a result of the intrinsic Finsler type geometrical structure, an interaction between matter and the dark energy sector is generated, which leads to a rich cosmological behavior. The total energy density balance equation of the Barthel-Randers geometric model, Eq.~(\ref{cons}), can be reformulated in a similar way to Eq.~(\ref{ncons1}) as
\begin{equation}\label{ncons3}
\dot{\rho}+3H\left( \rho +\frac{p}{c^2}\right) =-\frac{3}{2}\frac{\dot\phi}{\phi}\left( \rho +\frac{p}{c^2}\right).
\end{equation}
By assuming that the total density and pressure can be represented as $\rho=\rho_m+\rho_{DE}$ and $p=p_m+p_{DE}$, respectively,  then Eq.~(\ref{ncons3}) can be split as
\be
\dot{\rho}_m+3H\left(\rho _m+P_m\right)=-\frac{2\dot{\phi}}{\phi}\left(\rho _m+\rho_{DE}+\frac{p_m+p_{DE}}{c^2}\right),
\ee
and
\be
\dot{\rho}_{DE}+3H\left(\rho _{DE}+P_{DE}\right)=\frac{1}{2}\frac{\dot{\phi}}{\phi}\left(\rho _m+\rho_{DE}+\frac{p_m+p_{DE}}{c^2}\right),
\ee
respectively. Hence, we can obtain an alternative cosmological interpretation of the Barthel-Randers model as describing the interaction between the ordinary matter and the dark sector of the Universe, as initially suggested in the framework of Finsler cosmology in \cite{Fc19a}. It is also interesting to note that the matter decay and dark energy generation depend on the thermodynamic quantities describing both components, as well as on the scalar field $\phi$, and its derivative.

 One of the fundamental symmetries in nature is represented by the Lorentz invariance. In the limit of $\phi =1$, corresponding to $\eta =0$, we recover general relativity, and the corresponding Lorentz invariance. However, when $\phi \neq 1$, the deviation from Riemannian geometry induces a breaking of Lorentz invariance. As pointed out in \cite{Fc14}, the departures from Lorentz invariance are parameterized by the 1-form field $\beta$. Recently, bounds on the Lorentz invariance violation were obtained from the observations of gamma-ray burst GRB 190114C \cite{magic}, indicating that a lower bound of the quantum energy scale for the modification of the linear photon dispersion relation is of the order of $E_{QG}>0.59\times 10^{19}$ GeV. On the other hand, limits on the Lorentz invariance violation in the gravitational sector, obtained from Gravity Probe B, indicate an independent limit of $10^{-7}$ \cite{ProbeB}. Similar, or weaker bounds are obtained from the  analysis of the Solar System data using the Parameterized Post-Newtonian (PPN) formalism \cite{Will}. Hence, in order to obtain consistency with the observational data regarding Lorentz invariance, the zero component of the vector field $A_I$ must satisfy the constraint $a\eta <10^{-7}$. At the present time, corresponding to $z=0$ and $a(0)=1$, the constraint on $\eta$ can be formulated as $\eta (0)<10^{-7}$. In the case of the linear cosmological model, $\eta$ is given by $\eta(z)=\delta z\left(1+z\right)$, and the Lorentz violation constraint is satisfied in the limit $z\rightarrow 0$. A similar situation occurs in both logarithmic and exponential cases, with $\eta (z)=(1+z)\ln (1+\delta z)$, and $\eta (z)=(1+z)\left(e^{2\delta z}-1\right)$, and $\lim_{z\rightarrow 0}\eta (z)=0$. Hence, in the present day Universe Lorentz invariance is maintained, but its violation is expected in the earlier phases of the cosmological expansion.

 On the other hand, in the present approach we assume that the field $\phi=1+a\left(x^0\right)\eta \left(x^0\right)$ is a purely geometric quantity. From a classical point of view, the field $\phi$ encompasses the Finslerian geometric properties of the space-time. Geometrically, it is given by the time component of the 1-form field $\beta$ that appears in the Randers metric. The vector $A_I$ was initially interpreted by Randers \cite{Rand}, in a tentative to construct a unified theory of the gravitational and electromagnetic forces, as the electromagnetic four-potential. From a physical point of view $A_I$ can be seen as a U(1) gauge field. Hence, generally, the present approach corresponds physically to a {\it vector-tensor type theory}, and not to a scalar-tensor one. Since we are constructing the theory on a purely geometric basis, we do not impose any equation of motion for the field $A_I$, and, implicitly, to $\phi$, which in the present approach remains arbitrary.  Its main properties, as well as its functional form may be determined from observations, fixing in this way, the empirical form of the Finsler geometry of the space-time. On the other hand, a fundamental physical theory, based on a quantum gravitational approach, may provide an equation of motion for the gauge field $A_I$,  thus leading to the full determination of the Finsler function $F$.

From a cosmological point of view the generalized Friedmann equations have the important property of admitting a de Sitter type solution, which corresponds to a specific functional form of $\eta $. This opens the possibility of the explanation of the recent acceleration of the Universe by the existence of the Barthel-Randers geometry. Moreover, an effective cosmological constant can also be generated from the model, and it appears for all values of the scale factor. A cosmological constant in the Friedmann equations can be introduced in two ways, by modifying the first or the second  equation. However, the presence of the cosmological constant in one equation implies the introduction of an effective energy density, or pressure, in the other equation. Several models with a constant vector field $\eta$, which lead to exact solutions of the Friedmann equations, have been also investigated. An interesting model is obtained for $\eta$ satisfying the condition $(a\eta)'=\beta(1+a\eta)$, from which a complete description of a Universe beginning its evolution in a decelerating era, and ending in a de Sitter accelerating phase, can be obtained. For such a solution the field $\eta$ exponentially increases with the cosmological time, and this increase continues even when the matter density becomes negligibly small, but with the Universe expanding exponentially.

We have also considered a comparison of the Barthel-Randers model predictions with observations. For this we have adopted three simple functional forms of $\phi$ as a function of the redshift, and we have fitted the Hubble function of the theory with the observational data. It turns out that despite their simplicity,  the models can provide a satisfactory description of the observations.

In this work we have investigated some of the cosmological implications of a particular version of the Finsler geometry, and we have provided some basic theoretical tools that may help the in depth  investigation of the astrophysical and cosmological applications of this model, and, generally, of Finsler geometries.

\section{Acknowledgments}

We would like to thank the anonymous reviewer for comments and suggestions that helped us to significantly improve our manuscript. We thank Prof. Hideo Shimada for many useful suggestions and discussions. R. H. was financially supported by Office of the Permanent Secretary, Ministry of Higher Education, Science, Research and Innovation. Grant No. RGNS 63-241. The work of TH was partially supported by a grant of the Romanian Ministry of Education and
Research, CNCS-UEFISCDI, project number PN-III-P4-ID-PCE-2020-2255 (PNCDI III).

\appendix

\section{Computation of the coefficients $\hat{\gamma}_{ijk}$}\label{appa}

In the present Appendix we will present the details of the derivation of Eq.~(\ref{eq36}). Let $F=\alpha +\beta$ be the fundamental function of an $(\alpha, \beta)$ metric, with $\alpha=\sqrt{\epsilon g_{ij}(x)y^iy^j}$, $\epsilon=\pm 1$, and $\beta =A_i(x)y^i$, respectively. We introduce the following quantities,
\be
\hat{g}_{ij}(x,y)=\frac{F}{\alpha}h_{ij}+l_il_j,
\ee
where $h_{ij}$ is the angular metric of the Riemannian space $(M,\alpha)$, $l_i:=\dfrac{\partial F}{\partial y^i}$ and
\be
\hat{\gamma}_{ijk}(x,y):=\frac{1}{2}\left(\frac{\partial \hat{g}_{ik}(x,y)}{\partial x^j}+\frac{\partial \hat{g}_{ij}(x,y)}{\partial x^k}-\frac{\partial \hat{g}_{jk}(x,y)}{\partial x^i}\right),
\ee
\be
\Gamma _{ijk}:=\frac{1}{2}\left(\frac{\partial g_{ik}(x)}{\partial x^j}+\frac{\partial g_{ij}(x)}{\partial x^k}-\frac{\partial g_{jk}(x)}{\partial x^i}\right),
\ee
respectively.

{\bf Lemma A1}.
\be
\frac{\partial g_{ik}(x)}{\partial x^j}=\Gamma _{ikj}+\Gamma _{kij}.
\ee

\textbf{Lemma A2. }

a)
\be
\frac{\partial \alpha }{\partial x^{j}}=\frac{\epsilon \alpha }{2}%
\frac{\partial g_{st}(x)}{\partial x^{j}}\frac{y^{s}}{\alpha }\frac{y^{t}}{%
\alpha }={\epsilon \alpha }{\Gamma}_{nnj},
\ee
b) \be\frac{\partial \beta }{\partial x^{j}}=\alpha \frac{\partial A_{m}(x)}{%
\partial x^{j}}\tilde{l}^{m},\ee

c)
\bea\frac{\partial }{\partial x^{j}}\left( \frac{F}{\alpha }\right) &=&\frac{%
\partial }{\partial x^{j}}\left( 1+\frac{\beta }{\alpha }\right) =\frac{1}{%
\alpha }\frac{\partial \beta }{\partial x^{j}}-\frac{\beta }{\alpha ^{2}}%
\frac{\partial \alpha }{\partial x^{j}}\nonumber\\
&=&\frac{\partial A_{m}(x)}{\partial
x^{j}}\tilde{l}^{m}-\frac{\epsilon \beta }{\alpha }{\Gamma}_{nnj}.\eea

{\bf Lemma A3}.

a) \be
\hspace{-0.7cm}l_{i}:=\frac{\partial F}{\partial y^{i}}=\frac{\partial \alpha}{\partial y^{i}%
}+\frac{\partial \beta }{\partial y^{i}}=\epsilon \tilde{l}_{i}+A_{i}(x),\ee

b)
\bea
\frac{\partial \tilde{l}_{i}}{\partial x^{k}}&=&\frac{\partial }{\partial
x^{k}}\left( \frac{y_{i}}{\alpha }\right) \nonumber\\
&=&\frac{1}{\alpha ^{2}}\left(
\alpha \frac{\partial y_{i}}{\partial x^{k}}-\frac{\partial \alpha }{%
\partial x^{k}}y_{i}\right) =\frac{1}{\alpha }\frac{\partial g_{im}(x)}{%
\partial x^{k}}y^{m}\nonumber\\
&-&{\epsilon }{\Gamma} _{nnk}\frac{y_{i}}{\alpha }%
={\Gamma} _{ink}+{\Gamma} _{nik}-{\epsilon }{\Gamma} _{nnk}\tilde{l}%
_{i},
\eea

c) \bea
\hspace{-0.4cm}\frac{\partial l_{i}}{\partial x^{k}}&=&\frac{\partial }{\partial x^k}\left(\epsilon \tilde{l}_{i}+A_{i}(x)\right)\nonumber\\
\hspace{-0.4cm}&=&\frac{\partial  A_{i}(x)}{\partial x^{k}}%
+\epsilon \left({ \Gamma} _{ink}+{\Gamma} _{nik}\right) -{\Gamma} _{nnk}\tilde{l}_{i} .
\eea

{\bf Lemma A4}.

\begin{eqnarray}
\frac{\partial \hat{g}_{ij}\left( x,y\right) }{\partial x^{k}}
&=&\epsilon \frac{F}{\alpha }\frac{\partial g_{ij}(x)}{\partial x^{k}}%
+\left( \frac{\partial A_{m}(x)}{\partial x^{k}}\tilde{l}^{m}-\epsilon
\frac{\beta }{\alpha }{\Gamma}_{nnk}\right) h_{ij} \nonumber\\
&+&\left( {\Gamma}_{ink}+{\Gamma}_{nik}\right) \xi_j
+ \left( {\Gamma}_{jnk}+{\Gamma}_{njk}\right)\xi_i
 \nonumber\\
&-&\epsilon{\Gamma}_{nnk}
\left( \tilde{l}_{i}\xi_{j}+\tilde{l}%
_{j}\xi_{i}\right) 
 \nonumber\\
&+&\left( \frac{\partial A_{i}(x)}{\partial x^{k}}A_{j}(x)+\frac{\partial
A_{j}(x)}{\partial x^{k}}A_{i}(x)\right) \nonumber\\
&+&\epsilon \left( \frac{\partial
A_{i}(x)}{\partial x^{k}}\tilde{l}_{j}+\frac{\partial A_{j}(x)}{\partial
x^{k}}\tilde{l}_{i}\right) ,
\end{eqnarray}
where
\begin{equation}
\hspace{-2.5cm}\xi _{i}=\epsilon A_{i}(x)-\left( \beta /\alpha \right) \tilde{l}_{i},
\end{equation}
(compare with  \cite{Bao}, page 295).


\section{Derivation of the general relativistic Friedmann equations}\label{appb}

In order to derive the Friedmann equations in standard general relativity for a homogeneous and isotropic
Universe with metric given by Eq.~(\ref{metr}), we need to calculate the Ricci tensor and the Ricci scalar,
respectively. In Riemannian geometry the Christoffel symbols are defined as
\begin{equation}\label{Gamma for g}
\Gamma _{JI}^{L}=\frac{1}{2}g^{LM}\left( \frac{\partial g_{MI}}{\partial
x^{J}}+\frac{\partial g_{MJ}}{\partial x^{I}}-\frac{\partial g_{IJ}}{%
\partial x^{M}}\right) .
\end{equation}

For the case of the Friedmann-Lemaitre-Robertson-Walker metric (\ref{metr}),
the non-zero Christoffel symbols are
\begin{equation}
\Gamma _{ss}^{0}=aa^{\prime }\delta _{ss},\Gamma _{0s}^{s}=\frac{a^{\prime }}{a}\delta_s^s,s=1,2,3,
\end{equation}
where $a'=\frac{da}{dx^0}.$
After calculating the Christoffel symbols, we can obtain the Riemann tensor,
given by

\begin{equation}
R^A_{BCD}=\dfrac{\partial \Gamma^A_{BD}}{\partial x^C}-
    \dfrac{\partial \Gamma^A_{BC}}{\partial x^D}+\Gamma^E_{BD}\Gamma^A_{EC}
    -\Gamma^E_{BC}\Gamma^A_{ED},
    \end{equation}
where $A,B,C,D=0,1,2,3$,
and its contraction, the Ricci tensor, $R_{BD}=R_{BAD}^{A}$. For the
Friedmann-Lemaitre-Robertson-Walker metric the only nonzero components of
the Ricci tensor are
\begin{equation}
R_{00}=-3\frac{a^{\prime \prime }}{a}
,R_{ij}=\left(aa''+a'^2\right)\delta _{ij}.
\end{equation}

For the Ricci scalar we find
\begin{equation}
R=-6\left( \frac{a^{\prime \prime }}{a}+\frac{a^{\prime 2}}{a^{2}}\right) .
\end{equation}

Hence we arrive to the Friedmann equations, as given by
\begin{equation}
R_{0}^{0}-\frac{1}{2}R=3\frac{a^{\prime 2}}{a^{2}}=3\frac{1}{c^{2}}\frac{%
\dot{a}^{2}}{a^{2}}=\frac{8\pi G}{c^{4}}\rho c^{2},
\end{equation}
where $\dot{a}=\frac{da}{dt}$
and
\bea
R_{k}^{k}-\frac{1}{2}\delta _{k}^{k}R&=&2\frac{a^{\prime \prime }}{a}+\frac{%
a^{\prime 2}}{a^{2}}=\frac{1}{c^{2}}\left( 2\frac{\ddot{a}}{a}+\frac{\dot{a}%
^{2}}{a^{2}}\right) \nonumber\\
&=&-\frac{8\pi G}{c^{4}}p\delta _k^k, k=1,2,3,
\eea
respectively.


\section{The computation of the Ricci curvatures}\label{appc}

{\bf Lemma C1.}
$\Gamma_{nnI}(x,y)|_{y=A(x)}=0$, for all $I\in \{0,1,2,3\}.$

\smallskip

Indeed, observe that \eqref{special A_i} implies
$$
\Gamma_{nnI}(x,y)|_{y=A(x)}=\Gamma_{00I}.
$$

On the other hand, by writing in components formula \eqref{Gamma for g} for the metric $g_{IJ}$ in \eqref{metr}, it is elementary to see that the non-vanishing components are
\begin{equation*}
    \begin{split}
        \Gamma_{0ij} &=\frac{a^2}{c}H\delta_{ij},\\
    \Gamma_{ii0}&=-\frac{a^2}{c}H,\\
    \Gamma_{ij0}&=-\Gamma_{0ij}=-\frac{a^2}{c}H\delta_{ij},\\
    \Gamma_{ii0}&=\Gamma_{i0i}=-\frac{a^2}{c}H.
    \end{split}
\end{equation*}

{\bf Lemma C2.}
We have the following formulas\\

a)\\

     $\phi\left(x^0\right):=1+a\left(x^0\right)\eta\left(x^0\right)$,\\

b)

     \begin{eqnarray*}\phi '&:=&\frac{d\phi \left(x^0\right)}{dx^0}=a\left(\eta '+\mathcal{H}\eta \right)=a\eta '+(\phi-1)\mathcal{H}\nonumber\\
                   &=&(\phi-1)\left(\frac{\eta '}{\eta} +\mathcal{H}\right),\end{eqnarray*}
\hspace{0.2cm} c)
     \begin{eqnarray*}\phi''&:=&\frac{d^2\phi (\left(x^0\right)}{d\left(x^0\right)^2}=\phi'\left(\frac{\eta '}{\eta} +\mathcal{H}\right)\\
    &&+(\phi-1)\left(\frac{\eta ''}{\eta}-\frac{\eta '^2}{\eta ^2}+\mathcal{H}'\right).\end{eqnarray*}

{\bf Lemma C3.}
From the definition of the Ricci tensors, we have
\begin{equation*}
    \hat{R}_{00}=\sum_A\left(\frac{\partial \hat{\gamma}_{00}^A}{\partial x^A}-\frac{\partial \hat{\gamma}_{0A}^A}{\partial x^0}
    +\sum_E
    \hat{\gamma}^E_{00}\hat{\gamma}^A_{EA}-\sum_E \hat{\gamma}^E_{A0}\hat{\gamma}^A_{E0}
    \right).
\end{equation*}

{\bf Lemma C4.}
\begin{equation*}
    \sum_A\frac{\partial \hat{\gamma}_{00}^A}{\partial x^A}=
    \frac{\partial \hat{\gamma}^0_{00}}{\partial x^0}+\sum_{i=1}^3
    \frac{\partial\hat{\gamma}_{00}^i}{\partial x^i}=\frac{\partial \hat{\gamma}^0_{00}}{\partial x^0}=
    \frac{\phi \phi ''-(\phi ')^2}{\phi^2}.
\end{equation*}

{\bf Lemma C5.}

\begin{equation*}
  \hspace{-2.9cm}  \sum_A\frac{\partial \hat{\gamma}_{0A}^A}{\partial x^0}=3\mathcal{H}'+\frac{5}{2}\frac{\phi \phi ''-\left(\phi '\right)^2}{\phi ^2}.
\end{equation*}

{\bf Lemma C6.}\\

a)

\begin{equation*}
\hat{\gamma}_{00}^{E}\hat{\gamma}_{EA}^{A}=\hat{\gamma}_{00}^{0}\left( \hat{%
\gamma}_{00}^{0}+\sum _{i=1}^3\hat{\gamma}_{i0}^{i}\right),
\end{equation*}

b)

\begin{equation*}
\hat{\gamma}_{A0}^{E}\hat{\gamma}_{E0}^{A}=\left(\hat{\gamma}_{00}^{0}\right)^2+\sum _{i=1}^3\left(\hat{\gamma}_{i0}^{i}\right)^2,
\end{equation*}

c)

\begin{equation*}
\hat{\gamma}_{00}^{E}\hat{\gamma}%
_{EA}^{A}-\hat{\gamma}_{A0}^{E}\hat{\gamma}_{E0}^{A}=\sum_{i=1}^3 \hat{\gamma}_{i0}^{i}\left(\hat{\gamma}%
_{00}^{0}- \hat{\gamma}_{i0}^{i}\right) =-3\left( \mathcal{H}^{2}-\frac{\phi ^{\prime 2}}{4\phi
^{2}}\right).
\end{equation*}

{\bf Lemma C7.} Again from the definition of Ricci tensors, we have
\begin{equation*}
    \hat{R}_{ij}=\sum _A \frac{\partial \hat{\gamma}^A_{ij}}{\partial x^A}-\sum_A\left(\frac{\partial \hat{\gamma}^A_{iA}}{\partial x^j}
   +\sum_E\hat{\gamma}_{ij}^E\hat{\gamma}^A_{EA}-
    \sum_E\hat{\gamma}_{iA}^E\hat{\gamma}^A_{Ej}
    \right).
\end{equation*}

{\bf Lemma C8.}
\begin{equation*}
    \sum_A\frac{\partial \hat{\gamma}^A_{iA}}{\partial x^j}=
    \frac{\partial \hat{\gamma}^0_{i0}}{\partial x^j}+
    \sum_{k=1}^3\frac{\partial \hat{\gamma}^k_{ik}}{\partial x^j}=0.
\end{equation*}

{\bf Lemma C9.}
\begin{eqnarray*}
     \sum_A
    \frac{\partial \hat{\gamma}^A_{ij}}{\partial x^A}&=&
    \frac{\partial \hat{\gamma}^0_{ij}}{\partial x^0}+
    \sum_{k=1}^3\frac{\partial \hat{\gamma}^k_{ij}}{\partial x^k}=
    \frac{\partial\hat{\gamma}^0_{ij}}{\partial x^0}\nonumber\\
    &=&\frac{a^{2}}{\phi ^{2}}
    \left[
\phi\mathcal{H}'+2\phi\mathcal{H}^2+\frac{1}{2}\phi''-\frac{\phi'^2}{\phi}
    \right]\delta_{ij}.
\end{eqnarray*}

{\bf Lemma C10.}\\

a)

\begin{equation*}
\sum_{A,E}\hat{\gamma}_{ij}^{E}\hat{\gamma}_{EA}^{A}=\frac{a^{2}}{2\phi ^{2}}\left( \phi ^{\prime }+2\phi \mathcal{H}\right) \left( \frac{%
5}{2}\frac{\phi ^{\prime }}{\phi }+3\mathcal{H}\right)\delta _{ij} ,
\end{equation*}

b)

\begin{equation*}
\sum_{A,E}\hat{\gamma}_{iA}^{E}\hat{\gamma}_{jE}^{A}=\frac{a^{2}}{2\phi ^{3}}\left( \phi ^{\prime }+2\phi \mathcal{H}\right) ^{2}\delta _{ij},
\end{equation*}

c)

\begin{equation*}
\sum_{A,E}
    \left(\hat{\gamma}_{ij}^E\hat{\gamma}^A_{EA}-\hat{\gamma}_{iA}^E\hat{\gamma}^A_{Ej}
    \right)=\frac{a^{2}}{2\phi ^{2}}\left( \phi ^{\prime }+2\phi \mathcal{H}\right) \left( \mathcal{H}+%
\frac{3}{2}\frac{\phi ^{\prime }}{\phi }\right).
\end{equation*}
\smallskip


\end{document}